\documentclass[preprintnumbers, nofootinbib, aps, prd, superscriptaddress, 10pt, showkeys, notitlepage]{revtex4-1}
\usepackage[utf8]{inputenc}
\usepackage{graphicx,amssymb,amsmath,amsthm,amsfonts,epsfig,times,natbib}
\usepackage[usenames,dvipsnames]{color}
\usepackage{epstopdf}
\usepackage{aas_macros,amsmath,amssymb}
\usepackage{tensor}
\usepackage{mathtools}
\usepackage{amsbsy}
\usepackage{bm}

\definecolor{oxfordblue}{rgb}{0.0, 0.13, 0.28}
\definecolor{burgundy}{rgb}{0.5, 0.0, 0.13}
\definecolor{crimsonglory}{rgb}{0.75, 0.0, 0.2}
\definecolor{darkolivegreen}{rgb}{0.33, 0.42, 0.18}
\definecolor{darkblue}{rgb}{0,0,0.5}
\definecolor{richcarmine}{rgb}{0.84, 0.0, 0.25}
\definecolor{darkblue}{rgb}{0,0,0.5} 
\definecolor{venetianred}{rgb}{0.78, 0.03, 0.08}
\definecolor{skobeloff}{rgb}{0.0, 0.48, 0.45}

\def\go{\mathrel{\raise.3ex\hbox{$>$}\mkern-14mu
             \lower0.6ex\hbox{$\sim$}}}
\def\lo{\mathrel{\raise.3ex\hbox{$<$}\mkern-14mu
             \lower0.6ex\hbox{$\sim$}}}

\newcommand{\be}{\begin{equation}}
\newcommand{\ee}{\end{equation}}
\newcommand{\bear}{\begin{eqnarray}}
\newcommand{\eear}{\end{eqnarray}}

\newcommand{\cN}{{\cal N}}

\newcommand{\cC}{{\cal C}}

\begin{document}
\title{Universal Relations and Alternative Gravity Theories}

\begin{abstract}
This is a review with the ambitious goal of covering the recent progress in: 1) universal relations (in general relativity and alternative theories of gravity), and 2) neutron star models in alternative theories. We also aim to be complementary to recent reviews in the literature that are covering the topic of universal relations of neutron stars.
\end{abstract}

\author{Daniela D. Doneva}
\email{daniela.doneva@uni-tuebingen.de}
\affiliation{Theoretical Astrophysics, Eberhard Karls University of T\"ubingen,
T\"ubingen, D-72076, Germany}
\affiliation{INRNE - Bulgarian Academy of Sciences, 1784 Sofia, Bulgaria}

\author{George Pappas}
\email{georgios.pappas@tecnico.ulisboa.pt}
\affiliation{Departamento de F\'isica, CENTRA, Instituto Superior
T\'ecnico, Universidade de Lisboa, Avenida Rovisco Pais 1,
1049 Lisboa, Portugal}

\date{{\today}}

\pacs{
 04.40.Dg, 
 04.50.Kd, 
 04.80.Cc, 
 04.25.Nx, 
 97.60.Jd  
}

\maketitle

 \tableofcontents
 
\section{Introduction}
\label{sec:intro}

\begin{figure}[htb]
\includegraphics[width=0.495\textwidth]{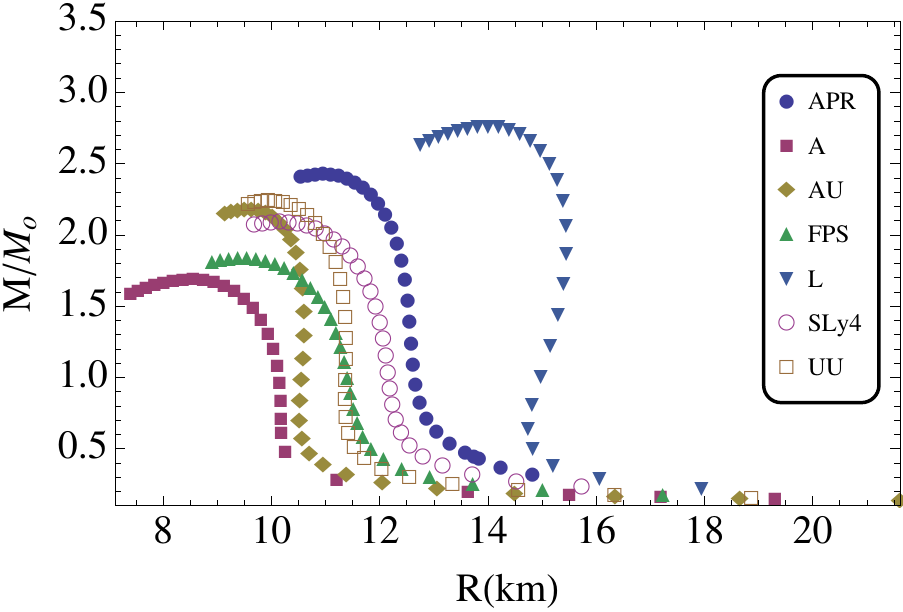}
\caption{ Mass-Radius relation of non-rotating neutron stars for different equations of state.}
\label{fig:eoss}
\end{figure}

Neutron stars are complicated astrophysical objects with a lot of physics involved in their description.  
But when it comes to their overall structure and their stationary properties, gravity is the most important player. To determine the structure of a neutron star one specifies an equation of state and then solves the Einstein field equations together with the hydrostatic equilibrium equation (see \cite{friedman2013rotating,Paschalidis2016arXiv}). The resulting models and their structure depend on the specific choice of the equation of state. At the moment there exists a large variety of realistic equations of state that come from different nuclear physics models (a result of our lack of knowledge with respect to the properties of matter at supra-nuclear densities), which in turn result to quite a large variety in neutron star models as Figure \ref{fig:eoss} shows. Specifying the equation of state is therefore a very hot topic from both astrophysical perspective as well as nuclear physics perspective.

This large variety was always considered to be a natural outcome of having a lot of degrees of freedom available in the description of the equation of state and for this reason neutron stars were not considered to be useful objects for testing our theories of gravity. The conventional wisdom was that neutron stars were infested by matter that would hide any possible signal coming from modifications from general relativity. Nevertheless, there always existed an interest in finding ways to describe neutron stars in ways that were not very sensitive on the specific choice of the equation of state. The motives behind such attempts are twofold. On the one hand, having observables that are insensitive to the specifics of the equation of state can be very advantageous in our attempts to measure astrophysically the various properties of neutron stars, by reducing for example modelling uncertainties. This can be also important in solving the inverse problem of determining the equation of state from measurements of the bulk properties of neutron stars. 
On the other hand, such observables would open the window to observing deviations that are due to gravity and possible modifications to general relativity and not the equation of state uncertainty. All the more so, since neutron stars can be in some cases a much more suitable object for testing such modifications, as we shall discuss shortly.   

Alternative theories of gravity attracted significant attention in the past decades. The reasons for this come both from theory and observations. Modifications of Einstein's theory of gravity for example are often employed as alternative explanation for the dark mater phenomenon or the accelerated expansion of the universe. The idea is that instead of attributing the accelerated expansion to unknown constituents of the Universe with rather unusual and strange properties such as dark energy, one can attribute it to our lack of understanding of gravity. Besides the observations, there are strong theoretical motivations for modifying general relativity. The standard Hilber-Einstein action is by no means the only possible one but instead the simplest. That is why different generalizations are possible coming for example from the theories trying to unify all the interactions. They predict the existence of a scalar field that can be considered as a new mediator of the gravitational interaction in addition to the spacetime metric. On the other hand the quantum corrections in the strong field regime which are introduced in order to make the theory renormalizable or to cure the singularities in the solutions, naturally lead to the fact that the Einstein-Hilbert action is supplemented with higher order terms. Last but not least, as the experience shows, studying in detail theories of gravity more general than Einstein's theory can often give as a better insight to general relativity itself.

Every viable theory of gravity should be consistent with 
the observations at all scales and regimes \cite{Will:2014xja}. The weak field experiments give strong constraints on the parameters of the theory, but  the strong field regime is essentially unconstrained and leaves a lot of space for modifications. Moreover, there are theories of gravity that are completely indistinguishable from general relativity for weak fields but can lead to large deviations for strong fields. Therefore, constraining the strong field regime of gravity or even detecting possible deviations from pure Einstein's theory is a difficult but very important task. 

Neutron stars and black holes in alternative theories of gravity have been explored for several decades because they offer the possibility to test the strong field regime of gravity at astrophysical scales. Black holes, though, have an important disadvantage because of the no-hair theorems. According to these theorems, the black hole solutions in some classes of alternative theories of gravity are the same as in general relativity\footnote{The perturbations, though, could be different and thus used for testing alternative theories of gravity.}. Naturally, this poses obstacles to testing such theories of gravity via the astrophysical observations of black holes. Neutron stars on the other hand do not fall in the scope of the no-hair theorems because of the presence of matter. Thus, alternative theories of gravity can lead to large deviations from Einstein's theory. The neutron star matter, though, is a double edged sword -- indeed it offers the possibility to have compact star solutions different from pure general relativity, but on the other hand the uncertainties in the high density nuclear matter equation of state are large. Moreover, there is a degeneracy between effects coming from modifying the equation of state and the theory of gravity. That is why testing the strong field regime of gravity via neutron star observations is also a subtle task. In order to address these problems one has to build a dense net of neutron star models and astrophysical predictions in various alternative theories that can help us either break the degeneracy or find effects that are stronger pronounced than the equation of state uncertainty. 
%
Again, as in general relativity, universal relations can play a significant part here by taking the subtleties of the equation of state out of the picture so that we can identify effects and deviations that are only due to modifications in gravity.

Neutron star solutions (both static and rotating) have been constructed in various alternative theories of gravity. The literature on the subject is vast and we do not aim 
to cover the subject thoroughly. 
This is why instead of reviewing the whole subject in detail, we will concentrate on certain classes of alternative theories of gravity. Since the goal r 
is to cover both neutron stars in alternative theories and universal relations, a natural choice would be to focus mainly on those theories, in which universal relations have been derived. In the present review we will concentrate on scalar-tensor theories (STT), $f(R)$ theories, Einstein-dilaton-Gauss-Bonnet (EdGB) theories and the Chern-Simons (CS) theories of gravity. As a matter of fact they are amongst the most natural and widely used generalizations of Einstein's gravity and most of the neutron star studies in the literature have been done exactly in these alternative theories. In addition all of them fall into the same class of modifications of Einstein's theory, i.e. theories for which dynamical scalar fields are included as mediators of the gravitational interaction in addition to the metric tensor. A very nice recent review that covers a much larger spectrum of theories can be found in \cite{Bertietal2015CQG}. \\

In what follows we will start by discussing in section \ref{sec:past1} the history and present status of universal relations for neutron stars in general relativity. We will proceed in section \ref{sec:past2} to briefly review neutron stars in various modifications of general relativity. In section \ref{sec:past3} we will discuss the current status on the various extensions of the universal relations to alternative theories of gravity. In section \ref{sec:future} we will talk about the work that needs to be done or could be done in the future as well as the various challenges that we will face in extending our current results. Finally we will close with some brief overview of the Chapter.

We will use geometric units ($G=c=1$) throughout, unless it is specifically mentioned otherwise.

\section{Review of past work}
\label{sec:past}

\subsection{Universal relations in GR}
\label{sec:past1}

\subsubsection{Prehistory}
\label{sec:pre}

Even though the subject of neutron star universal relations, i.e., equation of state independent or insensitive relations, has received a lot of attention in recent years, as was mentioned above it has a longer history. Some first results have been presented in the literature already since the 90s when it was recognised in \cite{Lattimer1989ApJ} that the binding energy $\textrm{BE}\equiv(\cN m_n-M)$ of a neutron star, where $\cN$ is the nucleons number and $m_n$ the corresponding nucleons mass, expressed in terms of the stellar compactness $\cC\equiv M/R$ (the mass here is in geometric units) is insensitive to the equation of state. A later improved expression between the two quantities was given by \cite{Lattimer2001ApJ} and reads,
\be \textrm{BE}/M=(0.60\pm0.05)\cC(1-\cC/2)^{-1}. \ee
Such an expression is motivated by the fact that analytic solutions like that of the incompressible fluid ($\rho=\textrm{const.}$), the Buchdahl solution (constructed using the equation of state $\rho=12\sqrt{p_*P}-5P$, where $p_*$ is some constant with dimensions of pressure) or the Tolman VII model (which is constructed by assuming that inside the star the density goes like $\rho=\rho_c\left(1-(r/R)^2\right)$, where $\rho_c$ is the central density of the star), give similar expansions in terms of the compactness (see \cite{Lattimer2001ApJ}).

On a different direction, on the topic of asteroseismology, in the late 90s it was recognised by \cite{Andersson1996,Andersson1998MNRAS} that the $f$-mode and $w$-modes exhibit some equation of state independent behaviour. In particular it was shown that the $f$-mode frequency (in kHz) is related to the average density of a neutron star following the relation 
\be \frac{\omega_f}{2\pi}=0.78+1.635\left(\frac{M}{R^3}\right)^{1/2}, \label{eq:fmode1}\ee 
where different equations of state have a relative small spread around this common fit. In the above expression the mass is measured in units of $1.4M_{\odot}$ and the radius of the star in units of 10km. The $f$-mode is related to fluid motion inside the star that takes place in dynamic time scales. It is reasonable therefore to assume that the $f$-mode frequency will be related to the characteristic dynamic time which is proportional to the square root of the average density, $\tau_{dyn}\sim \sqrt{\bar{\rho}}$. It was also shown that the damping time (in s) of the $f$-mode, when scaled in the right way, is related to the compactness in a linear way,
\be \left(\tau_f \frac{M^3}{R^4}\right)^{-1}=22.85-14.65\, \cC,  \label{eq:fmodeDamp1} \ee
where again the masses and radii are measured using the same units as above (this is the case also for $\cC$). Finally for the first $w$-mode the corresponding relations for the frequency (in kHz) and the damping time (in ms) are,
\bear 
          R \left(\frac{\omega_w}{2\pi}\right) &=&20.92-9.14\, \cC,\\
         M (\tau_w)^{-1}&=&5.74+103\, \cC-67.45\, \cC^2,
\eear
where the mass and radius units are as before. 
 
Returning to the topic of the structure of neutron stars, in the 2000s it was found by \cite{Lattimer2001ApJ,Bejger2002A&A,Lattimer2005ApJ} that there is an equation of state insensitive relation between a neutron star's moment of inertia and it's compactness,
\be \frac{I}{MR^2}=(0.237\pm0.008)(1+2.84\cC+18.9\cC^4), \ee
where the compactness is expressed in terms of the mass in geometric units. This expression is again motivated by the behaviour of the moment of inertia in the various analytic models, such as the Tolman VII model.
Finally, another instance where relations of this sort were studied is in \cite{Urbanec:2013fs} where for slowly rotating neutron stars, relations between the reduced moment of inertia $I/(MR^2)$ and the inverse compactness $x=1/(2\cC)$ where derived in addition to a new relation between the reduced quadrupole $\tilde{q}\equiv QM/J^2$ and $x$. This last relation could be of additional interest since it distinguishes between neutron stars and quark stars. We give here a fitting formula for neutron stars' $\tilde{q}$ in terms of the compactness,
\be \tilde{q}= -0.2588 \cC^{-1} +0.2274 \cC^{-2} +0.0009528 \cC^{-3} -0.0007747 \cC^{-4},\ee
as it is given in \cite{YagiYunes2016arXiv}. 

One will notice a common theme in all of these relations, which is that the various quantities are expressed in terms of the compactness. Using the compactness was a sensible first choice for parameterising the various properties of neutron stars, since it is a dimensionless quantity that is representative of the overall structure of the star. It was a sensible choice also because, in results from analytic models, relations between various quantities would usually be expressed in terms of the compactness which would be a measure of how relativistic a particular model is.\footnote{The newtonian limit is for $\cC\rightarrow 0$, while the relativistic limit is for $\cC\rightarrow 1/2$.} One could think of these results as precursors of the later proliferation of universal relations. In what follows we will give an incomplete list of some of these relations. The interested reader can complement the material presented here with a recently published review on the subject \cite{YagiYunes2016arXiv}.

\subsubsection{I-Love-Q}
\label{sec:iloveq}

Yagi~\&~Yunes in 2013 \cite{Yagi:2013bca,YagiYunes2013PRD}, while studying slowly rotating and tidally deformed neutron and quark stars, found that there exist relations between the various pairs of the following three quantities, the moment of inertia $I$, the quadrupole moment $Q$ and the quadrupolar tidal deformability or Love number $\lambda$, which are insensitive to the choice of the equation of state for both neutron as well as quark stars. We will present here a very brief outline of their calculation which comprises of two parts, first the calculation of the slowly rotating models and then the calculation of the tidally deformed models. 

{\bf Initial I-Love-Q formulation:} The slowly rotating models were calculated using the Hartle-Thorne slow rotation formalism \cite{Hartle1967ApJ,HT1968ApJ}. In order to construct a slowly rotating model, one first calculates a non-rotating spherically symmetric model that has some mass $M_*$ and radius $R_*$. The structure of the star is described by the Tolman-Oppenheimer-Volkoff (TOV) equation, while the metric has the form 
\be ds^2=-e^{\nu(r)}dt^2+e^{\lambda(r)}dr^2+r^2(d\theta^2+\sin^2\theta d\phi^2), \ee
where we can define the mass function $M(r)=r\left(1-e^{-\lambda(r)}\right)/2$. The spacetime outside the star's surface is the Schwarzschild spacetime. From the mass and the radius of the star one calculates an angular frequency scale $\Omega_K=\sqrt{M_*/R_*^3}$ which is then used to introduce a small expansion parameter $\epsilon\equiv\Omega_*/\Omega_K$ that characterises rotation. Then one introduces corrections to the geometry due to rotation in the form of perturbations up to second order in the rotation, 
\bear ds^2&=&-e^{\nu(r)}\left[1+2\epsilon^2\left(h_0+h_2P_2\right) \right]dt^2+\frac{1+2\epsilon^2\left(m_0+m_2P_2\right)/\left(r-2M(r)\right)}{1-2M(r)/r}dr^2\nonumber\\
&&
+r^2\left[1+2\epsilon^2K_2P_2 \right]\left[d\theta^2+\sin^2\theta \left(d\phi-\epsilon (\Omega_K-\omega_1) dt\right)^2\right], \eear
where $P_2=(3 \cos^2\theta-1)/2$ is the second order Legendre polynomial and $\omega_1\equiv\Omega_K-\omega$ is the angular velocity of the fluid relative to the local inertial frame. The equations for the perturbations are then solved on top of the background spherical non-rotating solution.  
The resulting perturbed configuration is deformed due to rotation and the deformation of the surfaces of constant density has the form, $\bar{r}=r+\epsilon^2\left(\xi_0(r)+\xi_2(r)P_2\right)$. Since all the perturbations scale with the angular velocity of the star, in practice one has only to calculate the model that rotates at $\Omega_K$ for $\epsilon=1$ and then all the models with values of $\epsilon \in[0,1]$ follow from that. Of course we should note that the results are accurate for values of $\epsilon$ that correspond to models that rotate as fast as a few milliseconds \cite{Berti:2004ny}.  

The various quantities of the rotating model will scale with rotation as, $\Omega=\epsilon \Omega_K$, $J=\epsilon J_K$, $R=R_*+\epsilon^2\delta R_K$, $M=M_*+\epsilon^2\delta M_K$, and $Q=\epsilon^2 Q_K$. The moment of inertia for a rigidly rotating configuration is defined as $I\equiv J/\Omega=J_K/\Omega_K$ which means that for a calculation at the order of $\epsilon$ in $J$, the moment of inertia is independent of $\epsilon$. The angular momentum and the quadrupole moment of the configuration can be evaluated from the form of the metric outside the star and in particular by it's asymptotic expansion in the appropriate coordinate system following Thorne's prescription \cite{ThorneRevModPhys1980,Quevedo1990}.\footnote{There will be a further discussion on multipole moments when we talk about the 3-hair relations.} Therefore, the angular momentum can be calculated by the fact that outside the star the metric function $\omega_1$ has the form 
\be \omega_1(r)=\Omega_K-\frac{2J_K}{r^3},\ee
where $J$ is a constant calculated by the matching of the interior metric to the exterior metric at the surface of the star. Similarly, in the exterior of the star the $h_2(r)$ perturbation functions has the form 
\be
  h_2(r)=\frac{J_K^2}{M_* r^3}\left(1+\frac{M_*}{r} \right)+A Q_2^2\left(\frac{r}{M_*}-1\right), 
\ee
where $Q_2^2(x)$ is the associated Legendre polynomial of the second kind\footnote{The polynomial here is defined as $Q^2_2(x)=\frac{3}{2} \left(x^2-1\right) \ln \left(\frac{x+1}{x-1}\right)+\frac{5 x-3 x^3}{x^2-1}$} and $A$ is an integration constant that is determined by the matching of the interior solution to the exterior solution. From the asymptotic expansion of $-(1+g_{tt})/2$, one can read the quadrupole as the coefficient in front of the $P_2/r^3$ term, which is $Q_K=-J_K^2/M_*-\frac{8}{5}AM_*^3$.

In a similar way to the slowly rotating models, the tidally deformed models are assumed to be slightly perturbed from sphericity due to the presence of an external deforming quadrupolar field. For an $l = 2$, static, even-parity perturbation, the perturbed metric will take the form, 
\be ds^2=-e^{\nu(r)}\left[1+2h_2P_2 \right]dt^2+\frac{1+2m_2P_2}{1-2M(r)/r}dr^2+r^2\left[1+2K_2P_2 \right]\left(d\theta^2+\sin^2\theta d\phi^2\right),  \label{eq:tid1}\ee
where we have introduced again the perturbation functions and assumed zero rotation \cite{Hinderer2008ApJ}. The tidal Love number is defined to be the deformability, i.e., the response, of a configuration for a given external deforming force. Specifically, for a given external tidal field $\mathcal{E}^{\textrm{tid}}$ that produces a quadrupolar deformation of the star $Q^{\textrm{tid}}$, the tidal Love number is defined to be, 
\be \lambda\equiv-\frac{ Q^{\textrm{tid}}}{\mathcal{E}^{\textrm{tid}}}. \ee
In addition to this definition, one can define the tidal apsidal constant and the dimensionless tidal Love number as
\bear k_2&\equiv&\frac{3}{2}\frac{\lambda}{R_*^5},\nonumber\\
         \bar{\lambda}&\equiv&\frac{\lambda}{M_*^5}=\frac{2}{3}k_2 \cC^{-5}, \label{eq:lovenumb}
\eear         
respectively. The quadrupolar response and the external quadrupole tidal field can both be extracted from the form of the metric perturbation $h_2(r)$ outside the star. One should be careful though in this case because the exterior to the star is not an asymptotically flat spacetime as it was in the rotating case. Instead, the asymptotic behaviour is determined by the external quadrupolar field that produces the deformation. In the exterior of the star the $h_2(r)$ perturbation function has in this case the form, 
\be
  h_2(r)=2c_1 Q_2^2\left(\frac{r}{M_*}-1\right)+c_2 \left(\frac{r}{M_*}\right)^2\left(1-\frac{2M_*}{r}\right), \label{eq:tid2}
\ee
where again the constants $c_1$ and $c_2$ are integration constants that are determined from the matching conditions at the surface of the star. To identify the external tidal field and the quadrupolar response of the star, one again expands $-(1+g_{tt})/2$ in powers of $r$. In this case, the expansion will have additional terms with positive powers of $r$ due to the external tidal field.\footnote{We should also note that the expansion in this case is not at infinity. It takes place at a buffer region outside the surface of the star and inside a radius given by the external field's characteristic curvature radius.} The expansion will have the form,
\be
-\frac{1+g_{tt}}{2}=-\frac{M_*}{r}-\frac{Q}{r^3}P_2+\cdots%
                               +\frac{1}{3}r^2 \mathcal{E}^{\textrm{tid}}P_2+\cdots,
\ee
and by comparing this to the corresponding expansion of $-(1+g_{tt})/2$, where $g_{tt}$ is evaluated from eqs. (\ref{eq:tid1}) and (\ref{eq:tid2}), one finds that $Q^{\textrm{tid}}=-\frac{16}{5}M_*^3c_1$ and $\mathcal{E}^{\textrm{tid}}=3M_*^{-2}c_2$. Therefore, from the definition of the tidal love number and the tidal apsidal constant, we have in terms of the integration constants $c_1$ and $c_2$ that,
\be \lambda=\frac{16}{15}M_*^5 \frac{c_1}{c_2}\Rightarrow \bar{\lambda}=\frac{16}{15}\frac{c_1}{c_2},\;\textrm{ and }\; k_2= \frac{8}{5} \frac{c_1}{c_2} \cC^5,\ee
where the ratio $c_1/c_2$ depends on the compactness and the quantity $y=R_*h_2'(R_*)/h_2(R_*)$ at the surface of the star. 

\begin{figure}[htb]
\begin{center}
\begin{tabular}{ccc}
\includegraphics[width=5.6cm,clip=true]{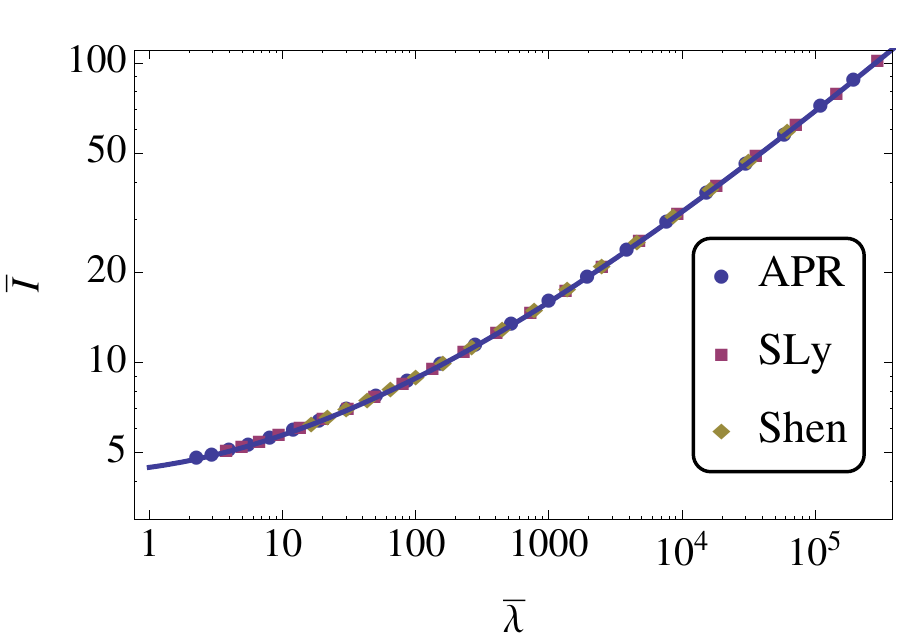}&\includegraphics[width=5.6cm,clip=true]{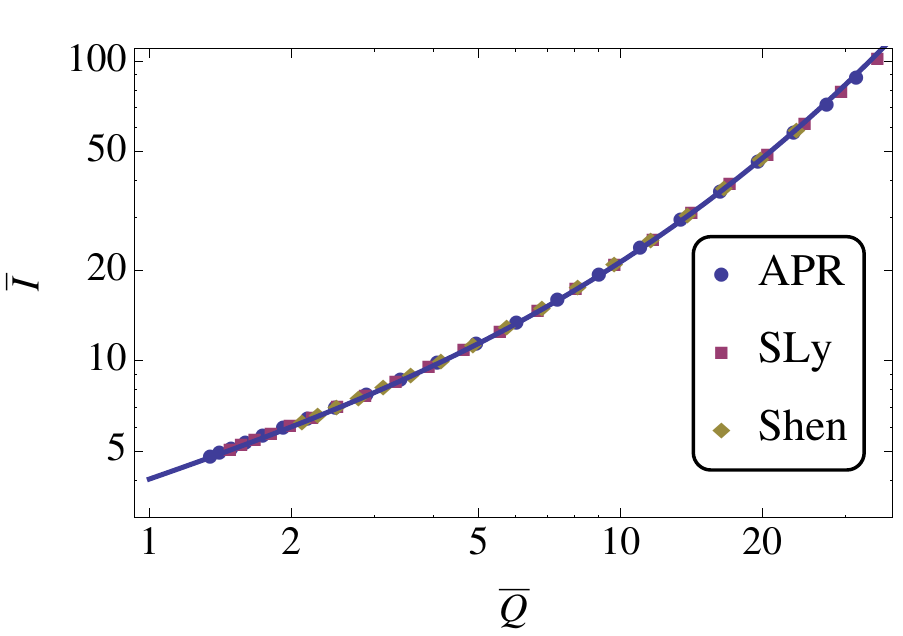} &\includegraphics[width=5.6cm,clip=true]{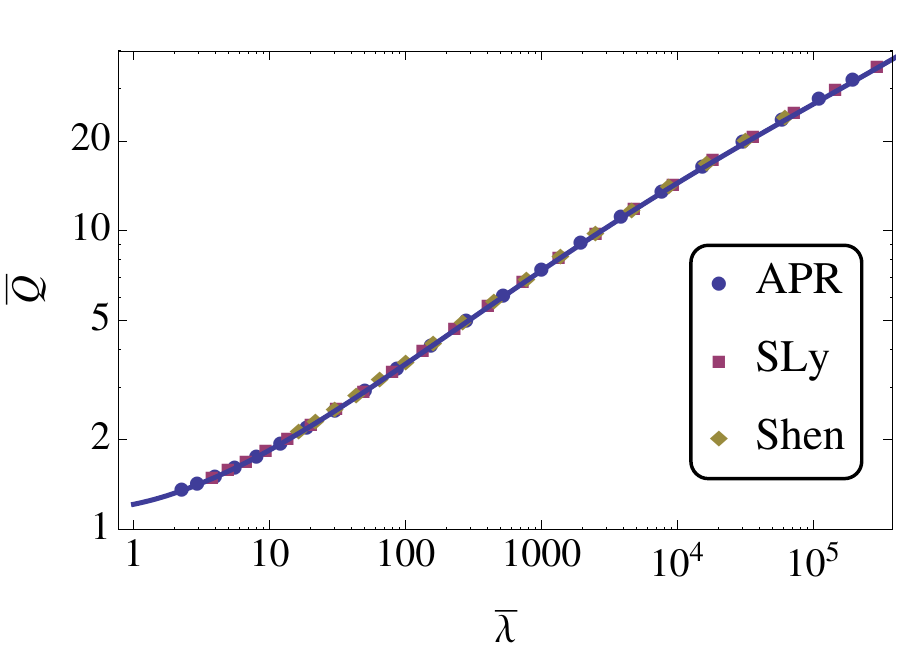} 
\end{tabular}
\caption{\label{fig:iloveq} I-Love, I-Q and Q-Love plots for three typical equations of state. The solid lines correspond to the fits given in Table \ref{tab:iloveq}.
}
\end{center}
\end{figure}

Having all the relevant quantities at hand, Yagi~\&~Yunes calculated sequences of neutron star models using various cold realistic equations of state\footnote{See also \cite{Lattimer2013ApJ} where the results were extended to a wider range of equations of state. We remind here that these results apply to quark stars as well.} and found that the quantities, normalised moment of inertia: $\bar{I}\equiv I/M_*^3$, normalised Love number: $\bar{\lambda}\equiv\lambda/M_*^5$, and normalised quadrupole: $\bar{Q}\equiv -Q/(\chi^2M_*^3)$,\footnote{The minus sign here is due to the fact that rotating neutron stars tend to be oblate due to rotation which gives a negative value for the quadrupole. On the other hand, an object that is prolate has a positive quadrupole.} where $\chi=J/M_*^2$, are related in an equation of state independent way following the relation 
\be \ln y_i=a_i+b_i \ln x_i + c_i (\ln x_i)^2 + d_i (\ln x_i)^3 +e_i (\ln x_i)^4 , \label{eq:iloveq} \ee
where the different coefficients depend on the pair of quantities to be related and are given in Table~\ref{tab:iloveq}. 
%
\begin{table}[htb]
\begin{centering}
\begin{tabular}{cccccccc}
\hline
\hline
\noalign{\smallskip}
$y_i$ & $x_i$ &&  \multicolumn{1}{c}{$a_i$} &  \multicolumn{1}{c}{$b_i$}
&  \multicolumn{1}{c}{$c_i$} &  \multicolumn{1}{c}{$d_i$} &  \multicolumn{1}{c}{$e_i$}  \\
\hline
\noalign{\smallskip}
$\bar{I}$ & $\bar{\lambda}$ && 1.496 & 0.05951  & 0.02238 & $-6.953\times 10^{-4}$ & $8.345\times 10^{-6}$\\
$\bar{I}$ & $\bar Q$ && 1.393  & 0.5471 & 0.03028  & 0.01926 & $4.434 \times 10^{-4}$\\
$\bar Q$ & $\bar{\lambda}$ && 0.1940  & 0.09163 & 0.04812  & $-4.283 \times 10^{-3}$ & $1.245\times 10^{-4}$\\
$\bar{\delta M}$ & $\bar{\lambda}$ && -0.703  & 0.255 & -0.045  & $-5.707 \times 10^{-4}$ & $2.207\times 10^{-4}$\\
\noalign{\smallskip}
\hline
\hline
\end{tabular}
\end{centering}
\caption{Numerical coefficients for the fitting formula given in eq.~\eqref{eq:iloveq}. This is an updated version of the table as it appears in \cite{YagiYunes2016arXiv}, where a very wide range of equations of state has been taken into account. In the table we also give the universal relation for the mass correction $\bar{\delta M}$ presented in \cite{Reina2017arXiv170204568R} .}
\label{tab:iloveq}
\end{table}
%
The relations presented here, i.e., the I-Love, I-Q and Q-Love relations, are equation of state independent in the sense that for any given realistic equation of state, the calculated quantities follow the corresponding fits given by eq.~\eqref{eq:iloveq} and Table~\ref{tab:iloveq} to an accuracy better than $\mathcal{O}(1\%)$ in the range of applicability of the fit, which is for neutron stars with masses slightly less than $1M_{\odot}$ up to the maximum mass of the given equation of state. The fits with the data from some typical equations of state are given in Figure \ref{fig:iloveq}.

As it was previously mentioned, the moment of inertia by definition (within the slow rotation scheme) does not depend on the rotation parameter $\epsilon$, while the normalised quadrupole $-Q/(\chi^2M_*^3)$ is defined in such a way that the rotation parameter is scaled out. Similarly, the tidal love number is defined in such a way that it is independent of the strength of the external tidal field and characterises the given neutron star model. This means that essentially the quantities $\bar{I}$, $\bar{\lambda}$, and $\bar{Q}$ form one parameter families characterised by the central density of the neutron star models. We should also note that the normalisation of the various quantities in the initial I-Love-Q formulation was performed with the masses $M_*$ of the corresponding spherical configurations. From a practical perspective this could be considered to be problematic, since when one observes a neutron star and measures it's mass, that mass is not the mass $M_*$ of the spherical configuration but instead it is the mass $M=M_*+\epsilon^2\delta M_K$. Nevertheless, in the slowly rotating and small deformations case $\delta M_K/M_*$ is of the order of 10\% and one can also assume values of $\epsilon$ up to 10\%, which make any deviations from the given relations to be of the order of $10^{-3}$. We will revisit this point later on when we will discuss rapidly rotating neutron stars. Nevertheless, recent work \cite{Reina2017arXiv170204568R} has shown that the mass correction, normalised as $\bar{\delta M}\equiv M_*^3 \delta M_K/J_K^2$, also follows a universal relation, extending in this way the original family of I-Love-Q relations to include the mass correction as well.   

{\bf Extensions beyond the initial I-Love-Q formulation:} The discussion so far has been about unmagnetised isolated neutron stars, in the sense that we have taken into account deformations that are only due to rotation or due to some static external tidal field, the latter of which is not what one would expect if we assumed that the star was part of some binary for example and the source of the tidal field were a companion star. The effects of adding a magnetic field were studied in \cite{Haskell2014MNRAS} while the effects of having the neutron star being part of a binary system were studied in \cite{Maselli2013PRD}. 

In \cite{Haskell2014MNRAS}, Haskell et al. investigated three magnetic field configurations. The first was a purely poloidal magnetic field configuration, the second was a purely toroidal magnetic field configuration and finally the third was a twisted-torus configuration. It is known that magnetic fields can cause deformations to neutron stars which depend on the magnetic field configuration. It is also known that neutron stars can be strongly magnetised with magnetic fields at the surface as high as $10^{12} G$ for pulsars and $10^{15} G$ for magnetars, while the magnetic field in the interior can be even stronger than that on the surface. Therefore, for high enough magnetic fields and sufficiently slow rotation, the deformations due to magnetic fields can dominate those due to rotation. 

When the magnetic field is purely poloidal, in the Newtonian limit and for an $n=1$ rotating polytrope, the normalised quadrupole $\bar{Q}$ can be expressed in terms of the normalised moment of inertia $\bar{I}$ as, 
\be \bar{Q} \approx 4.9 \bar{I}^{1/2} +10^{-3} \bar{I} \left(\frac{B_p}{10^{12} G}\right)^2 \left(\frac{P}{s}\right)^2, \ee
where the first term is the induced quadrupole due to rotation, while the second term is the induced quadrupole due to the magnetic field, $B_p$ is the field at the pole and $P$ is the rotation period of the star. Similarly for a purely toroidal magnetic field the reduced quadrupole is,  
\be \bar{Q} \approx 4.9 \bar{I}^{1/2} -3\times10^{-5} \bar{I} \left(\frac{\langle B \rangle}{10^{12} G}\right)^2 \left(\frac{P}{s}\right)^2, \ee
where $\langle B \rangle$ is the field average over the volume of the star.\footnote{We should note that the numerical coefficients in these expressions depend on the specific configuration.} There are a few things that one notices from the above equations. The first is that for the purely toroidal case the magnetic field tends to make the star more prolate in contrast to rotation and the effect the magnetic field has in the purely poloidal case. The second is that the induced quadrupole is proportional to the square of the product $B \times P$ and therefore the effect is more prominent for larger periods, i.e., slower rotation rates. Finally, one notices that the effect of the magnetic field is suppressed in relation to the effect of rotation by factors of $10^{-3}$ and $10^{-5}$ respectively. These observations also hold in the relativistic case studied in \cite{Haskell2014MNRAS} where it was found that the purely toroidal and purely poloidal magnetic field configurations give an approximately universal relation between $\bar{Q}$ and $\bar{I}$ which agrees with the unmagnetised case.

However, the purely toroidal or purely poloidal configurations are both known to be dynamically unstable (although, the crust could stabilise such configurations as long as it does not break). For this reason a more realistic configuration was also studied in \cite{Haskell2014MNRAS}, that of a twisted torus, where both toroidal and poloidal components of the magnetic field are present. In this case in addition to the strength of the magnetic field and the rotational period, the results also depend on the ratio of the toroidal-to-total magnetic field energy, i.e., on the particulars of the configuration of the magnetic field, while the $\bar{I}-\bar{Q}$ relation also acquires some equation of state dependence. Nevertheless, the I-Love-Q universality is preserved as long as the magnetic field is not too strong, i.e., $B\lo 10^{12}G$, and the neutron star is not rotating too slowly, i.e., $P\lo 10s$.

In \cite{Maselli2013PRD} Maselli et al. investigate what are the effects on the $\bar{I}-\bar{\lambda}$ relation if one were to assume the more realistic situation of having dynamic tides caused by a companion star in a binary system. The method used to model tidal deformations in compact binaries was the Post-Newtonian-Affine approach. It was shown that the $\bar{I}-\bar{\lambda}$ relation is not the same as the one in the stationary case and that the new relation depends on the inspiral frequency. However, for any given inspiral frequency the $\bar{I}-\bar{\lambda}$ relation is insensitive to the equation of state with an accuracy of a few \%. 
%
\begin{table}[htb]
\begin{centering}
\begin{tabular}{ccccccc}
\hline
\hline
\noalign{\smallskip}
$f_{GW}$ &&  \multicolumn{1}{c}{$a_i$} &  \multicolumn{1}{c}{$b_i$}
&  \multicolumn{1}{c}{$c_i$} &  \multicolumn{1}{c}{$d_i$} &  \multicolumn{1}{c}{$e_i$}  \\
\hline
\noalign{\smallskip}
170 && 1.54 & $-3.73\times10^{-2}$  &$5.49\times10^{-2}$ & $-4.78\times 10^{-3}$ & $1.87\times 10^{-4}$\\
300 && 1.58  & $-6.53\times10^{-2}$ & $6.26\times10^{-2}$  & $-5.68\times 10^{-3}$ & $2.26\times 10^{-4}$\\
500 && 1.60  & $-8.34\times10^{-2}$ & $6.83\times10^{-2}$  & $-6.39 \times 10^{-3}$ & $2.59\times 10^{-4}$\\
700 &&  1.64  & $-1.18\times10^{-1}$ & $7.89\times10^{-2}$  & $-7.69\times 10^{-3}$ & $3.18\times 10^{-4}$\\
800 &&  1.68  & $-1.46\times10^{-1}$ &$8.76\times10^{-2}$ & $-8.77\times 10^{-3}$ & $3.68\times 10^{-4}$\\
any &&  1.95  & $-3.73\times10^{-1}$ &$1.55\times10^{-2}$ & $-1.75\times 10^{-3}$ & $7.75\times 10^{-4}$\\
\noalign{\smallskip}
\hline
\hline
\end{tabular}
\end{centering}
\caption{Numerical coefficients for the binary $\bar{I}-\bar{\lambda}$ fitting formula, given in eq.~\eqref{eq:iloveq} with $y_i=\bar{I}$ and $x_i=\bar{\lambda}$. The different rows correspond to different inspiral frequencies, while the final row is an overall fit. The table is from Maselli et al. \cite{Maselli2013PRD}.
}
\label{tab:iloveBinary}
\end{table}
%
The fits for the $\bar{I}-\bar{\lambda}$ relations for the different gravitational wave frequencies $f_{GW}$ that the system would emit, that are related to the binary inspiral frequency as $f_{GW}=2f$, are given in Table \ref{tab:iloveBinary}. These fits are accurate to within $2\%$ for every frequency, while there is also an overall fit given which is valid up to a gravitational wave frequency of $\sim 900$Hz and accurate to within $5\%$ for any frequency in that range.

The tidal fields and the corresponding tidal deformations that we have discussed so far are of the so called ``gravito-electric'' type and they are the relativistic extensions of their Newtonian counterparts. In general relativity though there can exist ``gravito-magnetic'' tidal fields and deformations which result to gravitomagnetic Love numbers. In order to allow for the fullest possible effect in the response of a compact object under a gravitomagnetic tidal field one needs to go beyond configurations that are in strict hydrostatic equilibrium (no internal motion of the fluid), i.e., allow for the fluid to be in an irrotational state. This is because one expects that the external gravitomagnetic tidal field in a binary system would also drive internal fluid motion in the star. Allowing for irrotational fluid flows inside the stars gives a dramatically different behaviour for the Love numbers with respect to the restricted hydrostatic case. The magnetic Love numbers for irrotational stars were studied by Landry~\&~Poisson in \cite{Landry2015PhRvD} (where one can find further references to previous work). Furthermore Delsate explored in \cite{Delsate:2015wia} the existence of a universal relation between the $\ell=2$ gravitomagnetic Love number and the moment of inertia. 
As in the case of gravitoelectric Love numbers, the gravitomagnetic Love numbers can be defined as the coefficients $\sigma_{\ell}$ that relate the tidally induced response for a given external tidal field. 
Delsate found that in the case of irrotational stars there exists a $|\bar{k}_2^{mag}|-\bar{I}$ universal relation, where $\bar{k}_2^{mag}=\bar{\sigma}_2 (2\cC)^4$ is analogous to the tidal apsidal constant for the gravitoelectric Love number, which is equation of state independent with a variation less than $5\%$.

Another significant extension of the initial I-Love-Q analysis (performed in slow rotation) was the investigation of the $\bar{I}-\bar{Q}$ relation for rapidly rotating neutron stars with rotation rates as high as that of the mass shedding limit. Doneva et al. \cite{Doneva:2013rha} explored the $\bar{I}-\bar{Q}$ relation for neutron and quark stars rotating at different rotation rates, from a few hundred Hz up to kHz frequencies close to the Kepler limit, using models numerically constructed with the RNS numerical code \cite{RNSpaper}. They found that the $\bar{I}-\bar{Q}$ relation changes with rotation frequency $f$ and in addition for higher frequencies there is an increasing scattering of the different equations of state. Nevertheless Doneva et al. \cite{Doneva:2013rha} produced a general fit that captures the behaviour of neutron star models constructed with modern realistic equations of state and for different rotation rates that has the form,
\be \ln \bar{I}=a_0+a_1 \ln \bar{Q} + a_2 (\ln \bar{Q})^2, \quad \textrm{where }\; a_i=c_0+c_1\left(\frac{f}{\textrm{kHz}} \right)+c_2\left(\frac{f}{\textrm{kHz}} \right)^2+c_3\left(\frac{f}{\textrm{kHz}} \right)^3. \label{eq:iqrapidF} \ee 
%
\begin{figure}[htb]
\begin{center}
\begin{tabular}{ccc}
\includegraphics[width=5.6cm,clip=true]{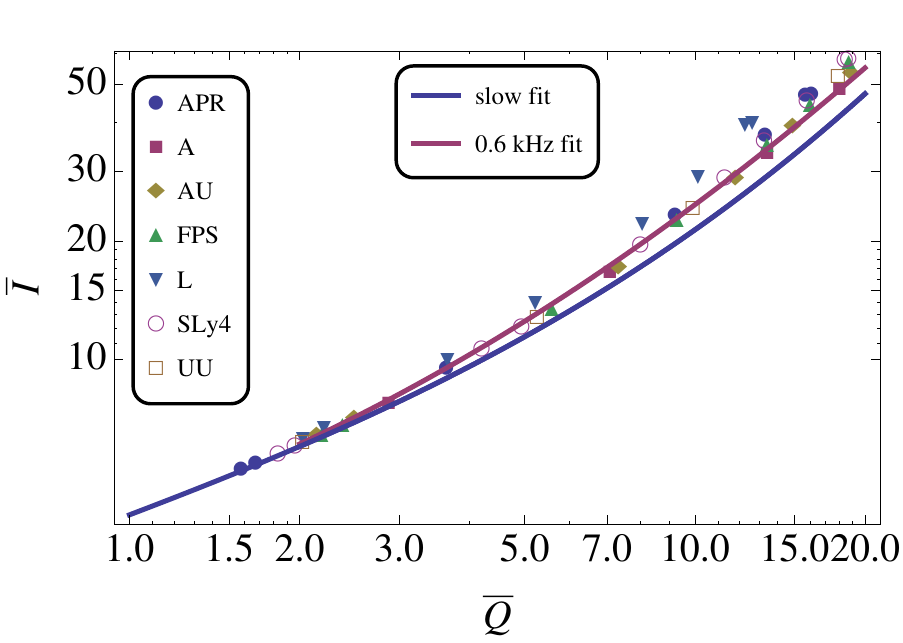}&\includegraphics[width=5.6cm,clip=true]{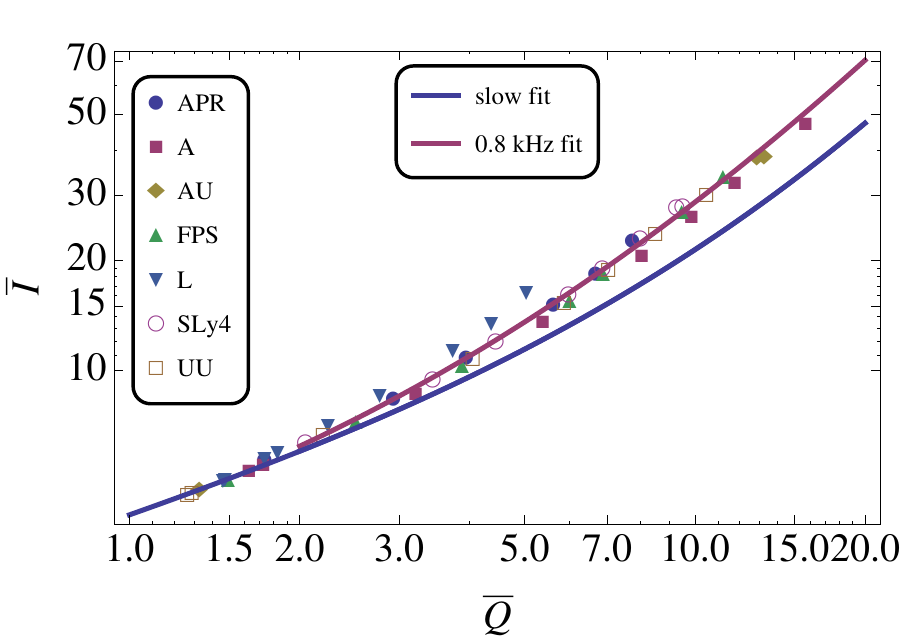} &\includegraphics[width=5.6cm,clip=true]{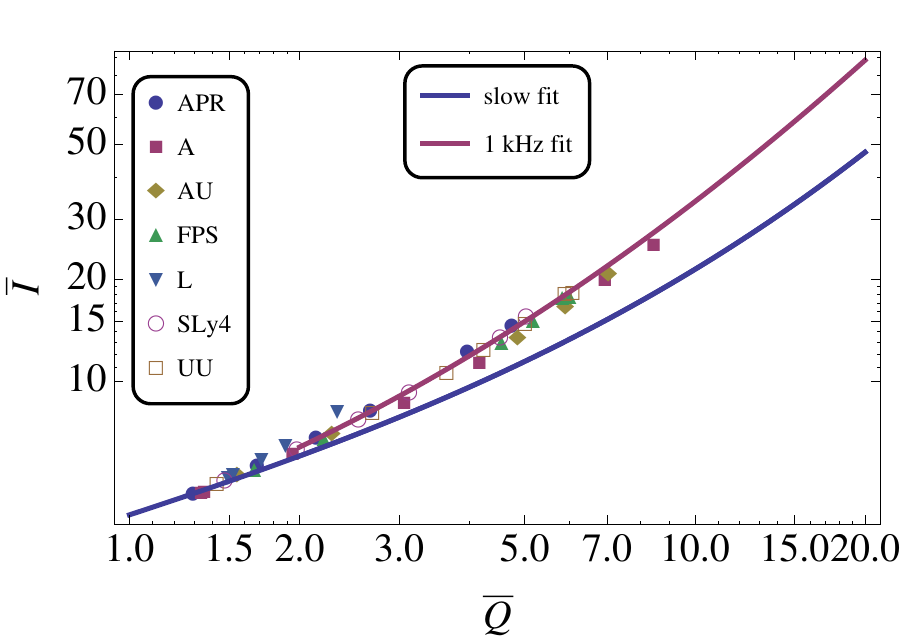} 
\end{tabular}
\caption{\label{fig:iqrapidF} I-Q plots for three rotation frequencies and various equations of state. The solid red lines correspond to the I-Q fit given in \cite{Doneva:2013rha}, while the solid blue correspond to the slow rotation I-Q fit given in Table \ref{tab:iloveq}. One can see that there is some scatter between the different equations of state that was not present in the I-Q relation of Figure~\ref{fig:iloveq}.
}
\end{center}
\end{figure}
%
%
\begin{table}[htb]
\begin{centering}
\begin{tabular}{cccccc}
\hline
\hline
\noalign{\smallskip}
$a_i$  &&  \multicolumn{1}{c}{$c_0$} &  \multicolumn{1}{c}{$c_1$}
&  \multicolumn{1}{c}{$c_2$} &  \multicolumn{1}{c}{$c_3$}   \\
\hline
\noalign{\smallskip}
$a_0$ && 1.406  & -0.051 & 0.154  & -0.131 \\
$a_1$ && 0.489  & 0.183 & -0.562  & 0.471 \\
$a_2$ && 0.098  & -0.136 & 0.463  & -0.273 \\
\noalign{\smallskip}
\hline
\hline
\end{tabular}
\end{centering}
\caption{Numerical coefficients for the fitting formula given in eq.~\eqref{eq:iqrapidF}. 
}
\label{tab:iqrapidF}
\end{table}
The coefficients of the fit are given in Table~\ref{tab:iqrapidF}, while in Figure~\ref{fig:iqrapidF} one can see the fit for the rapidly rotating models for different frequencies compared to the slow rotation fit. One can also notice the scattering around the fit for the different equations of state and the different rotation frequencies. 

\begin{figure}[htb]
\begin{center}
\begin{tabular}{ccc}
\includegraphics[width=5.6cm,clip=true]{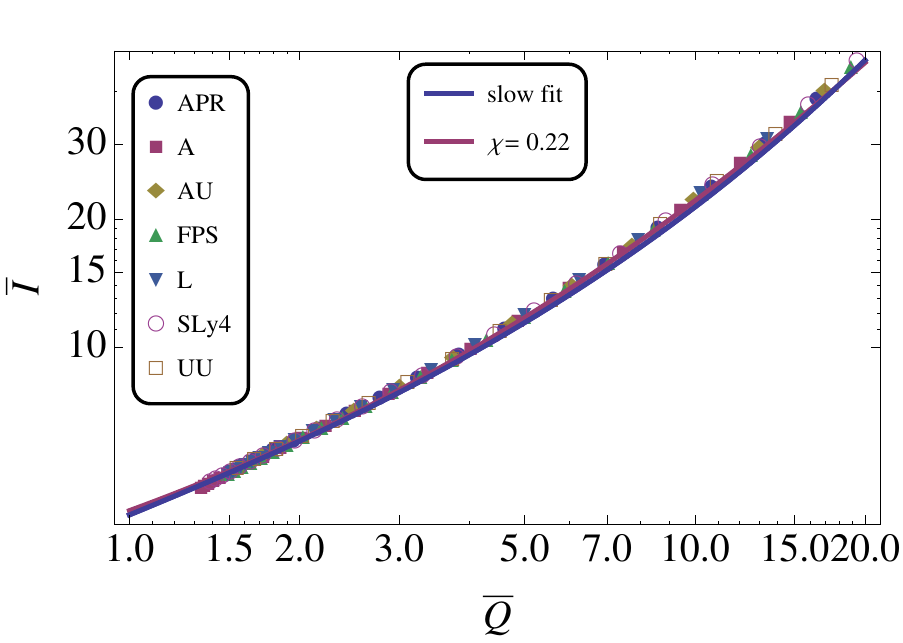}&\includegraphics[width=5.6cm,clip=true]{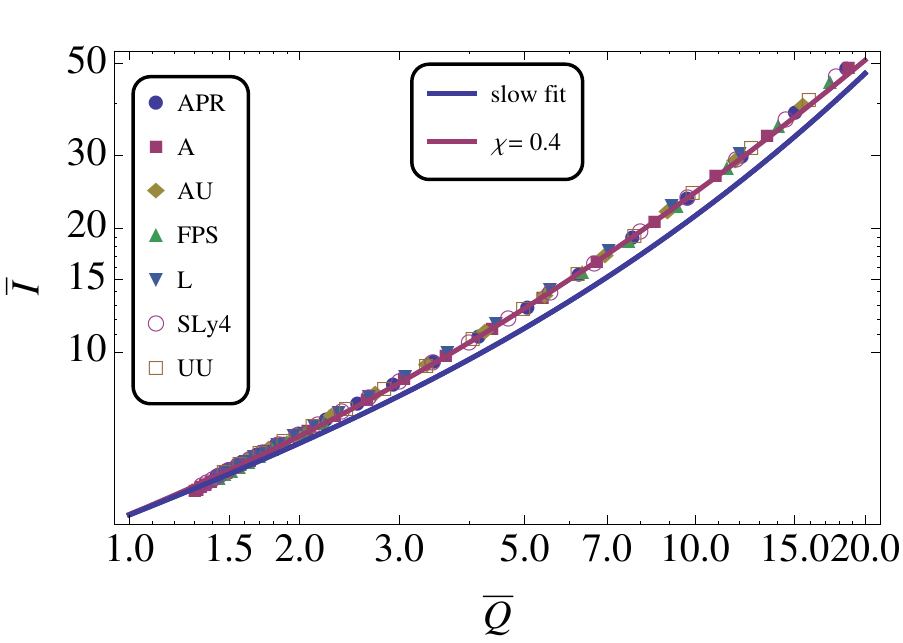} &\includegraphics[width=5.6cm,clip=true]{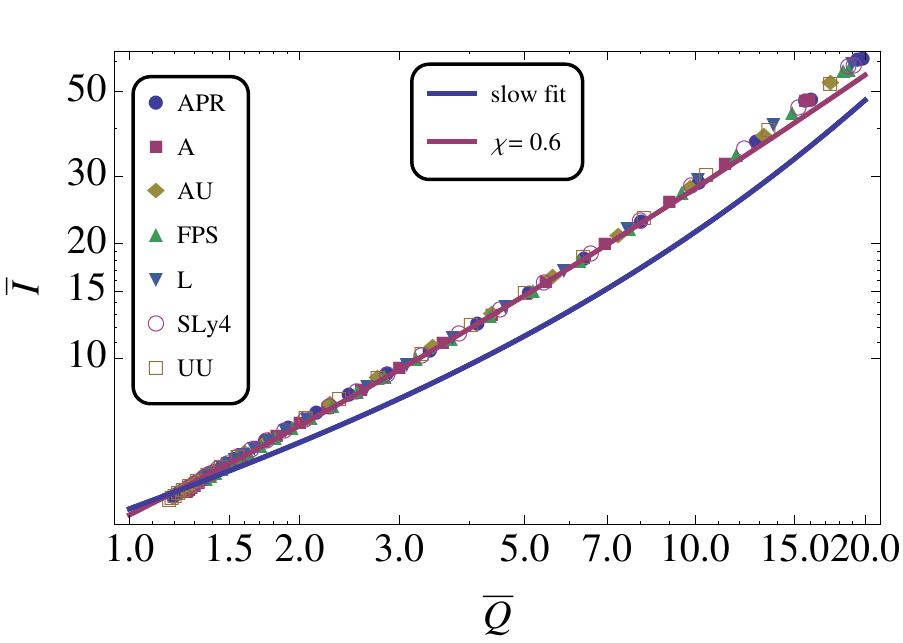} 
\end{tabular}
\caption{\label{fig:iqrapidJ} I-Q plots for three spin parameters and various equations of state. The solid red lines correspond to the I-Q fit given in \cite{Pappas2014PRL}, while the solid blue correspond to the slow rotation I-Q fit given in Table \ref{tab:iloveq}. Unlike in Figure \ref{fig:iqrapidF} there is no scattering between the equations of state when the parameterisation is done with the spin parameter $\chi=J/M^2$.
}
\end{center}
\end{figure}

A different approach in parameterising rotation for rapidly rotating models was taken by Pappas~\&~Apostolatos in \cite{Pappas2014PRL}, where the rotation was parameterised in terms of the spin parameter $\chi=J/M^2$, a dimensionless quantity, instead of the rotation frequency $f$. By comparing models of equal spin parameter, Pappas~\&~Apostolatos found that the different equations of state exhibit no noticeable scattering as one can see in Figure~\ref{fig:iqrapidJ}. They also produced a fit for the I-Q relation in terms of the spin parameter, which has the form,
 \be \sqrt{\bar{I}}=2.16+(0.97-0.14 \chi+1.6 \chi^2)\left(\sqrt{\bar{Q}}-1.13\right)+(0.09+0.23 \chi-0.54 \chi^2)\left(\sqrt{\bar{Q}}-1.13\right)^2, \label{eq:iqrapidJ} \ee 
and is accurate to better than $1\%$.
%
%
What becomes clear from the two different approaches is that the ``universality'' of the behaviour is sensitive to the choice that one makes in parameterising an effect. In this case, parameterising rotation with the spin parameter preserves the equation of state independence of the description, while the rotation frequency breaks it. Along these lines, Chakrabarti et al. explored in \cite{Chakrabarti2014PRL} different ways of parameterising rotation so as to preserve the equation of state independence of the I-Q relation. Specifically, in addition to using the spin parameter, Chakrabarti et al. also explored relations that are parameterised with respect to $R\times f$ and $M\times f$, where $R$ and $M$ are the radii and masses of the corresponding models, and found that these dimensionless quantities are also good parameters for preserving equation of state independence, although the best choice remains the spin parameter. 

At this point there is one more thing that we should note, with respect to an earlier discussion on the normalisation of the various quantities in the initial slow rotation approach and in view of the results from rapid rotation. As one can see in Figure~\ref{fig:iqrapidJ}, the models with spin parameter $\chi=0.22$ (left plot) are very close to the curve for the slow rotation I-Q fit.\footnote{In \cite{Chakrabarti2014PRL} the authors arrive at the same result with $\chi$ values as low as 0.1, which get even closer to the slow rotation fit}. But the normalisation for the moment of inertia and the quadrupole in this case is performed with the total mass of the numerically constructed model. Nevertheless, the agreement with the slow rotation results and the normalisation with respect to the non-rotating TOV mass is very good, providing thus more credibility to our earlier argument on the insensitivity of the result to whether one uses the TOV or the corrected mass in normalising the initial I-Love-Q relations.

 {\bf Applications:} The existence of the I-Love-Q universal relations is of theoretical interest by itself, but the most interesting aspect is the potential applications of such relations. There are two ways in which these relations can be useful: (1) one could use them to make indirect measurements of quantities that are difficult to measure, i.e., assume that the relations hold and use them as tools to do physics and astrophysics, or (2) test their validity by measuring more than one of these quantities and in this way use them to test the assumptions on which the I-Love-Q relations are based. Both of these are very interesting prospects. 
 
(1) Assuming the validity of the I-Love-Q relations, one could for example use a binary pulsar system to measure the moment of inertia (see for example \cite{Lattimer2005ApJ}) and from that, using the I-Q relation infer the quadrupole of the neutron star. Such a quadrupole measurement together with the simultaneous measurement of the mass and the angular momentum of the neutron star could be used to constrain the equation of state, by taking advantage of the fact that different equations of state constitute a different surface in an $(M,\chi,\bar{Q})$ parameter space, as described in \cite{Pappas2014PRL}. 
Another application could be to use the Q-Love relation to break degeneracies between individual spins and quadrupoles\footnote{The degeneracy is in the gravitational wave phase where a spin-spin coupling term has a contribution at the same order as the quadrupole term.} in the analysis of the waveforms of the gravitational waves emitted from the inspiral of neutron star binaries, along the lines given in \cite{Yagi:2013bca,YagiYunes2013PRD}.   
 
Returning to the equation of state measurement front, another application of I-Love-Q relations, as described by Silva et al. \cite{Silva2016MNRAS}, could be the estimation of parameters of the equation of state 
from electromagnetic observations of binary pulsars or gravitational wave observations of binary inspirals, from systems where the members are low mass neutron stars. Specifically, Silva et al. (using various equations of state) found that for low mass neutron stars, quantities like $\bar{I}$, $\bar{Q}$ and $\bar{\lambda}$ can be fitted by simple functions of the central density $\rho_c$ and of the dimensional parameter $\eta=\left(K_0 L^2\right)^{1/3}$, where $K_0$ is the incompressibility of symmetric nuclear matter and $L$ is the slope of the symmetry energy at saturation density (all three parameters have units of energy). A measurement of any two of $\bar{I}$, $\bar{Q}$ or $\bar{\lambda}$ could be used to constrain $\eta$ and $\rho_c$ and in this way constrain the parameters of the equation of state. Alternatively, the measurement of one of the quantities and the use of the I-Love-Q relations could also provide similar constraints. Finally, one could use the simultaneous measurement of two or more of these quantities, the I-Love-Q relations and the relations in \cite{Silva2016MNRAS} to perform consistency checks on the assumptions entering the modelling of the equation of state.    

(2) Testing the validity of the I-Love-Q relations offers another possibility for testing the equation of state. As it has been demonstrated \cite{Yiloveq2014PhRvD,Sham2015ApJ,Chan2015PhRvD}, the universality of the I-Love-Q relations comes from properties of the equation of state that are related to the fact that the equation of state for compact objects is close to being incompressible. Deviations from being incompressible would change the different relations and in principle could also introduce some spin dependence. Therefore, by measuring more than one of the $\bar{I}$, $\bar{Q}$ or $\bar{\lambda}$ one could test the validity of the I-Love-Q relations and in this way test whether the equation of state is close to our current models or not. 
One possibility for measuring more than one quantities is to combine astrophysical observations with gravitational wave observations. So for example, pulsar timing could provide a measurement of the moment of inertia while gravitational wave observations could provide a measurement of the tidal Love number \cite{Yagi:2013bca}. Another possibility has been recently proposed \cite{Chirenti2017ApJ}, where the analysis of gravitational waves from highly eccentric binary systems, where the $f$-mode is excited by close encounters between the members of the binary, could provide simultaneous measurements of the masses, moments of inertia, and tidal Love numbers of the members of the system. 

Another prospect is to test general relativity by testing the I-Love-Q relations. In principle one would expect that for different theories of gravity, neutron stars could follow different I-Love-Q relations or no I-Love-Q relations. Therefore by measuring combinations of the $\bar{I}$, $\bar{Q}$ or $\bar{\lambda}$, one could test for deviations from general relativity or even identify an alternative theory of gravity. These prospects will be further discussed after the discussion of neutron stars in alternative theories of gravity.
 
\subsubsection{Multipole moments 3-hair}

The spacetime around a rotating neutron star is a stationary and axisymmetric spacetime, i.e., there exist two Killing vectors, one timelike $\xi^a$ which characterises the spacetime's symmetry with respect to time translations and one spacelike $\eta^a$ which characterises the spacetime's symmetry with respect to rotations around an axis which in this case is the stars' axis of rotation. In addition, for isolated stars, it is assumed that the spacetime is asymptotically flat. Under these general assumptions the line element for the spacetime of a rotating neutron star can take the form 
\be ds^2=-e^{2\nu}dt^2+r^2(1-\mu^2) B^2 e^{-2\nu}(d\varphi-\omega dt)^2+e^{2a}(dr^2+\frac{r^2}{1-\mu^2}d\mu^2) \label{eq:NSspacetime}
\ee
where $\mu=\cos\theta$ and the metric functions $\nu$, $B$, $\omega$, and $a$ are all functions of $(r,\mu)$. For such a spacetime one can define relativistic multipole moments, which can characterise the structure and the properties of the spacetime, as the multipole moments in Newtonian theory characterise a Newtonian potential. As one would expect, the moments in the two cases are not completely equivalent, with one difference between relativistic and Newtonian moments being that in the relativistic case there exist angular momentum or mass current moments in addition to the usual mass moments. Another one is that the relativistic moments are also sourced by geometry in addition to masses and currents, a consequence of the non-linear nature of the theory. 

There exist a few different ways of defining the relativistic multipole moments of a spacetime, such as the Geroch~\&~Hansen algorithm for stationary spacetimes \cite{Geroch:1970cc,Geroch:1970cd,Hansen:1974zz} which was later customised to axisymmetric spacetimes by Fodor~et~al. \cite{Fodor1989JMP}, or alternatively the Thorne algorithm that was mentioned earlier \cite{ThorneRevModPhys1980}. These formalisms give essentially the same moments related by a multiplicative factor as,\footnote{ Multipole moments in general are tensorial quantities. The rank of the tensor is given by the order of the corresponding multipole moment. In the case that we also have axisymmetry, the moments are multiples of the symmetric trace-free tensor product of the axis vector $n^a$ with itself. In this case therefore one can define scalar moments as  $\mathcal{P}_{\ell}=1/(\ell!) \mathcal{P}_{i^1\ldots i^{\ell}} n^{i^1}\ldots n^{i^{\ell}}$, where $\mathcal{P}_{\ell}$ is the $\ell$-th order moment. Since we will discuss about stationary and axisymmetric spacetimes we will only talk about the scalar moments.}  
\be M_{\ell}=(2\ell -1)!! M_{\ell}^T\, \textrm{ and } \, J_{\ell}= \frac{2\ell (2\ell-1)!!}{2\ell+1} J_{\ell}^T\ee 
(for a review see \cite{Quevedo1990}). While the Thorne formalism seems easier and more intuitive, because it is based on identifying the coefficients of the asymptotic expansion of the metric functions, which reminds the Newtonian case, in practice finding the appropriate coordinate system for calculating higher order moments can be very difficult. On the other hand the Geroch-Hansen formalism, even though it is more involved mathematically, in practice it is more algorithmic. Furthermore, it's reformulation in terms of the Ernst potential in the case of stationary and axisymmetric spacetimes by Fodor~et~al., makes the calculation of multipole moments even more straightforward. For the different applications in the literature so far, where the multipole moments of numerical neutron star spacetimes have been calculated, people have used the properties of circular equatorial geodesics and their relation to the moments \cite{Laarakkers1999ApJ,Pappas2012PhRvL,Pappas2012arXiv,Pappas2013MNRAS.429,Yagi2014PRD}. This has been based on Ryan's formalism \cite{Ryan95} for relating the various relativistic precession frequencies between them and to other orbital properties, in terms of the multipole moments spectrum of the background spacetime. Although this approach has served us well in calculating the multipole moments of numerical spacetimes, it is somewhat limited and therefore here we will briefly present a more straightforward and rigorous way of calculating the moments of stationary and axisymmetric numerical spacetimes, that has the additional benefit of providing higher order moments in a less computationally expensive way.        

As mentioned earlier, the spacetime around a neutron star is a stationary, axisymmetric, vacuum spacetime that admits the Killing vectors $\xi^a$ (timelike) and $\eta^a$ (spacelike). Using the timelike Killing vector one can define the two scalar quantities $f$ and $\psi$ through the equations,
\be f=-\xi^a\xi_a, \quad \psi_{,a}=\varepsilon_{abcd}\xi^b\xi^{c;d}, \label{eq:scalars} \ee
where $f$ is related to the norm of the Killing vector, while $\psi$ is the scalar twist of the Killing vector. In the Ernst reformulation of the Einstein Field Equations these two scalar quantities define the complex Ernst potential $\mathcal{E}=f+i\psi$. %
If one has the Ernst potential for a given spacetime in terms of the Weyl-Papapetrou coordinates $(\rho,z)$ then one can define along the axis of symmetry $\rho=0$ the potential
\be \tilde{\xi}(\bar{z})=(1/\bar{z})\frac{1-\mathcal{E}(\bar{z})}{1+\mathcal{E}(\bar{z})}=\sum_{j=0}^{\infty} m_j\bar{z}^j, \ee
in terms of the coordinate $\bar{z}=1/z$ which is centred at infinity. The different coefficients in the expansion of $\tilde{\xi}(\bar{z})$ are the parameters $m_i$ 
that give the moments of stationary and axisymmetric spacetimes as they were calculated by Fodor~et~al. \cite{Fodor1989JMP}. The process of calculating the moments therefore involves initially two steps, first the calculation of the Ernst potential and then the transformation of the metric coordinates to Weyl-Papapetrou coordinates.

Returning to neutron star spacetimes, the metric functions in \eqref{eq:NSspacetime} have an asymptotic expansion outside the star of the form 
\bear 
\nu&=&\sum_{l=0}^{\infty}\left(\sum_{k=0}^{\infty}\frac{\nu_{2l,k}}{r^{2l+1+k}} \right)P_{2l}(\mu),\\
\omega&=&\sum_{l=1}^{\infty}\left( \sum_{k=0}^{\infty}\frac{\omega_{2l-1,k}}{r^{2l+1+k}} \right)\frac{dP_{2l-1}(\mu)}{d\mu},\\
B&=&1+\left(\frac{\pi}{2}\right)^{1/2}\sum_{l=0}^{\infty}\frac{B_{2l}}{r^{2l+2}}T_{2l}^{1/2}(\mu),
\eear
where $P_l(\mu)$ are the Legendre polynomials, $T_{2l}^{1/2}(\mu)$ are the Gegenbauer polynomials,\footnote{The Gegenbauer polynomials are given by the definition $T_l^{1/2}(\mu)=\frac{(-1)^l \Gamma(l+2)}{2^{l+1/2} l! \Gamma(l+3/2)} (1-\mu^2)^{-1/2} \frac{d^l}{d\mu^l}(1-\mu^2)^{l+1/2} $.} and the various coefficients are  not all independent, with the constraints coming from the field equations in vacuum, which in the frame of the zero angular momentum observers take the form, 
\bear &&\mathbf{D}\cdot(B\mathbf{D}\nu)=\frac{1}{2}r^2\sin^2\theta
B^3e^{-4\nu}\mathbf{D}\omega\cdot\mathbf{D}\omega
,\\
%
&&\mathbf{D}\cdot(r^2\sin^2\theta B^3e^{-4\nu}\mathbf{D}\omega)=0
, \\
%
&&\mathbf{D}\cdot(r\sin\theta\mathbf{D}B)=0
,\eear
where $\mathbf{D}$ is a flat space 3-dimensional derivative operator in spherical coordinates (see \cite{BI1976ApJ}). As an indicative example of how the coefficients are constrained, we have from the field equations that $\nu_{0,1}=0$, which means essentially that there is no ``mass dipole'' contribution in the metric function $\nu$, while the next coefficient is constrained to be $\nu_{0,2}=-\frac{1}{3}B_0\nu_{0,0}$. We should note here that the coefficient $\nu_{0,0}$, as one can see from the asymptotic expansion of $\nu$, gives the mass, i.e., $\nu_{0,0}=-M$, while similarly the angular momentum comes from the asymptotic expansion of $\omega$, which gives $\omega_{1,0}=2J$. Coefficients such as $\nu_{0,0}$, $\nu_{2,0}$, $\omega_{1,0}$, $\omega_{3,0}$, and so on, as well as all the $B_{2l}$ coefficients, are not constrained by the field equations and are free parameters of the external spacetime that are determined by the characteristics of the fluid configuration. 

Having the metric in terms of $\mu$ and as an expansion in $1/r$ one can proceed to calculate the scalar twist from eq.~\eqref{eq:scalars}. The definition of the scalar twist results in two equations, one for $\psi_{,r}$ and one for $\psi_{,\mu}$.  
Since the calculation is done using an expansion in $1/r$, the resulting scalar twist will be accurate up to some order in $1/r$ and can be evaluated as
\be \psi(r,\mu)=-\int_r^{\infty}\psi_{,r}|_{\mu=\textrm{const.}} dr, \ee 
where the asymptotic condition is that $\psi(r\rightarrow \infty,\mu)=0$. With the scalar twist at hand, the Ernst potential will be given in terms of the angular coordinate $\mu$ and an expansion in inverse powers of $r$. By further setting $\mu=1$ we have the Ernst potential along the axis. What remains is to express the $r$ coordinate in terms of the Weyl-Papapetrou coordinate $z$ along the axis of symmetry. This is done following the procedure given in \cite{Pappas2008CQG}. By integrating along curves of constant $r$ from the equatorial plane up to the axis of symmetry and inverting the resulting expansion to solve for $r$ we have, 
\be  z=\int_0^{1} d\mu (r^2 B_{,r} + r B) 
%
%
\quad \Rightarrow \quad r=z+\frac{B_0}{z}+\frac{B_2-B_0^2}{z^3}+\frac{2B_0^3-4B_0B_2+B_4}{z^5}+\ldots \ee
In this way we can calculate the Ernst potential along the axis of symmetry for a spacetime given in the form of a quasi-isotropic metric as in eq~\eqref{eq:NSspacetime}. From that Ernst potential one can calculate the moments from the coefficients of the expansion of $\tilde{\xi}$. The resulting first few multipole moments are,
\begin{align}
   M_0&=
               -\nu_{0,0}, \quad M_2=
                                 \frac{1}{3} (4 B_0 \nu_{0,0} + \nu_{0,0}^3 - 3 \nu_{2,0}), \nonumber\\
        M_4&=
                                  -\nu _{4,0} -\frac{32}{21} B_0 \nu _{0,0}^3-\frac{16}{5} B_0^2 \nu _{0,0}+\frac{64}{35} B_2 \nu _{0,0}+\frac{24}{7} B_0 \nu _{2,0}+\frac{3}{70} \nu
   _{0,0} \omega _{1,0}^2-\frac{19}{105} \nu _{0,0}^5+\frac{8}{7} \nu _{2,0} \nu _{0,0}^2 ,  \nonumber       \\
   J_1&=
               \frac{\omega_{1,0}}{2}, \quad J_3=
                            -\frac{3}{10}  ((4 B_0 + \nu_{0,0}^2) \omega_{1,0} - 5 \omega_{3,0}) , \nonumber \\ 
   J_5&=
                 \frac{5}{2}  \omega _{5,0} +  \frac{104}{63}  B_0 \nu _{0,0}^2 \omega _{1,0}+\frac{24}{7}  B_0^2 \omega _{1,0}-\frac{32}{21}  B_2 \omega _{1,0}-\frac{20}{3}  B_0
   \omega _{3,0}+\frac{25}{126}  \nu _{0,0}^4 \omega _{1,0}-\frac{5}{3}  \nu _{0,0}^2 \omega _{3,0}\nonumber \\
   &-\frac{5}{21}  \nu _{2,0} \nu _{0,0}
   \omega _{1,0}-\frac{1}{28}  \omega _{1,0}^3     .  
\end{align}

{\bf Multipole moments of neutron stars and 3-hair relations:} The calculation of the moments up to the mass hexadecapole $M_4$ has been numerically implemented so far for a wide variety of equations of state and for both slowly and rapidly rotating neutron (and quark) stars  \cite{Pappas2012PhRvL,Pappas2012arXiv,Urbanec:2013fs,Pappas2014PRL,Yagi2014PRD}. One very interesting property of the neutron star multipole moments that has been discovered is that they follow a simple scaling with the spin parameter $\chi$ and the mass $M$ that is the same as the one that the moments of rotating black holes follow, i.e., the higher than the angular momentum moments behave as 
\be M_2=  a \chi^2 M^3, \quad J_3 =  \beta \chi^3 M^4, \quad M_4 = \gamma \chi^4 M^5, \ee
 where for Kerr black holes the mass moments behave as $M_{2n}=(i\chi)^{2n} M^{2n+1}$ and the mass current moments behave as $iJ_{2n+1}= (i\chi)^{2n+1} M^{2n+2}$, where $i$ is the imaginary unit. Table \ref{tab:moments} gives the values of the coefficients $a$, $\beta$, and $\gamma$ for some typical equations of state.  
\begin{table}[htb]
\begin{centering}
\begin{tabular}{lrclrclrclrclrclrcl}
\hline
\hline
\noalign{\smallskip}
   &   & APR &   &   & AU &   &   & SLy4 &   &   & FPS &   &   & L &   &   & UU &   \\ \hline
 $M/M_{\odot}$ & $a$ & $\beta$  & $\gamma$  & $a$ & $\beta$  & $\gamma$   & $a$ & $\beta$  & $\gamma$  & $a$ & $\beta$  &
   $\gamma$  & $a$ & $\beta$  & $\gamma$  & $a$ & $\beta$  & $\gamma$  \\ \hline
 0.9 & -10.9 & -23.8 & 295.2 & -8.2 & -17.2 & 160.9 &  -10.2 & -21.9 & 252.7 & -9.1 & -19.7 & 205.6 &  -14. & -30.3 & 466. & -9.2 & -19.5 &  203.1 \\
 1. & -9.3 & -20.1 & 216.8 & -6.9 & -14.4 & 115.1 & -8.6 & -18.4 & 180.8 & -7.8 & -16.5 & 149. &  -12.3 & -26.5 & 359. & -7.8 & -16.5 &  147.6 \\
 1.1 & -7.7 & -16.6 & 149.4 & -6. & -12.2 & 85.1 &  -7.6 & -16.1 & 141.5 & -6.6 & -13.9 & 108.3 &  -10.9 & -23.1 & 280.3 & -6.8 & -14.1 &  110.6 \\
 1.2 & -6.8 & -14.2 & 112.9 & -5.2 & -10.5 & 63.9 &  -6.4 & -13.4 & 99.8 & -5.8 & -12.2 & 85.1 &  -9.3 & -19.8 & 206.4 & -5.9 & -12.1 & 83.6  \\
 1.3 & -6. & -12.6 & 89.6 & -4.4 & -8.7 & 45.8 &  -5.6 & -11.6 & 77.8 & -5. & -10.3 & 61.8 & -8.7 & -18.2 & 177.8 & -5.1 & -10.3 & 61.6 \\
 1.4 & -5.2 & -10.5 & 64.9 & -3.9 & -7.8 & 37.2 & -5. & -10.2 & 61.3 & -4.3 & -8.6 & 45.1 &  -7.6 & -15.8 & 136. & -4.6 & -9.2 & 50.4 \\
 1.5 & -4.7 & -9.5 & 54. & -3.5 & -6.6 & 27.4 &  -4.3 & -8.7 & 46.2 & -3.7 & -7.3 & 33.5 &  -6.6 & -13.8 & 104.7 & -3.9 & -7.7 & 37.1 \\
 1.6 & -4.1 & -8.1 & 40.2 & -3.1 & -5.8 & 22.3 &  -3.8 & -7.4 & 34.4 & -3.2 & -6.3 & 26.5 &  -6.2 & -12.7 & 90.9 & -3.4 & -6.5 & 26.9 \\
 1.7 & -3.8 & -7.4 & 34.5 & -2.7 & -4.9 & 16.1 &  -3.4 & -6.6 & 28.1 & -2.8 & -5.4 & 19.9 &  -5.5 & -11.2 & 71.5 & -3.1 & -5.8 & 22. \\
 1.8 & -3.3 & -6.2 & 25.1 & -2.5 & -4.4 & 13.7 &  -3. & -5.7 & 21.9 & -2.4 & -4.5 & 14.5 &  -5. & -10.1 & 59.2 & -2.7 & -4.9 & 16.3 \\
 1.9 & -2.9 & -5.4 & 19.7 & -2.2 & -3.7 & 10.2 &  -2.6 & -4.8 & 16.2 & {\it-1.9} & {\it -3.3} & {\it 9.5} &  -4.6 & -9.1 & 49.3 & -2.5 & -4.5 & 14.3 \\
 2. & -2.7 & -4.9 & 16.3 & -2. & -3.4 & 8.6 & -2.3 & -4.1 & 12.3 & {\it -1.6} & {\it -2.4} & {\it 5.1} &  -4.3 & -8.5 & 43.5 & -2.2 & -3.8 & 10.5 \\
 2.1 & -2.4 & -4.1 & 12.3 & -1.7 & -2.8 & 6.2 & -1.9 & -3.4 & 9. & {\it -1.5} & {\it -2.2} & {\it 4.2} & -3.9 & -7.7 & 36.1 & -1.9 & -3.2 & 8. \\
\noalign{\smallskip}
\hline
\hline
\end{tabular}
\end{centering}
\caption{Multipole moments of some typical equations of state. The table has included equations of state that give neutron star models with masses above the maximum mass observed, i.e., $2M_{\odot}$, while the quantities shown are the coefficients $a=M_2/(\chi^2 M^3)$, $\beta=J_3/(\chi^3M^4)$, and $\gamma=M_4/(\chi^4 M^5)$. The last three sets of coefficients for FPS (in italics) correspond to models with masses above the maximum mass of the non-rotating configurations. One can easily verify that the coefficients presented here fit the curves in Figure \ref{fig:3-hair}.}
\label{tab:moments}
\end{table}
%
 As one can see, while for Kerr black holes these coefficients are equal to $\pm1$, in the case of neutron stars their magnitude can be quite larger than that. An immediate implication of this is that the spacetime around neutron stars can be quite different from that of Kerr black holes. One should notice from Table \ref{tab:moments} that softer equations of state, like AU, FPS and UU, produce smaller values for the quadrupole and the higher order moments than the stiffer equations of state, like Sly4, APR and L. Also, as one increases the central density and approaches the models close to the maximum mass, then the deviations of the moments from their corresponding Kerr values become smaller, which means that in these cases the neutron star spacetime behaves more like a Kerr spacetime. We should note though that neutron star models never quite reach to the Kerr point where $-a=-\beta=\gamma=1$. 
 
\begin{figure}[htb]
\begin{center}
\begin{tabular}{cc}
\includegraphics[width=7.1cm,clip=true]{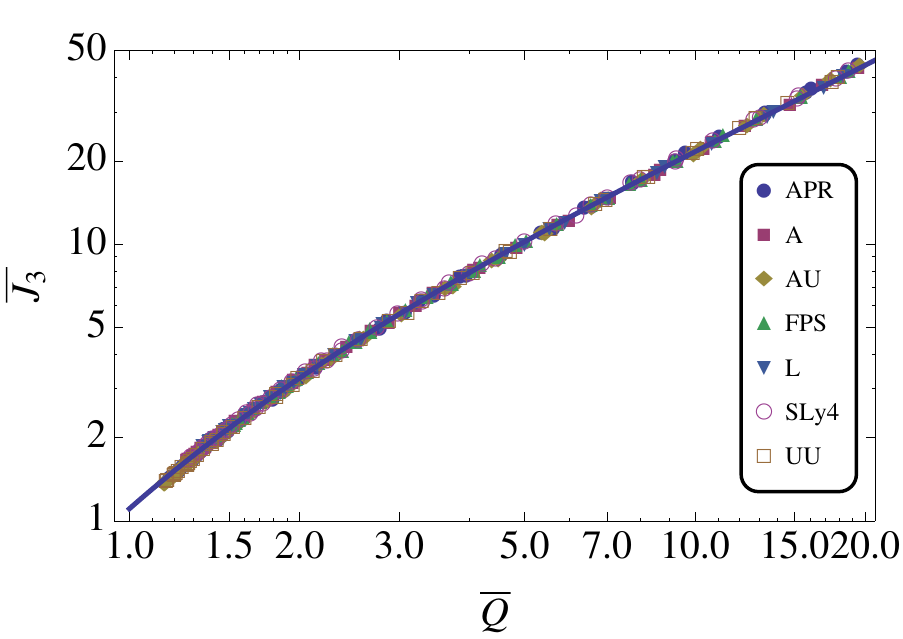}&\includegraphics[width=7.5cm,clip=true]{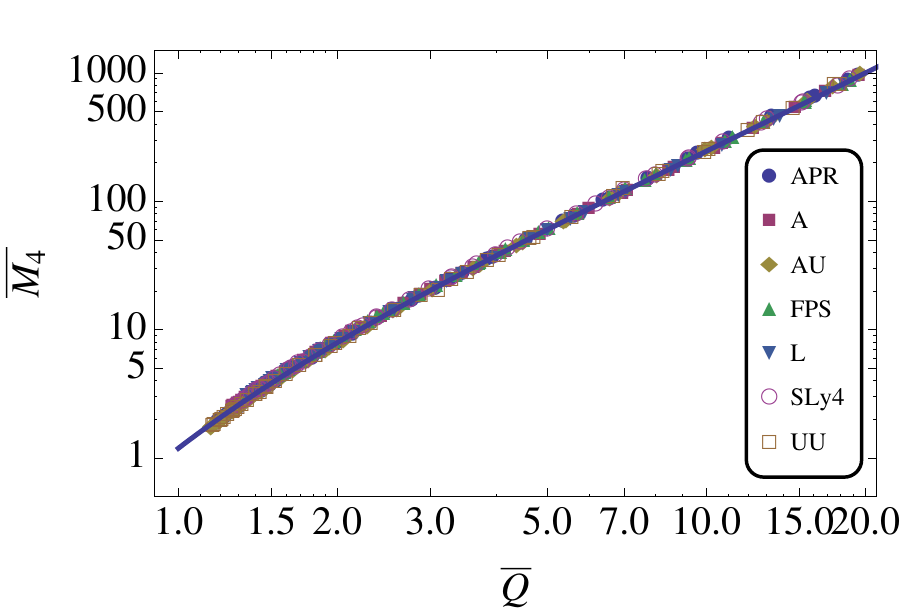} 
\end{tabular}
\caption{\label{fig:3-hair} Relations between the higher order moments and the mass quadrupole for different equations of state. The solid lines correspond to the best fits using a double power-law given in Table \ref{tab:3-hair}. The $\bar{J_3}$-$\bar{Q}$ fit comes from \cite{Pappas2014PRL}, while the $\bar{M_4}$-$\bar{Q}$ fit comes from \cite{Pappas2017MNRAS}. 
}
\end{center}
\end{figure}

If we further define the reduced moments, $\bar{Q}\equiv \bar{M_2}\equiv -\frac{M_2}{\chi^2 M^3}$, $\bar{J_3}\equiv -\frac{J_3}{\chi^3 M^4}$, and $\bar{M_4}\equiv \frac{M_4}{\chi^4 M^5}$, then it was shown in \cite{Pappas2014PRL,Yagi2014PRD} that both $\bar{J_3}$ and $\bar{M_4}$ are related to $\bar{Q}$ following universal relations, i.e., relations that are equation of state independent. These relations can be seen plotted in Figure \ref{fig:3-hair}.  
 %
%
As it can be seen from the two plots, the points seem to follow a power-law with some break appearing towards the lower values of the reduced quadrupole, which correspond to the most compact models close to the maximum mass limit. The two curves can be fitted with an expression of the form, 
\be y_i = A+B_1 x^{n_1}+ B_2 x^{n_2}, \label{eq:3-hair}\ee
where $y_i$ can be either $(\bar{J_3})^{1/3}$ or $(\bar{M_4})^{1/4}$, while $x$ is $(\bar{Q})^{1/2}$. 
%
\begin{table}[htb]
\begin{centering}
\begin{tabular}{cccccccc}
\hline
\hline
\noalign{\smallskip}
$y_i$ & $x$ &&  \multicolumn{1}{c}{$A$} &  \multicolumn{1}{c}{$B_1$}
&  \multicolumn{1}{c}{$B_2$} &  \multicolumn{1}{c}{$n_1$} &  \multicolumn{1}{c}{$n_2$}  \\
\hline
\noalign{\smallskip}
$(\bar{J_3})^{1/3}$ & $(\bar{Q})^{1/2}$ && -4.82 & 5.83  & 0.024 & 0.205 & 1.93\\
$(\bar{M_4})^{1/4}$ & $(\bar{Q})^{1/2}$ && -4.749 & 0.27613 & 5.5168  & 1.5146 & 0.22229\\

\noalign{\smallskip}
\hline
\hline
\end{tabular}
\end{centering}
\caption{Numerical coefficients for the fitting formula given in eq.~\eqref{eq:3-hair}.}
\label{tab:3-hair}
\end{table}
%
The two fits presented in Table \ref{tab:3-hair} have been performed for neutron star models only and deviate from the models by less than 4-5\%. In \cite{Yagi2014PRD,YagiYunes2016arXiv} one can find fits for the relations $\bar{J_3}-\bar{Q}$ and $\bar{M_4}-\bar{Q}$ that also include quark star models. Including quark stars results in a slightly wider spread of the data points, but nevertheless the relations between the moments remain remarkably equation of state independent. We should note here that the range of quadrupoles plotted in Figure \ref{fig:3-hair} is between the most compact neutron stars close to the maximum mass limit down to neutron stars with masses a little less than $1 M_{\odot}$, i.e., within the entire observed mass range.  

The 3-hair universal relations have been studied in both the slow rotation limit, using a Hartle-Thorne expansion in terms of rotation up to the 4th order so that $M_4$ can be included in the calculation, as well as for rapidly rotating neutron stars in full numerical relativity using both LORENE and RNS numerical codes. The results from all approaches have been in full agreement. 

{\bf Newtonian insights:} 3-hair relations where also studied in Newtonian theory by Stein et al. \cite{Stein2014ApJ}, providing some insight on the possible origin of this universality, as well as a very elegant result. Stein et al. studied the Newtonian limit of multipole moments for a rotating neutron star by using the definitions,
\bear
   M_{\ell} &=& 2\pi \int_0^{\pi} \int_0^{R(\theta)} \rho(r,\theta) P_{\ell}(\cos\theta) \sin\theta d\theta r^{\ell+2}dr,\\
   J_{\ell} &=& \frac{4\pi}{\ell+1} \int_0^{\pi} \int_0^{R(\theta)} \Omega \rho(r,\theta) \frac{d P_{\ell}(\cos\theta)}{d(\cos\theta)} \sin^3\theta d\theta r^{\ell+3}dr,
\eear  
where $R(\theta)$ is the surface of the star as a function of the polar angle $\theta$ and $\rho(r,\theta)$ is the density inside the star. Since the star is rotating it is assumed to have axial symmetry and in addition we also assume reflection symmetry with respect to the equatorial plane. These symmetries require that the odd mass moments and the even angular momentum moments are zero. Of course the angular momentum moments have no physical meaning in Newtonian theory but could be formally defined as the Newtonian limit of their relativistic counterparts. 

Stein et al. in order to make progress with the analytic calculation of the moments introduced the following two assumptions, (1) the isodensity surfaces inside the star are self-similar ellipsoids with a constant eccentricity, and (2) the density as a function of the isodensity radius $\tilde{r}$ for a rotating configurations is the same as the corresponding radius of a non-rotating configuration with the same volume. For these two assumptions, the eccentricity is defined as $e=\sqrt{1-(\textrm{semi-minor axis})^2/(\textrm{semi-major axis})^2}$, while the isodensity radius is defined as $\tilde{r}\equiv r/\Theta(\theta)$, where 
 \be \Theta(\theta)=\sqrt{\frac{1-e^2}{1-e^2\sin^2\theta}}. \ee
The first assumption is strictly true only for constant density stars, i.e., for $n=0$ polytropes,\footnote{A polytropic equation of state has the form $P=k\rho^{\Gamma}$, where the exponent can be written as $\Gamma=1+1/n$ in terms of the polytropic index $n$.} which give Maclaurin spheroides. In any other case the eccentricity of the iso-density surfaces varies with the radius inside the star. Nevertheless, for slowly rotating and compact objects, where the the deviations from sphericity are not large and the equation of state is close to the incompressible limit, these assumptions turn out to be good approximations.  

Under the aforementioned assumptions, the integrals for the multipole moments separate to an angular and a radial part and can be given by the expressions, 
 \be M_{\ell}=2\pi I_{\ell,3} R_{\ell}, \quad \textrm{ and } \quad J_{\ell}= \frac{4\pi \ell}{2\ell+1} \Omega (I_{\ell-1,5}-I_{\ell+1,3}) R_{\ell+1}, \ee
 where the radial and angular integrals are,
 \be R_{\ell}\equiv \int_0^a \rho(\tilde{r})\tilde{r}^{\ell+2} d\tilde{r}, \quad  \textrm{ and } \quad I_{\ell,k}\equiv \int_{-1}^1\Theta(\mu)^{\ell+k}P_{\ell}(\mu) d\mu,  \ee
where $\mu\equiv\cos\theta$ and $a$ is the equatorial radius of the surface. The angular integral depends only on the eccentricity $e$, while the radial integral depends on the eccentricity and the mass distribution. Using the second assumption, the density profile can be expressed in terms of a non-rotating model. Assuming a polytropic equation of state, the density will be given by solving the Lane-Emden equation
\be \frac{1}{\xi^2}\frac{d}{d\xi}\left( \xi^2\frac{d\vartheta}{d\xi}\right) +\vartheta^n=0, \ee
where $n$ is the polytropic index, the function $\vartheta$ is related to the density as $\rho=\rho_c\vartheta^n$, and the radial coordinate is $r=\alpha\xi$ with $\xi$ being dimensionless and $\alpha$ being a length scale that depends on the equation of state.\footnote{For $n=0$ Lane-Emden admits an exact solution which is $\vartheta=1-\frac{1}{6}\xi^2$ with a surface at $\xi_1=\sqrt{6}$, while for $n=1$ it admits the solution $\vartheta=\frac{\sin\xi}{\xi}$ with a surface at $\xi_1=\pi$.}  
For any choice of the polytropic index corresponds a solution of the Lane-Emden $\vartheta(\xi)$ that has a surface when $\vartheta(\xi_1)=0$. Substituting such a solution in the integrals for the moments, we have that
\be M_{2\ell+2}=\frac{(-1)^{\ell+1}}{2\ell+3} \frac{e^{2\ell+2}}{(1-e^2)^{\frac{\ell+1}{3}}} \frac{\mathcal{R}_{n,2+2\ell}}{\xi_1^{2\ell+4}|\vartheta'(\xi_1)|} \frac{M^{2\ell+3}}{C^{2\ell+2}}, \quad   J_{2\ell+1}=\frac{(-1)^{\ell}}{2\ell+3} \frac{2\Omega e^{2\ell}}{(1-e^2)^{\frac{\ell+1}{3}}}  \frac{\mathcal{R}_{n,2+2\ell}}{\xi_1^{2\ell+4}|\vartheta'(\xi_1)|} \frac{M^{2\ell+3}}{C^{2\ell+2}}, \ee
where we have defined the integral $\mathcal{R}_{n,\ell}\equiv \int_0^{\xi_1}\vartheta^n(\xi)\xi^{\ell+2}d\xi,$ and the compactness $C=M/\bar{R}$ in terms of the mean radius $\bar{R}=a(1-e^2)^{1/6}$. The expressions for the multipole moments can be combined so as to eliminate $\Omega$ and $C$ giving the final result in terms of the normalised moments $\bar{M}_{\ell}=(-1)^{\ell/2}\frac{M_{\ell}}{\chi^{\ell}M^{\ell+1}}$ and $\bar{J}_{\ell}=(-1)^{(\ell-1)/2}\frac{J_{\ell}}{\chi^{\ell}M^{\ell+1}}$. The 3-hair relations for the Newtonian moments will be,
 \be \bar{M}_{2\ell+2}+i\bar{J}_{2\ell+1}=B_{n,\ell}\bar{M}_2^{\ell}(\bar{M}_2+i \bar{J}_1), \ee
where we note that $\bar{M}_0=\bar{J}_1=1$ by definition. In this expression all the dependence on the equation of state is incorporated in the coefficient $B_{n,\ell}$ which has the form,
\be B_{n,\ell}\equiv \frac{3^{\ell+1}}{2\ell+3} \frac{\mathcal{R}_{n,0}^{\ell} \mathcal{R}_{n,2\ell+2}}{\mathcal{R}_{n,2}^{\ell+1}} ,\ee
where we can see that everything depends on the polytropic index $n$ and the corresponding $\vartheta(\xi)$ as well as the order $\ell$ of the moment. Therefore the universality of the 3-hair relations will depend on how sensitive the coefficients $B_{n,\ell}$ are to different choices of the equation of state. The numerical analysis in \cite{Stein2014ApJ} as well as analytic investigations performed by Chatziioannou et al. \cite{2014PhRvD..90f4030C} (where the coefficients $B_{n,\ell}$ are expanded around the $n=0$ solution) show that the variation of the coefficients for $\ell \leq 2$ and for polytropic indices in the range of $0\leq n \leq 1$ is less than $10\%$ around a fiducial value obtained for $n=0.6$. This means that the Newtonian moments up to $M_6$ and $J_5$ satisfy an approximate universal relation that expresses them in terms of the mass, the angular momentum and the quadrupole. Unfortunately, the variation in the coefficients $B_{n,\ell}$ increases with increasing $\ell$, therefore for higher order moments the coefficients become more sensitive to the equation of state. These results have been also tested for piecewise polytropes by Chatziioannou et al. \cite{2014PhRvD..90f4030C} and have been found to be robust. 

The question now is, what does the Newtonian analysis tells us about the relativistic 3-hair relations? The first thing we should note is that the Newtonian results are in good agreement with the relativistic results at the low compactness limit \cite{Stein2014ApJ,Yiloveq2014PhRvD} for a polytropic index $n\approx 0.5$ and the agreement extends to the relativistic polytrope as well, which means that the Newtonian calculation captures the basic elements of the relativistic problem. Furthermore it is known \cite{Lattimer2001ApJ} that realistic equations of state behave as polytropes with an effective polytropic index in the range $0\leq n \leq 1$. As it was shown in \cite{Yiloveq2014PhRvD}, the assumption in the Newtonian calculation to which the result is most sensitive is that of the isodensity surfaces being self-similar ellipses. If one were to violate this assumption by introducing a radial dependence to the eccentricity of the form $e(\tilde{r})=e_0 f(\tilde{r})$, then the result would be to change the form that the $B_{n,\ell}$ coefficients would have, introducing a dependence on the function $f(\tilde{r})$. Specifically, it was shown in \cite{Yiloveq2014PhRvD} that the larger the variation of the eccentricity inside the star, the larger the variation in $B_{n,\ell}$ in the range $n\in(0,1)$. For Newtonian polytropes with $n\in(0,1)$, one can calculate the radial profile of the eccentricity and see that the variation between the centre and the surface of the star ranges from $0\%$ for incompressible $n=0$ models up to $20\%$ for $n=1$ models, while for models with $n>1.5$ it is larger than $40\%$. Therefore the small variation we see in the Newtonian case for $B_{n,\ell}$ is related to a less than $20\%$ variation in the eccentricity profile. These results, i.e., regarding the eccentricity variation, generally also hold for relativistic polytropic stars, as well as realistic equations of state. The only difference with realistic equations of state is that lower compactness models can have a lower density atmosphere where the equation of state differs from the interior higher density equation of state and in this region the eccentricity varies more drastically. In these cases though, the outer parts of the star contribute a very small percentage of the mass of the star and have a small contribution to the various quantities. Therefore, having almost self-similar ellipses seems to be a key element in having universal 3-hair relations. 

The above results can be seen from a slightly different point of view. As it has been presented in \cite{Yiloveq2014PhRvD,2014PhRvD..90f4030C,Sham2015ApJ,Chan2015PhRvD}, one could alternatively consider the 3-hair universal relations as well as the I-Love relation in terms of how close are realistic equations of state to being incompressible, as it was discussed earlier. Specifically, in \cite{2014PhRvD..90f4030C,Sham2015ApJ,Chan2015PhRvD} various universal relations were expanded in terms of deviations from the incompressible equation of state ($n=0$ limit) and it was shown that the results were insensitive to these deviations. We remind here that there is a relation between having an incompressible equation of state or one close to incompressible and the zero or small eccentricity variation inside the star, so in a sense statements about eccentricity variation and incompressibility are equivalent. These conclusions are further strengthened by the analysis of proto-neutron stars constructed with hot equations of state that was performed by Martinon et al. \cite{Martinon2014PhRvD}. Martinon et al. found that in the first moments of the proto-neutron stars, where the equations of state have large entropy gradients, which correspond to models with high effective polytropic indices $n$ and therefore deviate significantly from being incompressible, the ellipticity profiles inside the star vary significantly while the corresponding I-Love-Q relations differ from the relations that hold for cold starts.  

{\bf Applications:} As with the I-Love-Q relations, the 3-hair relations could be used either as tools, by assuming their validity, to infer the higher order multipole moments from the mass, the angular momentum, and the quadrupole, or alternatively could be tested for their validity against observations.  

In the context of general relativity, one could use the 3-hair relations to construct more accurate spacetimes for describing the exterior of rotating neutron stars \cite{twosoliton,Berti2004MNRAS,Pappas2013MNRAS,2011CQGra..28o5015T,Pappas2015MNRAS,Tsang2016ApJ,Pappas2017MNRAS}. There exist general algorithms for constructing stationary and axisymmetric neutron star spacetimes that can be parameterised by the multipole moments. Having the moments of these spacetimes prescribed using the 3-hair relations one could have an accurate description of the neutron star spacetime that in addition depends on only three essential parameter, i.e., the first three multipole moments. This is of relevance to astrophysical observations in the electromagnetic spectrum. There is a plethora of astrophysical phenomena in the environment of neutron stars. One can observe phenomena related to matter accreting onto neutron stars that are members of X-ray binaries, such as quasi-periodic oscillations of the X-ray flux coming from the disc, the reflection spectra also coming from the disc, X-ray pulse profiles from matter accreting to the surface of the neutron star and so on. The accurate modelling of these phenomena depends on having an accurate description of the spacetime so that the motion of both matter and photons is described accurately. Furthermore, the analytic modelling in terms of only three parameters could improve our capability to solve the inverse problem of determining the equation of state from observations. As it has been discussed for the I-Love-Q case, one could in principle use combinations of observables to estimate the parameters $(M,J,Q)$ for a given neutron star spacetime, which in turn can be used to constrain the equation of state for the matter inside neutron stars \cite{Pappas2012MNRAS,Pappas2014PRL,Pappas2015MNRAS}.          

The alternative use of the 3-hair relations will be to test their validity. As in the case of I-Love-Q relations this could provide a test for our models of realistic equations of state, but the most interesting possibility will be to test general relativity itself. The last prospect will be further discussed after we have discussed neutron stars in alternative theories of gravity.

\subsubsection{Universality in Oscillation Frequencies}
\label{sec:seismology}

An extended discussion of asteroseismology falls outside the scope of our presentation. Here we will only briefly present results that are related to the equation of state independent description of the various oscillation frequencies of neutron stars. We have already mentioned some early work \cite{Andersson1996,Andersson1998MNRAS} that indicated that there could be some equation of state independent description of $f$-modes and $w$-modes and their damping times in terms of some neutron star parameters. These results were later extended by including other modern realistic equations of state \cite{Benhar2004}. As it has become clear from our discussion so far, a key element to having a universal description is the appropriate choice of parameters.  

In an attempt to extend earlier work, Tsui~\&~Leung \cite{Tsui:2004qd} and Lau et al. \cite{Lau2010ApJ} used a different type of normalisation, more similar to the I-Love-Q type of normalisations, that has demonstrated a higher degree of equation of state independence\footnote{One should always keep in mind that at the end of the day what matters most is not exactly the errors of the fitting function but instead the accuracy of solving the inverse problem, i.e. determining the stellar parameters from the observed gravitational wave frequencies and damping times.}. The approach by Lau et al. \cite{Lau2010ApJ}, inspired by \cite{Lattimer2005ApJ} where a universal behaviour for the normalised moment of inertia was observed, turned out to be more useful since both neutron and quark stars could be fitted with a single relation. Lau et al. \cite{Lau2010ApJ} investigated the relation between the QNM frequency (both real $\omega_r$ and imaginary $\omega_i$ parts) of the (fundamental) $f$-mode oscillations to the mass and moment of inertia of compact stars. In contrast though to \cite{Tsui:2004qd} and earlier work, Lau et al. convincingly argue that a better quantity instead of the compactness $\cC$ to characterise models of different internal mass profiles would be the inverse square root of the normalised moment of inertia, $\eta\equiv\bar{I}^{-1/2}=\sqrt{M^3/I}$, which they call effective compactness. The reason is that this quantity is on the one hand a measure of the compactness\footnote{If we approximate the moment of inertia by $MR^2$ then we can see that $\eta=M/R=\cC$.} while on the other hand it also takes into account the distribution of mass inside the star and it could in this way counterbalance differences coming from having a stiffer or softer equation of state. 
The universal relations that Lau et al. found to describe the $f$-mode in terms of the effective compactness are,\footnote{See the relevant comment in \cite{Chan2014PRD} for correcting some typos in the numerical coefficients given in \cite{Lau2010ApJ}.} 
\be M\omega_r = -0.0047 + 0.133\eta + 0.575\eta^2, \quad \textrm{ and }\quad \bar{I}^2 \omega_i M =0.00694-0.0256\eta^2, \label{eq:fmode}\ee 
and describe the real and imaginary parts of the $\ell=2$ $f$-mode frequency with an accuracy that is better than $1-2\%$ for both neutron and quark stars. These two relations could be used to determine the mass and the moment of inertia of a compact object, if the $f$-mode was detected by gravitational waves. We should note here that this discussion is for non-rotating compact stars. Combining this work to the I-Love-Q results, Chan et al. \cite{Chan2014PRD} extended the relation between the $f$-mode frequency and the moment of inertia to include also the Love number. They further proceeded to produce relations between higher order oscillation modes and higher order Love numbers (for a review see \cite{YagiYunes2016arXiv}).

The $f$-mode asteroseismology relations were further investigated and more equations of state were included by Bl\'azquez-Salcedo et al. \cite{Blazquez-Salcedo2014}  and Chirenti et al. \cite{Chirenti2015PRD}, where in addition an alternative relation for the $f$-mode damping time was proposed in terms of the compactness. The asteroseismology relations with $w$-modes were investigated further by Tsui~\&~Leung \cite{Tsui2005,Tsui2005a} and Bl\'azquez-Salcedo et al. \cite{Blazquez-Salcedo2013} with a special emphasis on the influence of the presence of hyperons and quarks in the neutron star core in the latter papers.

Up to this point the discussion has been about the quasi-normal modes of non-rotating stars. In the case of rotating stars things become a little more complicated. Due to rotation the frequencies of modes with the same spherical mode number $\ell$ but opposite azimuthal index $m = \pm|m|$, which correspond to co- and counterrotating modes, separate and become distinct. In this case, one needs to express the modes in the comoving frame in order to remove the complications that the rotation is creating. 
Rapidly rotating models in the so-called Cowling approximation, where the spacetime is assumed to be frozen (see App.~\ref{sec:app:Cowling}), were studied by Gaertig~\&~Kokkotas in \cite{Gaertig:2010kc} for polytropic equations of state were the aforementioned behaviour was observed. 
For a star rotating at a frequency $\Omega$, the relation between the mode frequencies in the inertial frame and the comoving frame is $\omega_{in}=\omega_{co}- m\Omega$. Gaertig~\&~Kokkotas observed in \cite{Gaertig:2010kc} that even though the mode frequencies in the inertial frame showed a large variation with the equation of state, the corresponding frequencies in the comoving frame were quite insensitive to the equation of state. 
The real part of the quasi-normal modes in the comoving frame for the $m=-2$ stable modes, $\omega^s_{co}$, can be fitted for polytropic equations of state with the expression
\be \frac{\omega^s_{co}}{\omega_0}=1.0-0.27\left(\frac{\Omega}{\Omega_K} \right) - 0.34 \left(\frac{\Omega}{\Omega_K}  \right)^2,\ee
where the frequency is normalised with respect to the frequency of the non-rotating models $\omega_0$, while the rotation frequency $\Omega$ is normalised with the Kepler limit frequency $\Omega_K$. Similarly, the $m=2$ potentially unstable modes, $\omega^u_{co}$, frequency is fitted with the expression,
\be \frac{\omega^u_{co}}{\omega_0}= 1.0 + 0.47 \left(\frac{\Omega}{\Omega_K} \right) - 0.51\left(\frac{\Omega}{\Omega_K} \right)^2.\ee 
The non-rotating frequency $\omega_0$ can be expressed in terms of the average density and has the form,
\be \frac{\omega_0}{2\pi} = 0.498 + 2.418 \left(\frac{M}{R^3}\right)^{1/2}, \label{eq:fmode2}\ee 
where the mass is measured in units of $1.4M_{\odot}$ and the radius of the star in units of 10km. One will notice that eq.~\eqref{eq:fmode2} is slightly different than eq.~\eqref{eq:fmode1} from \cite{Andersson1998MNRAS}. This is because in \cite{Gaertig:2010kc}, the calculations are done in the Cowling approximation which tends to overestimate the values of the frequencies. Also, the reference to stable and potentially unstable modes here is with respect to the Chandrasekhar, Friedman \& Schutz (CFS-)instability of non-axisymmetric pulsation modes (see App.~\ref{sec:app:CFS}). 
The imaginary part of the quasi-normal modes will give the damping times for the different models. Things in this case are a little more complicated and damping times can be better expressed in terms of the frequencies in the inertial or the comoving frame. For the retrograde branch ($m>0$) the fit for the damping times has the form,
\be \frac{\tau_0}{\tau}= sgn(\omega_{in}^u)\times 0.256 \left(\frac{\omega_{in}^u}{\omega_0}\right)^4\times\left[1+0.048 \left(\frac{\omega_{in}^u}{\omega_0}\right)+0.359 \left(\frac{\omega_{in}^u}{\omega_0}\right)^2\right]^4,  \ee
where $\tau_0$ is the damping time of the non-rotating models and $sgn(x)$ is the sign function. Similarly for the prograde branch ($m<0$) the fit for the damping times has the form,
\be \frac{\tau_0}{\tau}=-0.656\times\left[1-7.33 \left(\frac{\omega_{co}^s}{\omega_0}\right)+14.07 \left(\frac{\omega_{co}^s}{\omega_0}\right)^2-9.26 \left(\frac{\omega_{co}^s}{\omega_0}\right)^3\right].  \ee
The damping time of the non-rotating models can be given by the equation 
\be \left(\tau_0 \frac{M^3}{R^4}\right)^{-1}=22.49-14.03 \left(\frac{M}{R}\right),  \label{eq:fmodeDamp2} \ee
where again the mass is measured in units of $1.4M_{\odot}$ and the radius of the star in units of 10km. One will notice that eq.~\eqref{eq:fmodeDamp2} is much closer to eq.~\eqref{eq:fmodeDamp1}. 
These results were extended to higher order modes and realistic equations of state by Doneva et al. \cite{Doneva:2013zqa}, where it was found that higher modes exhibit the same behaviour.
We refer the reader to \cite{Doneva:2013zqa} for a detailed presentations of these relations.

A more recent approach that is closer to the I-Love-Q perspective using a parameterisation in terms of the effective compactness $\eta$ was investigated by Doneva~\&~Kokkotas \cite{DonevaKokkotas2015PRD} and was found to be even more universal with respect to different equations of state.
The frequencies for unstable modes in the inertial frame for realistic equations of state can be expressed for different values of $\ell$ as, 
\bear
\omega_{in}^u M&=&\left[(-0.332-0.725 \ell)+(0.085-0,111\ell)\hat{\Omega}+(0.0112-0.00903\ell)\hat{\Omega}^2\right]\nonumber\\
            &&+\left[(1.755+0.955\ell)+(-0.0165-0.0149\ell)\hat{\Omega}+(-0.00501+0.00355\ell)\hat{\Omega}^2\right]\eta,
\eear
where $\hat{\Omega}=M\Omega$ and is given in units of $[M_{\odot}\, \textrm{kHz}]$ and $\eta$ is expressed in units of $\sqrt{\frac{M_{\odot}^3}{10^{45}g cm^2}}$. Similarly, the stable $\ell=2$ mode in the comoving frame is given by the equation,
\be
\omega_{c}^s M=\left(-1.66-0.249\hat{\Omega}\right)+\left(3.66+0.0633\hat{\Omega}\right)\eta.
\ee
The corresponding $f$-mode damping times for the potentially unstable branches can be given by the second order polynomial,
\be
    \eta \left(\frac{M}{\tau \eta^2}\right)^{1/2\ell}=c_1(M\omega_{in}^u)+c_2(M\omega_{in}^u)^2,
\ee
where the coefficients $c_1$ and $c_2$ are constants that depend on $\ell$. More details and tables for $c_1,c_2$ can be found in \cite{DonevaKokkotas2015PRD}.

We should again stress at this point that these results are in the Cowling approximation and should be taken with a grain of salt. An obvious necessary extension of these results is to go beyond the Cowling approximation. Moreover, some improvement in the universality might also come, as we have seen in the case of I-Love-Q relations for rapidly rotating stars, if the rotation is parametrised in terms of the spin parameter $\chi=J/M^2$ instead of $\Omega$. 

\subsubsection{Other Universal properties}

Having reviewed the three main classes of universal relations we will briefly mention here some additional results that either extend those that we have already discussed or go in a different direction. 

One of the extensions of the I-Love-Q results was the multipole Love relations by Yagi \cite{Yagi2014PhRvD..89d3011Y}, where he produced universal relations among various $\ell$-th (dimensionless) electric, magnetic and shape tidal deformabilities for neutron stars and quark stars, in the same sense as we have the various universal relations between the quadrupole and higher order multipole moments. His results are also reviewed in \cite{YagiYunes2016arXiv}. We should note that the tidal Love numbers were calculated for non-rotating compact objects. 

Another extension of the I-Love-Q and 3-hair relations, with a more theoretical interest, is the one for anisotropic stars by Yagi~\&~Yunes   \cite{YagiYunes2015PhRvD..91j3003Y,YagiYunes2015PRD,YY2016CQGra..33i5005Y}. As we have noted, when we calculate the multipole moments for example of compact objects we see that as we increase the compactness up to the maximum possible value for these objects, we never quite get to the Black Hole limit of $\bar{Q}=\bar{J}_3=\bar{M}_4=\ldots=1$. Instead there seems to be a gap between fluid configurations and Black Holes. Of course this is expected with respect to compactness, because a matter configuration cannot be more compact than the Buchdahl limit, i.e., $\cC<4/9\approx0.\bar{4}$ \cite{BuchdahlPhysRev.116.1027}. The Buchdahl limit though is not a hard physical limit. One can go around it by using anisotropic pressure, which can produce objects with a compactness arbitrarily close to the Black Hole value of $\cC=1/2$ \cite{1974ApJ...188..657B}. For this reason Yagi~\&~Yunes \cite{YagiYunes2015PhRvD..91j3003Y} used anisotropic stars to study how one approaches the Black Hole limit for the multipole moments and the other neutron star parameters. For anisotropic stars one has the radial pressure $p$ and a tangential part of the pressure $p_{\perp}$ and the difference defines the anisotropy $\sigma=p-p_{\perp}$. In the model that they used, neutron stars with negative anisotropy reached continuously the Black Hole limit. 
The exact way that the Black Hole limit is reached though depends on the value of the anisotropy. The very interesting result was that there is a critical value of anisotropy above which the models reach the Black Hole limit directly, while bellow that value the models circle around the Black Hole limit until they almost spiral to it. Finally, in order to get to the Black Hole values one needs to have an ``extreme'' value of anisotropy. 
On a more practical side, there is also the question of how some anisotropy would affect the various universal relations of regular neutron stars. On that front Yagi~\&~Yunes \cite{YagiYunes2015PRD} find that anisotropy affects the universal relations only weakly, i.e., the relations become less universal by a factor of 1.5-3 relative to the isotropic case when anisotropy is maximal, but even then they remain approximately universal to 10\%. They further find that this increase in variability is strongly correlated to an increase in the eccentricity variation of isodensity contours.

The results presented so far are mostly about isolated neutron stars, with the exception of the work by Maselli et al. \cite{Maselli2013PRD} where they studied I-Love-Q for a binary system. Extending this work Yagi~\&~Yunes investigated the existence of universal relations between tidal parameters for binaries \cite{YY2016CQGra..33mLT01Y,YagiYunes2017CQG} which they called binary Love relations. These relations are between the individual tidal Love numbers of the members of the binary $\bar{\lambda}_1,\bar{\lambda}_2$, as they are combined to form the symmetric $\bar{\lambda}_s$ and antisymmetric $\bar{\lambda}_a$ tidal parameters, and the so-called dimensionless chirp tidal deformability $\bar{\Lambda}$ and it's companion parameter $\delta\bar{\Lambda}$. The idea behind this is that the parameters $\bar{\Lambda}$ and $\delta\bar{\Lambda}$ enter the analysis of the waveforms from binary inspirals with the dimensionless chirp tidal deformability $\bar{\Lambda}$ being the dominant tidal parameter in analogy with the chirp mass being the dominant mass parameter in the waveform. Yagi~\&~Yunes find universal relations that relate $\bar{\Lambda}$ to $\delta\bar{\Lambda}$ as well as $\bar{\lambda}_s$ to $\bar{\lambda}_a$, that depend on the symmetric mass ratio $X\equiv\frac{q}{(1+q)^2}$, where $q$ is the mass ratio of the members of the binary. These results are also reviewed in \cite{YagiYunes2016arXiv}. 

Continuing with binary inspirals, there is another recent interesting result. Initially Read et al. \cite{Read2013PRD} found, by modelling numerically the waveforms from binary inspirals and coalescences (spinning, equal mass, $q=1$), that the instantaneous frequency at the moment of maximum amplitude in the waveform, which indicates the transition from the inspiral phase to the merger or coalescence phase, is related in an equation of state independent way with a parameter which is essentially the tidal Love number $\bar{\lambda}$ of one of the neutron stars defined in eq.~\eqref{eq:lovenumb}. 
Specifically they find the relation 
\be \log_{10}(f_{GW}/\textrm{Hz})=8.51155-0.303350 \,\bar{\lambda}^{1/5}.  \ee
Since in their numerical calculations they considered only equal mass inspirals, essentially one tidal Love number characterises both neutron stars.  
Expanding on this result, Bernuzzi et al. \cite{Bernuzzi2014PRL} showed that for more general inspirals, where $q\neq1$, a more appropriate quantity should be used, which is essentially an effective tidal Love number, called tidal coupling constant and is defined as, 
\be \kappa_{\ell}\equiv 2 \left[ \frac{1}{q}\left(\frac{X_A}{\cC_A}  \right)^{2\ell+1} k_{\ell}^A+ q\left( \frac{X_B}{\cC_B} \right)^{2\ell+1} k_{\ell}^B\right],\ee
where the definition is with respect to different $\ell$-orders in tidal deformabilities and we define the mass ratio to be $q\equiv M_A/M_B\geq1$, and for each member of the binary with respect to the total mass $X_A\equiv M_A/M$ and $X_B\equiv M_B/M$, where $M=M_A+M_B$ is the total mass, while $\cC_{A,B}$ are the respective compactnesses and $k_{\ell}^{A,B}$ the respective $\ell$th order tidal apsidal constants.\footnote{Here we are using the naming convention employed in subsection \ref{sec:iloveq}, which is different than the one used in \cite{Read2013PRD,Bernuzzi2014PRL}.} For $\ell=2$ one can see that $\kappa\equiv\kappa_{\ell}\propto \bar{\lambda}$ and the results of  Bernuzzi et al. reproduce the results of Read et al. quite accurately. Using the tidal coupling constant, Bernuzzi et al. produced an empirical relation for the mass scaled instantaneous frequency at peak amplitude in terms of $\kappa$, which is 
\be M\omega_{GW}=0.360\frac{1+2.59\times 10^{-2}\kappa-1.28\times10^{-5}\kappa^2}{1+7.49\times10^{-2} \kappa} , \ee
and also produced a relation for the binding energy per reduced mass $\mu=M_AM_B/M$ in terms of $\kappa$, which is 
\be E_b=-0.120\frac{1+2.62\times 10^{-2}\kappa-6.32\times10^{-6}\kappa^2}{1+6.18\times10^{-2} \kappa} . \ee
These relations were found to be insensitive to variations in the mass ratio $q$ and showed small sensitivity to the spin of the members of the binary, probably due to spin-orbit coupling (the calculations were performed for aligned spin binaries). 
These results were reproduced by Takami et al. \cite{Takami2014PRL,Takami2015PRD}, where they also find by doing spectral analysis of the waveforms, that there is an equation of state independent relation between the lowest observed frequency $f_1$ of the spectrum and the average compactness $\bar{\cC}=\bar{M}/\bar{R}$, where $\bar{M}=(M_A+M_B)/2$ and $\bar{R}=(R_A+R_B)/2$, which is
\be \frac{f_1}{\textrm{kHz}}= (-22.0717\pm6.477)+(466.616\pm31.2)\bar{\cC} +(-3131.63\pm878.7)\bar{\cC}^2+(7210.01\pm1947)\bar{\cC}^3. \ee

Returning to isolated neutron stars, there are some more results that we would like to conclude with. 
Baub\"ock et al. \cite{Baubock2013ApJ} produced in the slow rotation approximation (using the Hartle~\&~Thorne approach) analytic formulae that relate the ellipticity and eccentricity of the stellar surface to the compactness, the spin parameter, and the quadrupole moment of the neutron star. Defining the ellipticity $\varepsilon_s$ and the eccentricity $e_s$ as,
\be \varepsilon_s\equiv 1-\frac{R_p}{R_{eq}},\quad \textrm{ and } \quad e_s\equiv \sqrt{\left(\frac{R_{eq}}{R_p}\right)^2-1}, \ee
where $R_p$ and $R_{eq}$ are the polar and equatorial radius respectively, they found in terms of the compactness $\cC$, the spin parameter $\chi_K$ and the quadrupole $\bar{q}_K$, the relations
\bear
e_s^K(\cC,\chi_K,\bar{q}_K)&=&\left[1-4\chi_K\cC^{3/2}+\frac{15(\chi_K^2-\bar{q}_K)(3-6\cC+7\cC^2)}{8\cC^2}+\cC^2\chi_K^2(3+4\cC)\right.\nonumber\\
&&\left.+\frac{45}{16 \cC^2}(\bar{q}_K-\chi_K^2)(\cC-1)(1-2\cC+2\cC^2)\ln\left(1-2\cC\right)\right]^{1/2},\\
\varepsilon_s^K(\cC,\chi_K,\bar{q}_K)&=&\frac{1}{32\cC^3}\left\{2\cC\left[8\cC^2-32\chi_K\cC^{7/2}+(\chi_K^2-\bar{q}_K)(45-35\cC+60\cC^2+30\cC^3)+24\chi_K^2\cC^4+8\chi_K^2\cC^5-48\chi_K^2\cC^6\right]\right.\nonumber\\
&&+\left. 45(\chi_K^2-\bar{q}_K)(1-2\cC)^2\ln\left(1-2\cC\right)\right\}.
\eear
In these relations, the compactness is defined in terms of the non-rotating models' masses and radii, while the spin parameter $\chi_K=J_K/M_*^2$, the quadrupole $\bar{q}_K=-Q_K/M_*^3$ and the ellipticity $\varepsilon_s^K$ and eccentricity $e_s^K$ correspond to the models that rotate at the frequency $\Omega_K$, as it is discussed in section \ref{sec:iloveq}. The ellipticities and eccentricities of models rotating with a rotation $\Omega<\Omega_K$ will be $\varepsilon_s=\epsilon^2 \varepsilon_s^K$ and $e_s=\epsilon e_s^K$ respectively, where $\epsilon=\Omega/\Omega_K$. The above exact expressions hold independently of the equation of state. All the equation of state information is encoded in the various parameters. AlGendy~\&~Morsink \cite{AlGendy2014ApJ} also produce an equation of state independent fit for the ellipticity, where they use a slightly different parameterisation for the surface than the one used in the Hartle~\&~Thorne approach. They parameterise the surface with a function of the form,\footnote{The radius here is expressed in terms of Schwarzschild-like coordinates which reduce to being circumferential at the non-rotating limit. The radial coordinates in the Hartle~\&~Thorne approach also have this property.}   
\be R(\theta)=R_{eq}\left(1+o_2(\cC,\epsilon)\cos^2\theta\right) ,\ee
where the function $o_2(\cC,\epsilon)$, which is minus the ellipticity defined above, has the form 
\be o_2(\cC,\epsilon)=\epsilon^2(-0.788+1.030 \cC),\ee
where $\epsilon=\Omega/\Omega_K$ as in the Hartle~\&~Thorne case. Their results are in good agreement with previous results for rapidly rotating neutron stars. Additionally they find that the effective gravity at the surface of a rotating neutron star can be written as the simple function $g(\theta)/g_0 = c(\cC,\epsilon^2,\theta)$, where $g_0$ 
is the acceleration due to gravity on the surface of a non-rotating relativistic star while the function $c$ can be written in an equation of state independent form. 

Finally, Breu~\&~Rezzolla \cite{Breu2016MNRAS} find that the mass of rotating configurations on the turning-point line shows a universal behaviour when expressed in terms of the spin parameter at the Kepler limit. In particular they find that the mass at the turning point, $M_{crit}$ normalised by the maximum non rotating mass $M_{TOV}$, is expressed in terms of the spin parameter, $\chi$ normalised by the spin parameter at the Kepler limit $\chi_K$, as 
\be \frac{M_{crit}}{M_{TOV}} = 1+ 0.1316 \left( \frac{\chi}{\chi_K} \right)^2 + 0.07111 \left( \frac{\chi}{\chi_K} \right)^4.\ee
This expression implies that the maximum mass for any given equation of state will be, for $\chi=\chi_K$, $M\simeq 1.203 M_{TOV}$.\footnote{The interested reader might want to also have a look at the review by Hartle \cite{Hartle1978PhR....46..201H}.}  Of course $M_{TOV}$ will depend on the equation of state.
In addition, they further explore the $\bar{I}=I/M^3$ relation to inverse powers of $\cC$, where they find that there is a universal relation (that holds up to 10\%) that also depends on the spin $\chi$, as expected. In \cite{Breu2016MNRAS} they give fitting coefficients for different values of $\chi$, while here we will give a relations that is spin dependent, i.e., 
\be I/M^3 =(1.471 +0.448 \chi) -\frac{0.0802+0.27289 \chi }{\cC}+ \frac{0.438 -0.0346 \chi}{\cC^2}-\frac{0.01694+0.0056 \chi}{\cC^3}+\frac{(3.316+1.57 \chi)\times10^{-4}}{\cC^4} \label{eq:icfit} \ee
which is also accurate to 10\% and is in agreement with the relations given in \cite{Breu2016MNRAS,Staykov2016PhRvD} .
%
\begin{figure}[htb]
\begin{center}
\begin{tabular}{ccc}
\includegraphics[width=5.55cm,clip=true]{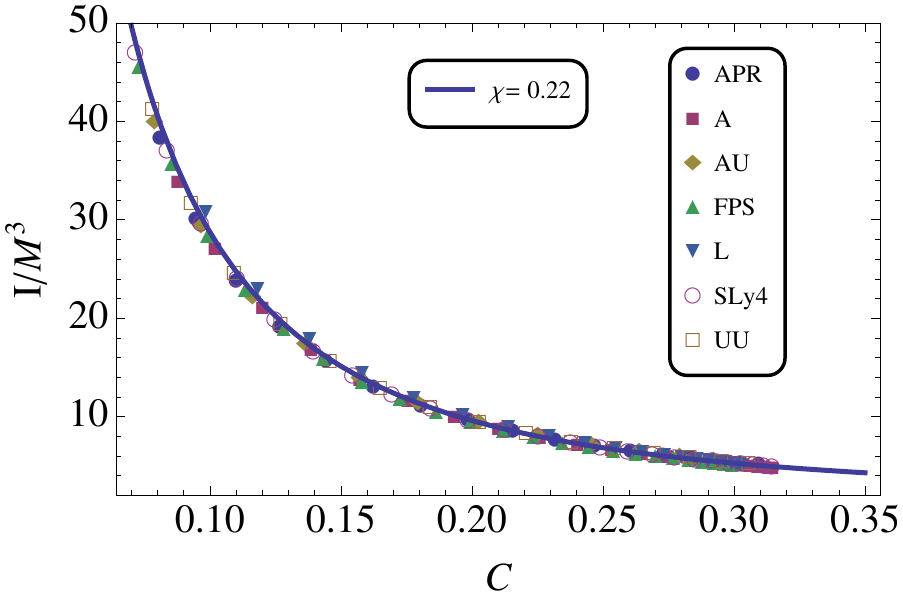}&\includegraphics[width=5.5cm,clip=true]{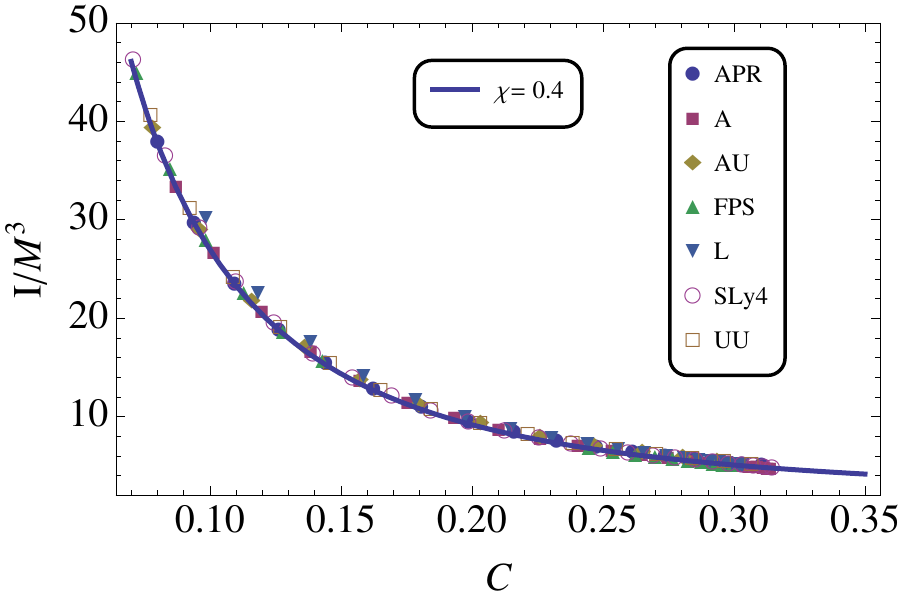} &\includegraphics[width=5.5cm,clip=true]{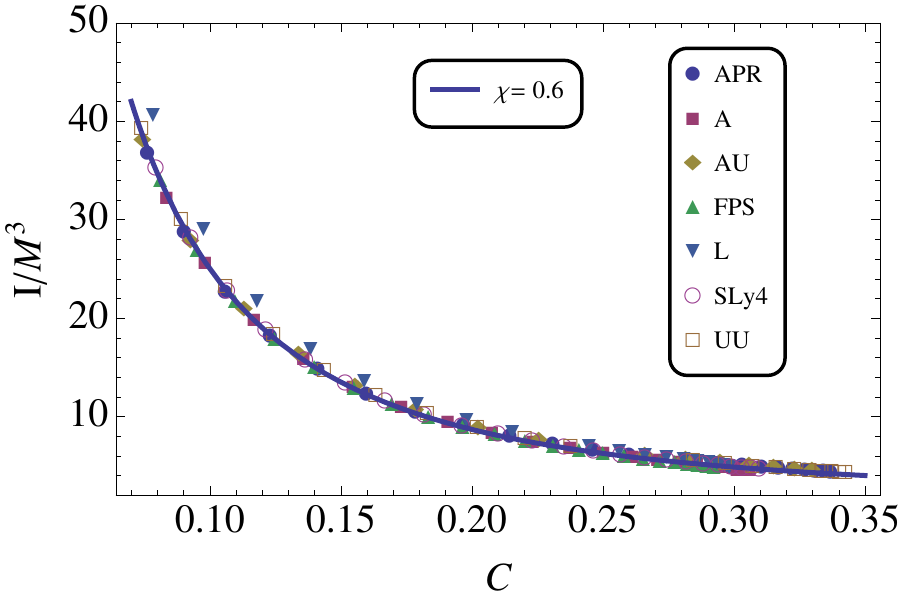} 
\end{tabular}
\caption{\label{fig:icrapidJ} I-C plots for three spin parameters and various equations of state. The solid lines correspond to the I-C fit given in eq.~\eqref{eq:icfit}.
}
\end{center}
\end{figure}

\subsection{Neutron stars in alternative theories}
\label{sec:past2}
Below we will review the basic theory behind the alternative theories of gravity we have chosen to concentrate on, the neutron star models in these theories and their astrophysical implications. We will consider the scalar-tensor theories, $f(R)$ theories, Einstein-dilaton-Gauss-Bonnet theories and Chern-Simons theories of gravity. These are all theories for which neutron stars universal relations have been considered. The strongest emphasis will be on the scalar-tensor theories that are ones of the simplest and most natural extensions of general relativity. 

\subsubsection{Alternative theories of gravity: mathematical setup and overview}

\textbf{\emph{Scalar-tensor theories of gravity: }} 
The simplest representative of gravitational theories with an additional scalar field are the scalar-tensor theories of gravity. Historically, these are some of the first alternative theories to be constructed. The reason lies on one hand in their simplicity and the fact that such generalization of Einstein's theory seems quite natural. On the other hand there is a deep theoretical motivation coming from the theories trying to unify all the interactions such as Kaluza-Kelin theories, string theories, etc. It can be easily shown for example that after a dimensional reduction of the five dimensional Kaluza-Klein theory action to four dimensions, an additional scalar field appears as a mediator of the gravitational interaction. The scalar-tensor theory of gravity can be formulated also independently and that was done first by Brans and Dicke on the basis of Mach's principle \cite{Brans1961,Dicke1962}. According to it, the inertia of a particle is a consequence of the total mass distribution in the Universe. Therefore, the inertial mass is not a constant but also depends on the mass distribution, i.e. it depends on the interaction of the particle with some cosmological field $\Phi$. According to the weak equivalence principle, which is verified extremely precisely, the interaction with this scalar field should be the same for all the particles up to some constant. Thus the mass of the particles should be
\begin{equation}
m=m_0 f(\Phi),
\end{equation}
where $f(\Phi)$ describes the interaction of the particles with the scalar field. Using the above considerations and also the fact that the absolute scale of the masses of the particles can be measured only by their gravitational acceleration, we can conclude that the gravitational constant should depend on the total distribution of the matter in the universe, i.e. from the cosmological field $\Phi$. Using all this, one can derive the scalar-tensor theory action and a very interesting fact is that its vacuum sector  coincides with the action coming from the dimensional reduction of the Kaluza-Klein theory. 

The most general form of the action in the scalar-tensor theories is \cite{Damour1992}
\begin{eqnarray}
S=\frac{1}{16\pi G_*} \int{d^4 x \sqrt{-\tilde{g}}\left(F(\Phi)\tilde{R} - Z(\Phi)\tilde{g}^{\mu\nu}\partial_\mu \Phi
	\partial_\nu \Phi - 2U(\Phi)\right)}+ S_{\rm matter}(\tilde{g}_{\mu\nu}, \chi), \label{eq:Action_JF}
\end{eqnarray}
where $G_*$ is the bare gravitational constant and as a matter of fact the scalar field $\Phi$ can be interpreted as a ``variable gravitational constant''. The second and the third them in the action are  the kinetic and potential terms for the scalar field respectively. The choice of functions $F(\Phi)$, $Z(\Phi)$, and $U(\Phi)$ determine the specific class of scalar-tensor theory. The requirement that the gravitons carry
positive energy leads to the following condition $F(\Phi)>0$, while the non-negativity of the scalar field
energy requires that $2F(\Phi)Z(\Phi) + 3[dF(\Phi)/d\Phi]^2\geq 0$. It is important to point out, that the action of the matter $S_{\rm matter}(\tilde{g}_{\mu\nu}, \chi)$,  where $\chi$ denotes collectively the matter fields, is the same as in general relativity. Therefore, there is no direct coupling between the matter and the scalar field, and the scalar field influences the matter only via the spacetime metric $\tilde{g}_{\mu\nu}$. Thus, the equation of motion is the same as in general relativity ${\tilde \nabla}_\alpha {\tilde T}^{\alpha}_{\mu}=0$ and the weak equivalence principle is satisfied.

The action (\ref{eq:Action_JF}) is in the so-called Jordan frame which is the physical frame where distances, time, etc. are measured. The field equations, though, are quite complicated and difficult to work with in this frame. That is why a common practice is to introduce the so-called Einstein frame which simplifies the  field equations considerably. The transition to the Einstein frame is made by performing a conformal transformation of the metric\footnote{For the interesting case of disformal coupling we refer the reader to \cite{Minamitsuji2016}.} and a re-definition of the scalar field:
\begin{eqnarray}\label{CONFTRANS}
&g_{\mu\nu}=F(\Phi) \tilde{g}_{\mu\nu}, \\ \notag \\
&\left(\frac{d\varphi}{d\Phi}\right)^2=\frac{3}{4}\left(\frac{d \ln(F(\Phi))}{d\Phi}\right)^2 +
\frac{Z(\Phi)}{2F(\Phi)},
\end{eqnarray}
where $\varphi$ and $g_{\mu\nu}$ are the Einstein frame scalar field and metric. After we introduce the functions ${\cal A} (\varphi)$ and $V(\varphi)$ defined as
\begin{eqnarray}
&{\cal A} (\varphi) = F^{-1/2}(\Phi), \label{eq:A(phi)}\\
&V(\varphi) = \frac{1}{2} U(\Phi) F^{-2} (\Phi),
\end{eqnarray}
the Einstein frame action becomes
\begin{equation}
S=\frac{1}{16\pi G_*} \int{d^4 x \sqrt{-g}\left(R - 2g^{\mu\nu}\partial_\mu \varphi \partial_\nu \varphi -
	4V(\varphi)\right)} + S_{\rm matter}({\cal A}^2(\varphi) \tilde{g}_{\mu\nu}, \chi), \label{eq:Action_EF}
\end{equation}
where $R$ is the Ricci scalar curvature with respect to the spacetime metric $g_{\mu\nu}$. As we can see the action in the Einstein frame is much simpler and easier to work with, but everything comes with a price and the price we pay for this simplicity is that a direct coupling between the sources of gravity and the scalar field appears through the function  ${\cal A}(\varphi)$. Nevertheless, the resulting field equations are easier to handle and most of the studies of compact objects in scalar-tensor theories adopt the Einstein frame. The specific choice of the scalar-tensor theory is completely determined by the function ${\cal A}(\varphi)$ and by the potential of the scalar field $V(\varphi)$.

The transformation between the two frames is regular practically for all of the physically relevant scalar-tensor theories. Thus, it is irrelevant in which frame we are going to perform our calculations as long as the final observable quantities are expressed in the physical Jordan frame. 

The field equations in the Einstein frame are much simpler as well
\begin{eqnarray}
&R_{\mu\nu}=8\pi G_*\Big( T_{\mu\nu}-\frac{1}{2} g_{\mu\nu} T \Big)+ 2\partial_\mu \varphi \partial_\nu \varphi -\frac{1}{2}g_{\mu\nu} V(\varphi), \label{eq:FieldEqEF_g} \\
&\nabla_\alpha \nabla^\alpha \varphi -\frac{1}{4}V'(\varphi)=-4\pi\alpha(\varphi) T, \label{eq:FieldEqEF_phi}
\end{eqnarray}
where $\nabla_\alpha$ is the covariant derivative with respect to the metric $g_{\mu\nu}$ and $T$ is the trace of the Einstein frame energy momentum tensor $T_{\mu\nu}$. The function $\alpha(\varphi)$ is called coupling function and it is defined as
\begin{equation}
\alpha(\varphi)=\frac{d\ln A(\varphi)}{d\varphi}. \label{eq:alpha_phi}
\end{equation}
The connection between the Einstein frame and the Jordan frame energy momentum tensor is
\begin{equation}
T_{\mu\nu}=A^4(\varphi)\tilde{T}_{\mu\nu}.
\end{equation}
Since we have a direct coupling between the matter and the scalar field in the Einstein frame, the equations of motion for the matter fields differ from that in pure general relativity:
\begin{equation}
\nabla_\alpha T^{\alpha}_{\mu}=\alpha(\varphi)T\partial_\mu \varphi.
\end{equation}
Therefore, there will be an additional force acting on the particles in the Einstein frame which is proportional to the gradient of the scalar field and the
particles will not move on the geodesic of the metric $g_{\mu\nu}$.

Let us briefly comment on the field equations and more precisely under what circumstances nontrivial scalar field can develop. For simplicity we will consider the case with zero scalar field potential. The right hand side of equation  \eqref{eq:FieldEqEF_phi} is nonzero only for nonzero trace of the energy momentum tensor $T$. For isolated black holes $T=0$ which roughly speaking leads to the fact that the solutions are the same as in pure general relativity, but for neutron stars $T$ is nonzero and thus nontrivial scalar field can develop.

Let us now discuss briefly the parametrized post-Newtonian (PPN) expansion of the metric in scalar-tensor theories and the observational constraints one can impose, since in general different theories of gravity predict different values of the post-Newtonian parameters. In what follows we will again assume that the scalar field potential is zero for simplicity. The coupling function $\alpha(\varphi)$ on the other hand can be expanded in power series of the scalar field $\varphi$
\begin{equation}
\alpha(\varphi) = \alpha_0(\varphi-\varphi_0) +\frac{1}{2}\beta (\varphi-\varphi_0)^2 + O(\varphi-\varphi_0)^3,
\label{eq:Aphi_expansion}
\end{equation}
where $\varphi_0$ is the background value of the scalar field and $\alpha_0$ and $\beta$ are constants. The case with $\beta=0$ corresponds to the well known Brans-Dicke scalar-tensor theory. The PPN expansion of the Schwarzschild metric in isotropic coordinates takes the form \cite{Damour1992,Will:2014xja}:
\begin{eqnarray}
 &-g_{00}= 1 - 2\frac{Gm}{rc^2} + 2 \beta^{\rm PPN} \left(\frac{Gm}{rc^2}\right)^2 +
 O {\left({1 \over {c}^{6}}\right)}, \label{eq:gmunu_PPN1}\\
 &g_{ij}=\delta_{ij}\left(1+2\gamma^{\rm PPN}\frac{Gm}{rc^2}\right) + O\left(\frac{1}{c^4}\right)\,.
 \label{eq:gmunu_PPN2}
\end{eqnarray}
As one can see, in scalar-tensor theories only two of the PPN parameters differ from GR -- $\beta^{PPN}$ and $\gamma^{PPN}$ which were introduced for the first time by Eddington for the Schwarzschild metric in isotropic coordinates. They are connected to the coefficients in the expansion of the coupling function \eqref{eq:A(phi)} in the following way:
\begin{equation}
 \gamma^{\rm PPN}-1 = - \frac{2 \alpha_0^2}{1+\alpha_0^2}\,, \qquad \beta^{\rm PPN}-1 = \frac{1}{2}\,
 \frac{\alpha_0\beta\alpha_0}{(1+\alpha_0^2)^2}\,. \label{PPN}
\end{equation}
Clearly, for $\alpha_0=0$ and $\beta=0$ we have $\gamma^{\rm PPN}=0$ and $\beta^{\rm PPN}=0$ and the theory reduces to pure general relativity. 

\begin{figure}
	\includegraphics[width=8.5cm]{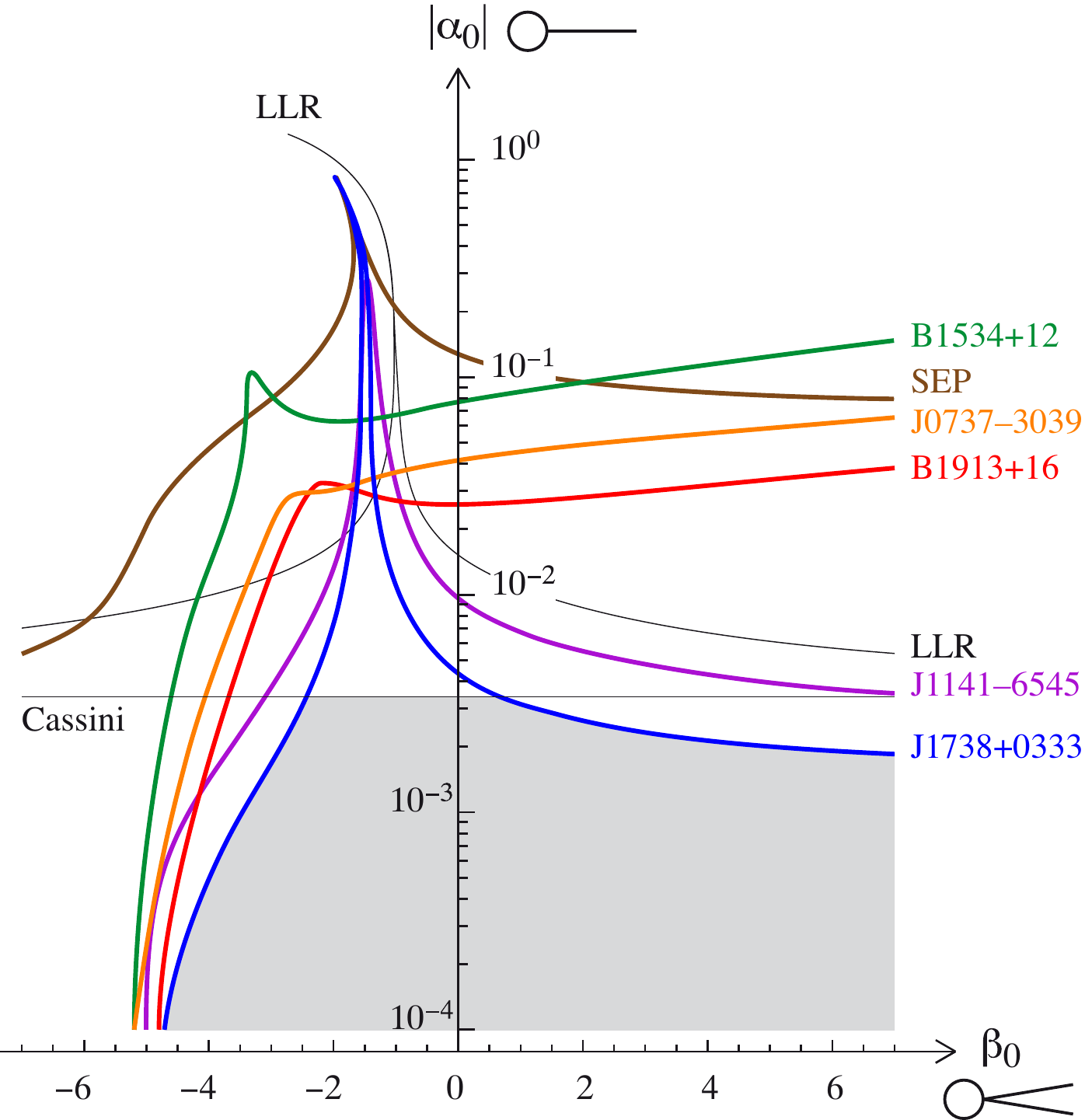}
	\caption{\label{fig:STT_observations}
		Solar-system and binary pulsar constraints
		on the constant $\alpha_0$ and $\beta$ in the expansion of the coupling function \eqref{eq:Aphi_expansion}. LLR stands for 
		lunar laser ranging, Cassini for the measurement of a Shapiro time-delay
		variation in the Solar System, and SEP for tests of the strong 
		equivalence principle using a set of neutron star-white dwarf low-eccentricity 
		binaries. [Credit Ref. \cite{Freire2012}].}
\end{figure}

Different observations can impose constraints on the PPN parameters  $\gamma^{\rm PPN}$ and $\beta^{\rm PPN}$ in the weak field regime, such as the rate of precession of the Mercury perihelia, deflection of light by the Sun, delay of the light travel time in the vicinity of the Sun and  Lunar Laser Ranging experiment (for a review on this subject see \cite{Will:2014xja}). One can easily show, though, that all these experiments set constraints only on the parameter $\alpha_0$, while $\beta$ remains essentially unconstrained. This is natural, since the case of $\alpha_0=0$ and  $\beta \neq 0$ corresponds to a scalar-tensor theory that is perturbatively equivalent to general relativity, i.e. all the weak field experiments are automatically satisfied, but large deviations can be observed for strong fields. Currently, only the pulsars in close binary systems  can impose constraints on $\beta$ through observations of the shrinking of their orbits due to gravitational wave emission because the strong field effects in this case are non-negligible. More precisely, the shrinking would be different in general relativity and in scalar-tensor theories because in the latter case we have scalar gravitational radiation in addition to the standard gravitational waves. It turns out that the observations match very well to the predictions of pure general relativity and limit severely the emitted scalar radiation and thus the value of the scalar field. At the end, taking into account both the weak and the strong field experiments, we have $\alpha_0<10^{-4}$ and  $\beta > -4.5$ \cite{Freire2012,Antoniadis:2013pzd}. The observational constraints on STT coming from both the weak field and the strong field experiments are plotted in Fig. \ref{fig:STT_observations}. Such small values of $\alpha_0$ make the neutron stars in Brans-Dicke scalar tensor theories practically indistinguishable from general relativity and leave little space for deviation even for the class of scalar-tensor theory with $\alpha_0=0$ and  $\beta \neq 0$ \footnote{As we will discuss below, this is true only  in the static case. The rapid rotation magnifies the differences significantly and offers new possibilities for probing scalar-tensor theories of gravity}.

The discussion above was concentrated on the case with zero scalar-field potential. The picture changes drastically if we consider a nontrivial potential. One of the simplest cases is to assume the following standard form of the Einstein frame potential: 
\begin{equation}
V(\varphi)=2m^2_{\varphi}\varphi^2 \label{eq:Vphi}
\end{equation}
that is equivalent to considering a scalar field with nonzero mass $m_{\varphi}$. Even though this might seems like a simple and straightforward extension, it has dramatic influence on the observational constraints of the theory because of the following reasons. Introducing a scalar-field mass means that we have introduces a new characteristic scale, namely the Compton wave-length of the scalar field $\lambda_\varphi=2\pi/m_{\varphi}$, and the scalar field is loosely speaking confined inside $\lambda_\varphi$. More precisely, it drops exponentially at infinity and has non-negligible values only inside its  Compton wave-length. This help us reconcile the theory with the observations for a much larger range of parameters compared to the massless case. For example let's consider the binary pulsar observations. If the Compton wavelength of the scalar field is smaller than the orbital separation between the two stars, no scalar gravitational radiation will be emitted and the predictions for the shrinking of the orbit would be the same as in pure general relativity. For such values of the scalar field mass the parameter $\beta$ is essentially unconstrained \cite{Ramazanoglu:2016kul,Yazadjiev:2016pcb}.

Below we will briefly discuss other alternative theories of gravity that are either equivalent to scalar-tensor theories, or fall into the same class of modifications of general relativity, i.e. we have inclusion of a dynamical scalar field. Exactly in these theories universal relations were built.

\vspace{0.3cm}
\textbf{\emph{$f(R)$ theories of gravity: }} We will continue with the $f(R)$ theories, since they are equivalent to a particular class of scalar-tensor theories with nonzero scalar field potential. The essence of  these theories is that the Ricci scalar $R$ in the Hilbert-Einstein actions is substituted with a function of $R$, thus $f(R)$ theories:
\begin{equation}\label{A}
S= \frac{1}{16\pi G} \int d^4x \sqrt{-g} f(R) + S_{\rm matter}(g_{\mu\nu}, \chi).
\end{equation}
The viable $f(R)$ theories have to be free of tachyonic instabilities and the appearance of ghosts which is equivalent to \cite{Sotiriou:2008rp,DeFelice:2010aj}
\begin{equation}
\frac{d^2f}{dR^2}\ge 0,  \;\;\; \frac{df}{dR}>0
\end{equation}
respectively. One can easily show that after certain transformations (see e.g. \cite{Yazadjiev:2014cza}) this actions is equivalent to a scalar-tensor theory with zero kinetic term in the Jordan frame and nonzero potential. The gravitational scalar $\Phi$  and the  potential $U(\Phi)$ are connected to the $f(R)$ function via the relations 
\begin{equation}
\Phi=\frac{df(R)}{dR},  \;\;\; U(\Phi)=R \frac{df}{dR} - f(R).
\end{equation} 
If one wants to express these quantities in terms of the Einstein frame function that appear in the action \eqref{eq:Action_EF}, then the coupling function becomes $\alpha(\varphi) = -1/\sqrt{3}$ and the potential $V(\varphi)=A^4(\varphi)U(\Phi(\varphi))=[R (df/dR) - f(R)]/(df/dR)^2$. For example, for one of the simplest case -- the $R^2$ gravity, where $f(R) = R + aR^2$, we have
\begin{equation}
V(\varphi)= \frac{1}{4a}\left(1-e^{-\frac{2\varphi}{\sqrt{3}}}\right)^2.
\end{equation}

$f(R)$ theories of gravity were mainly explored in cosmological context as an alternative explanation of the accelerated expansion of the Universe. Every viable theory of gravity should pass the observational test on astrophysical scales as well and that is why neutron stars in $f(R)$ theories attracted particular interest recently. The form of the $f(R)$ function, though, that is normally used to explain the dark energy phenomenon would not give significant influence on astrophysical scales. That is why most of the authors considered the problem the other way around -- assume that indeed the gravitational theory is not the pure Einstein's theory, but instead we have an $f(R)$ type of modification of the action. Then one can ask the question what would be the additional terms in the $f(R)$ function that would give the dominant contribution on astrophysical scales. It is expected that this is exactly the $R^2$ term and that is why most of the compact star solutions were constructed in $R^2$  gravity.

\vspace{0.3cm}

\textbf{\emph{Quadratic gravity: }} The idea of the quadratic gravity is to supplement the Hilbert-Einstein action with all the possible algebraic curvature invariants of second order. These invariants are: $R^2$, $R^2_{\mu\nu}\equiv R_{\mu\nu}R^{\mu\nu}$, $R^2_{\mu\nu\rho\sigma}\equiv R_{\mu\nu\rho\sigma}R^{\mu\nu\rho\sigma}$ and the Pontryagin scalar $*RR \equiv 1/2 R_{\mu\nu\rho\sigma} \epsilon^{\nu\mu\lambda\kappa} R^{\rho\sigma}\mbox{}_{\lambda\kappa}$, where $\epsilon^{\nu\mu\lambda\kappa}$ is the Levi-Civita tensor. An extra dynamical scalar field is included as well that couples non-minimally to the second order curvature corrections. 

The motivation behind such modifications lies in the fact that  pure general relativity is not a renomalizable theory which naturally poses severe obstacles to the efforts of quantizing gravity. The modification of the action proposed by the quadratic gravity makes the theory renormalizable \cite{Stelle:1976gc}. The price we pay is the appearance of ghosts. A way to circumvent this problem is just by assuming that the quadratic gravity is a truncated effective theory of a more general one (such as the string theory where the effective action contains infinite series of higher curvature corrections and it is ghost free).  

A general form of the action is \cite{Yunes2011,Pani2011}:
\begin{eqnarray}
S=&&\frac{1}{16\pi}\int\sqrt{-g} d^4x 
\Big[R - 2\nabla_\mu \varphi \nabla^\mu \varphi - V(\varphi) \\
&& + f_1(\varphi) R^2 + f_2(\varphi) R_{\mu\nu} R^{\mu\nu} + f_3(\varphi) 
R_{\mu\nu\rho\sigma}R^{\mu\nu\rho\sigma}+f_4(\varphi){}^{*}\!RR\Big]+S_{\rm matter}\left[\chi, \gamma(\varphi) g_{\mu\nu}\right] ,\label{eq:quadratic}
\end{eqnarray}
where $V(\varphi)$ is the scalar field potential, the coupling functions  $f_1..f_4$ depend only on the scalar field, $\chi$ denotes collectively the matter fields in the matter action $S_{\rm matter}$ and we have a nonminimal coupling between the scalar field and the matter via the function $\gamma(\varphi)$.

The field equations derived from this action in their most general form are of order higher that two. This leads to problems in the theory such as the appearance of ghosts as we mentioned above. In some special cases, though, the field equations remain of second order as discussed below.

Two sectors of the quadratic gravity attracted particular interests -- the Einstein-dilaton-Gauss-Bonnet gravity and the Chern-Simons gravity. As a matter of fact almost all of the studies of compact objects in quadratic gravity were made exactly in these sub-theories.

\vspace{0.3cm}

\textbf{\emph{Einstein-dilaton-Gauss-Bonnet gravity : }}  In the EdGB theory the function $f_4(\varphi)=0$ and one choses a special combination between the other three, namely $f_1(\varphi)=f_3(\varphi)=-1/4 f_2(\varphi) \equiv f(\varphi)$ \cite{Kanti1996}. This combination is chosen in such a way that the resulting field equations are of second order. 
The action has the form
\begin{equation}
S=\frac{1}{16\pi }\int\sqrt{-g} d^4x \left[R - 2\nabla_\mu \varphi \nabla^\mu \varphi - V(\varphi) + f_{GB}(\varphi) (R^2-4R_{\mu\nu}^2+R_{\mu\nu\rho\sigma}^2)\right], \label{eq:EdGB}
\end{equation}
where $R^2_{GB}\equiv R^2-4R_{\mu\nu}^2+R_{\mu\nu\rho\sigma}^2$ is called Gauss-Bonnet scalar. Clearly the function $f(\varphi)$ has dimensions of inverse curvature and thus a characteristic scale can be introduced in the EdGB theory equal to $\sqrt{\alpha_{\rm GB}}$ where we assumed that $f(\varphi)$ is proportional to $\alpha_{\rm GB}$ times a dimensionless function of the scalar field. The strongest constraint on the parameters of the theory comes form the observations of black hole low-mass X-ray binaries \cite{Yagi:2012gp}, namely $ \sqrt{|\alpha_{\rm GB}|} \lesssim 5 \times 10^6 {\rm cm}$. Another purely theoretical constraint comes from the requirement for the existence of black hole solutions, that is fulfilled when  $\sqrt{\alpha_{BD}}$ is smaller than the black hole horizon size \cite{Kanti1996}. This leads to $\alpha_{\rm GB}/M^2\lesssim0.691$ \cite{Pani2009}. One should note that the scalar field potential is neglected in most of the studies of compact objects in EdGB gravity.

\vspace{0.3cm}

\textbf{\emph{Chern-Simons gravity : }}  The CS gravity considers a different sector of the quadratic gravity when $f_1=f_2=f_3=0$ and only the function $f_4$ is nontrivial (for a review on the CS theory see e.g.  \cite{Alexander:2009tp}). This means that only the term proportional to the Pontryagin scalar ${}^{*}\!RR$ remains. 

There are  two versions of the theory. The first one is a non-dynamical version where the kinetic and the potential terms for the scalar field are omitted and the scalar field is externally prescriber, i.e. it does not evolve dynamically. This case is simpler and it was the first one to be considered. It turned out, though, that in this case the theory has certain problems and restrictions, such as the Pontryagin constraint ${}^{*}\!RR=0$ \cite{Grumiller2008,Alexander:2009tp} That is why in the last several years another version of the theory, where the scalar field is dynamical, attracted much more attention. In order to distinguish the dynamical CS gravity from the non-dynamical version, the abbreviation dCS is used.
 
Thus the dCS action takes the form
\begin{equation}
S=\frac{1}{16\pi }\int\sqrt{-g} d^4x \left[R - 2\nabla_\mu \varphi \nabla^\mu \varphi - V(\varphi) + f_{CS}(\varphi) ({}^{*}\!RR)\right]. \label{eq:CS}
\end{equation}
A very interesting property of the dCS gravity is that the static spherically symmetric solutions do not differ from Einstein's theory but rotation can introduce large deviations. As a matter of fact the dCS gravity is almost the only theory with such property which makes it very interesting to study.
 
In most of the studies, the function $f_{CS} = \alpha_{\rm CS} \varphi$ and the potential of the scalar field $V(\varphi)$ is zero, which introduces a length scale of the theory $\sqrt{\alpha_{\rm GB}}$. As we discussed above the general form of the quadratic gravity is prone to problems such as the appearance of ghost, because of the fact that the field equations contain higher order derivatives. This problem is circumvented in the EdGB theory because of the special combination of the curvature invariants, but this is not true for the dCS theory. Instead, in order to have a well posed theory, one can consider the decoupling limit, where the field equation are still of second order. Thus, in most paper dCS gravity is studied perturbatively in the coupling constant $\alpha_{\rm GB}$. In order to be able to apply the perturbative approach, the following condition should be fulfilled, $\alpha_{\rm GB}/M^2 \ll 1$. Due to the fact that the static solutions are the same as in GR, the only weak field experiments  that can impose constraints on $\alpha_{\rm GB}$ are the ones measuring the frame dragging effects, such as the Gravity Probe B, which gives $\sqrt{\alpha_{\rm GB}} \lesssim 10^{13} {\rm cm}$ \cite{AliHaimoud:2011fw} that is also in agreement with the results in \cite{Yagi2012}.

\subsubsection{Alternative theories of gravity: neutron star models and astrophysical implications}
Here, we will review the neutron star models constructed in the above discussed alternative theories of gravity and their astrophysical implications. We are not aiming towards an exhaustive review, but instead we will discuss the most important results with an emphasis on the recent achievements in the field.  

\vspace{0.3cm}

\textbf{\emph{Neutron stars in (massive) scalar-tensor theories }} Historically, the compact star models in STT where some of the first to be considered (see e.g. \cite{Horbatsch2011} and the references therein) and they still attract significant attention, because of the fact that STT are ones of the most natural and widely used alternative theories of gravity. Most of the results fall into two categories based on the exact form of the Einstein frame coupling function $\alpha(\varphi)$ (see eqs. \eqref{eq:alpha_phi} and \eqref{eq:Aphi_expansion}) -- neutron stars in Brand-Dicke scalar-tensor theories with constant coupling function $\alpha(\varphi) = \alpha_0$ and neutron stars in a theory with $\alpha(\varphi) = \beta \varphi$ that is perturbatively equivalent to general relativity in the weak field regime but can lead to large deviations for strong fields. As we discussed in the previous section, $\alpha_0$ is severely limited by the weak field observations which leaves practically no space for any measurable deviations in the neutron star properties. This is not the case, though, with the second type of STT, where the only constraints come from the strong field experiments such as the observations of pulsars in binary systems. The current constraint on  $\beta$, i.e. $\beta>-4.5$,  still leaves space for non-negligible deviations from the pure Einstein's theory especially in the rapidly rotating case. That is why we will consider only the second class of scalar-tensor theory.

\begin{figure}
	\includegraphics[width=0.40\textwidth]{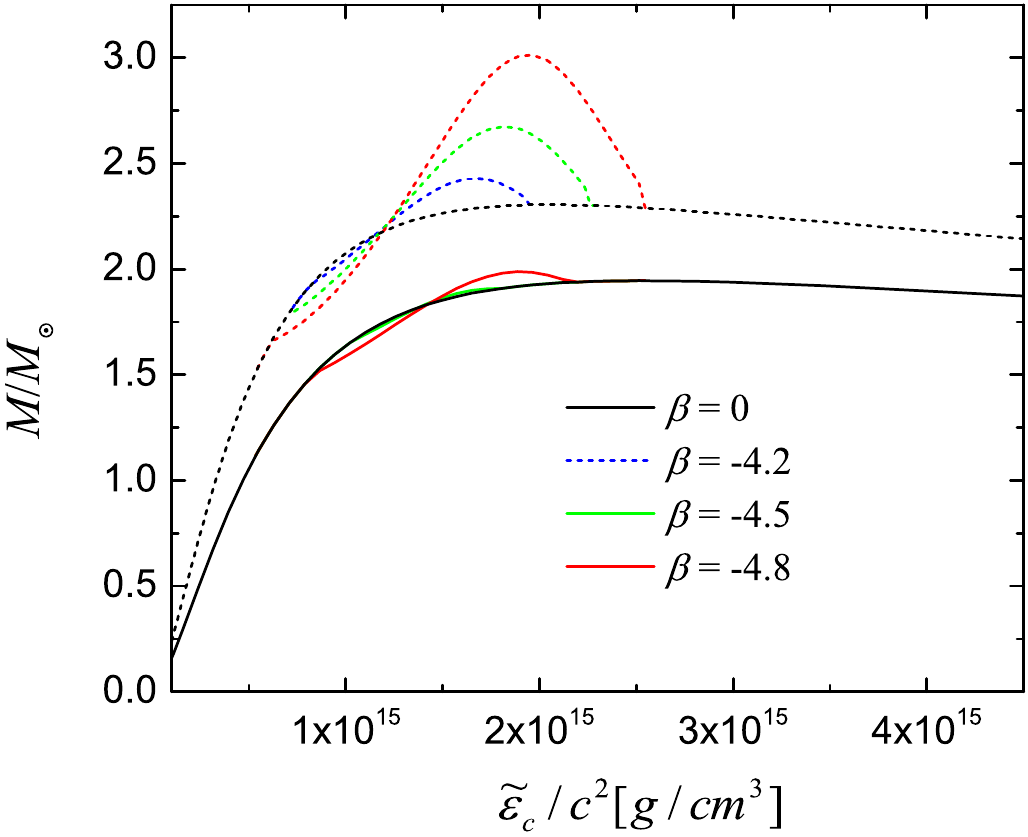}
	\includegraphics[width=0.42\textwidth]{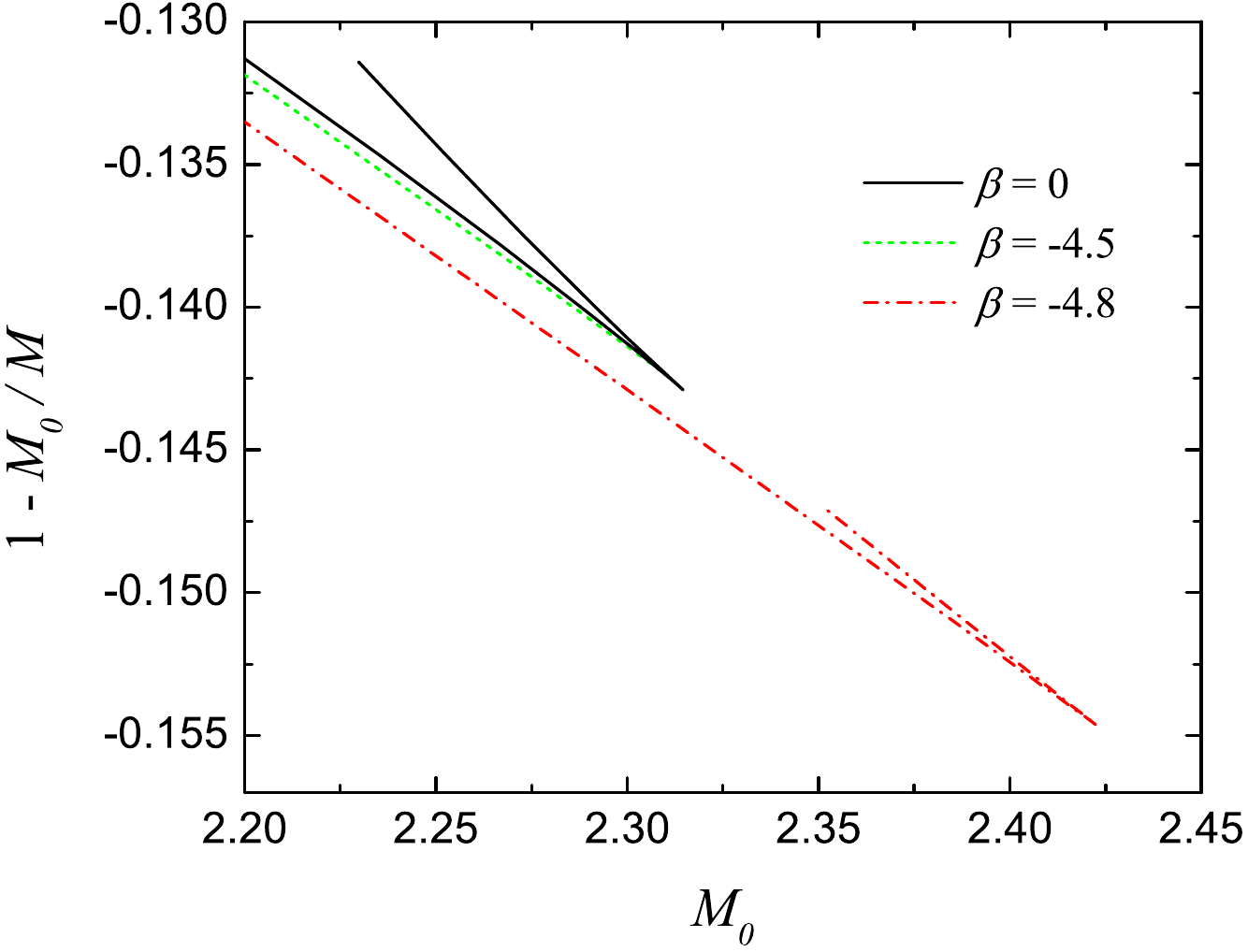}
	\caption{[Left panel] The neutron star mass as a function of the central energy density for	static sequences of neutron stars (solid lines) and sequences of stars rotating at the mass-shedding limit
		(dotted lines) with  different values of the coupling constant $\beta$. A polytropic equation of state with $N = 0.7463$ and
		$K = 1186$ is used. [Right panel] The quantity $1-M_0/M$, connected to the gravitational binding energy, as a function of the baryon rest mass $M_0$ for a fixed value of the angular momentum $cJ/(GM_\odot^2)= 1.38$. [Credit Ref. \cite{Doneva:2013qva}]. }	\label{fig:NS_STT}
\end{figure}

Neutron stars  in scalar-tensor theories with the Einstein frame coupling function $\alpha(\varphi) = \beta \varphi$ were considered for the first time in \cite{Damour:1993hw} where an effect, called sponteneous scalarization, was found that consists of the following. There exists a range of stellar parameters where neutron stars can develop nontrivial scalar field even if the background cosmological value of the scalar field $\varphi_0$ is zero. As a matter of fact very similar phenomenon is observed also for black holes in the presence of nonlinear fields, such as charged black holes described by nonlinear electrodynamics \cite{Stefanov2007,Stefanov2007a,Stefanov2009,Stefanov2008,Doneva2010a}, in the presence of a complex scalar field \cite{Kleihaus2015} or black holes surrounded by matter \cite{Cardoso2013,Cardoso2013a}. Sequences of scalarized solutions are plotted in Fig. \ref{fig:NS_STT} for several values of the parameter $\beta$ and for both nonrotating star and stars rotating at the Kepler (mass-shedding) limit. Let us now discuss first the static case. This corresponds to  the black, blue and green tick lines in Fig. \ref{fig:NS_STT}. Note that the case with $\beta=-4.8$ is actually outside of the observational limit $\beta>-4.5$ but we plotted it so that one can have a better intuition about the qualitative behavior with the increase of $\beta$. One of the most important properties of this class of scalar-tensor theories is the nonuniqueness of the solutions. First, one should note that for $\alpha(\varphi)=\beta \varphi$ the field equations \eqref{eq:FieldEqEF_g} and \eqref{eq:FieldEqEF_phi} always admit the solutions with zero scalar field (we will call them trivial solutions). This means that the pure general relativistic solutions are also solutions for this class of STT. As one can see from the left panel in Fig. \ref{fig:NS_STT}, at some critical central energy density a new sequence of neutron star solutions with nontrivial scalar field (we will call them nontrivial solutions) branch out of the sequence of trivial solutions. This is the so-called spontaneous scalarization with the name coming from the analogy to the spontaneous magnetization in ferromagnets. Thus, for a certain range of parameters nonuniqueness of the solutions is present. It is even more interesting that the scalarized neutron stars are energetically more favorable over the neutron stars with zero scalar field \cite{Damour:1993hw,Doneva:2013qva} and they will be the ones that will be realised in practice. This is demonstrated also in the right panel of Fig. \ref{fig:NS_STT} where the quantity $1-M_0/M$, that is connected to the gravitational binding energy, is plotted as a function of the baryon rest mass $M_0$. The sequences are calculated for the more general case of rotating stars with a fixed value of the angular momentum. As one can see, there is a cusp at the point where the mass reaches maximum which represents a turning point along the fixed-J sequence. This is the point where secular instability to collapse sets in. Moreover, the scalarized neutron stars have lower values of $1-M_0/M$ and therefore, higher binding energy compared to the pure GR case, which makes them energetically more favorable. The stability of the scalarized neutron stars was examined in \cite{Harada:1997mr,Harada:1998ge} and it was found that nontrivial scalar field develops for $\beta<-4.35$ in the non-rotating case. This result was derived in \cite{Harada:1998ge} for one particular polytropic EOS, but it turns out that the threshold value of $\beta$ is quite similar for other realistic EOS \cite{Novak:1998rk}. Since the current observational constraint is $\beta>-4.5$  there is not much space for deviations from pure general relativity. 

The scalarization was further examined in \cite{Salgado:1998sg} and the maximum mass limit was studies in \cite{Sotani:2017pfj}. The effect of different equations of state on the scalarization for two different coupling functions was examined in \cite{Zahra2017}. Slowly rotating neutron stars in STT where constructed in \cite{Damour:1996ke} where also the constrains on $\beta$ coming from the binary pulsar observations where discussed for the first time. Rotational corrections up to first order in the rotational frequency were studied later in \cite{Sotani:2012eb} and up to second order in \cite{Pani:2014jra}. What gives the largest difference, though, is the inclusion of rapid rotation \cite{Doneva:2013qva} that enhances the differences between the pure general relativistic solutions and the scalarized ones, especially close to the Kepler limit, and also increases the range of parameters where nontrivial scalar field can developed. It was shown for example that scalarized solutions exist even for $\beta<-3.9$ close to the Kepler limit.  As a matter of fact these were the first rapidly rotating models in alternative theories of gravity. The effect of rapid rotation is demonstrated in Fig. \ref{fig:NS_STT} where the dotted lines correspond to sequences of models rotating at the mass-shedding limit. As one can see for $\beta=-4.2$ no scalarized branch of solutions exists in the nonrotating limit, but such branch clearly appears close to the Kepler frequency with significant deviations from pure general relativity. Moreover, while the differences between the pure general relativistic solutions and the scalarized ones is quite small for the maximum allowed value $\beta=-4.5$ in the static case, the rapidly rotating models can reach large deviations for the same value of $\beta$. This opens a completely new window towards testing scalar tensor theories. Currently the fastest known pulsar rotates with a frequency of about 700 Hz \cite{Hessels2006} where the rotational effects are important but they would not lead to significant enhancement of the differences with pure general relativity. Other objects, though, such as the binary neutron star merger remnants, will rotate with frequencies close to the Kepler limit shortly after their birth and they are supposed to be observed in the near future via their gravitational wave emission. This makes them perfect candidates for further testing of scalar-tensor theories and the strong field regime of gravity in general.

All the solutions discussed so far are in the negative $\beta$ regime. It was argued recently that scalarization can occur for positive $\beta$ as well for a limited set of equations of state that admit negative values of trace of the energy-momentum tensor inside the star \cite{Mendes:2014ufa,Palenzuela:2015ima,Mendes:2016fby}. This condition practically means that there are region inside the star where the pressure  surpasses
one third of the energy density that can be translated to a threshold value of the compactness weakly dependent on the particular equation of state, i.e. $(M/R)_{\rm min} \sim 0.265$ \cite{Mendes:2014ufa}, where the mass of the star is in solar masses and the radius is in kilometers.

A lot of work has been done on  astrophysical implications of the scalarized neutron stars. Possible ways to constrain scalar-tensor theories through measurements of surface atomic line redshifts was considered in \cite{DeDeo:2003ju}. The orbital and epicyclic frequencies, and the innermost stable circular orbit for particles orbiting around scalarized neutron stars were calculated in \cite{DeDeo:2004kk} in the non-rotating case and in \cite{Doneva:2014uma} for  rapidly rotating models, while general expressions in terms of the multipole moments have been produced in \cite{PappasSotiriou2015MNRAS}. These quantities are related to the properties of accretion discs around compact stars and the observations of quasi-periodic oscillations in the emitted X-ray flux, and that is why they can be used to put constraints on the theory. The process of dynamical scalarization resulting for example after accretion of matter on the neutron star that brings it above the scalarization threshold, was considered in \cite{Novak:1998rk}. The collapse in scalar-tensor theories was examined in \cite{Harada:1996wt,Novak:1999jg,Gerosa2016} and the collapse of a neutron star to black holes was considered in \cite{Novak:1997hw}. Almost all of these studies, with the exception of \cite{Doneva:2014uma}, are in the nonrotating limit. Even though some of them offered promising ways of constraining the coupling parameter $\beta$ at that time, the very recent observations of pulsars in binary systems set very tight limit on $\beta$ which makes the deviations from pure general relativity in these astrophysical scenario practically unmeasurable. Only the rapidly rotating case leads to larger deviations. If we restrict ourselves to objects rotating with frequencies up to 700Hz, though, the deviations in the epicyclic frequencies and the position of the innermost stable circular orbit considered in \cite{Doneva:2014uma}, would be still difficult to measure. That is why one needs to go to rotation close to the Kepler limit expected for example for the binary neutron star merger remnants. As a matter of fact the neutron star merger in scalar-tensor theories was studied for the first time a few years ago in  \cite{Barausse:2012da,Shibata:2013pra}. In these studies a phenomenon called ``dynamical scalarization'' was observed similar to the spontaneous scalarization. The essence of the phenomenon is that even though the two neutron stars would not be scalarized when they are isolated, they can develop nontrivial scalar field if they are close enough, i.e. as the orbital separation shrinks the stars undergo dynamical scalarization. This can have significant effect on the stellar dynamics and leads to some observable effects.  Later a semi-analytical approach based on a modification of the post-Newtonian formalism was developed that takes into account the presence of a nontrivial scalar field and was proven to be in agreement with the previous fully nonlinear relativistic results \cite{Palenzuela:2013hsa}. This approach was recently improved in \cite{Sennett:2016rwa}. The advantage of the semi-analytical approach is that a much larger parameter space can be explored and template of waveforms can be generated.  The problem of binary neutron star mergers in scalar-tensor theories was examined also via calculation of quasi-equilibrium sequences of equal-mass, irrotational binary neutron stars in \cite{Taniguchi:2014fqa}. The question of detectability of possible (dynamically) scalarized stated by the Advanced LIGO, VIRGO and KAGRA was discussed in \cite{Sampson2014} and the results show that the gravitational wave signal is indeed detectable in certain case that are in agreement with the current observational constraints. In addition, the electromagnetic counterpart of the binary neutron star mergers was studied \cite{Ponce:2014hha} and it was concluded that the differences with pure general relativity are small but if they are combined with gravitational wave observations, constraints can be imposed  on the deviations from Einstein's theory. An analytic model of dynamical scalarization using an effective action approach was developed in \cite{Sennett2017}.

The oscillations of neutron stars in scalar-tensor theories were examined for the first time in \cite{Sotani:2004rq} where the non-radial polar oscillations modes ($f$- and $p$-modes) were calculated in the Cowling approximation, i.e. when the spacetime and the scalar field perturbations are neglected. Even though this is a crude approximation it gives us good qualitative picture of the possible deviations from pure general relativity. Later, the full perturbation equations for both the axial and polar modes were derived and the frequencies of the axial modes were calculated \cite{Sotani:2005qx}. In  \cite{Sotani:2014tua} a different approach was undertaken -- the radial oscillations of scalarized neutron stars were considered in the Cowling approximation, but without neglecting the scalar field perturbations. The radial perturbations of course do not lead to gravitational wave emission but due to the presence of a scalar field, scalar waves are emitted \cite{Morganstern1967}. All these studies showed that the scalarization indeed changes significantly the oscillations spectrum leading to non-negligible deviations from general relativity for small enough values of $\beta$. The most recent constraints on $\beta$, though, limit the possible deviations from Einsteins' theory considerably similar to the other astrophysical scenarios.  Torsional oscillations of scalarized neutron stars were examined in \cite{Silva:2014ora}.
 The results in \cite{Silva:2014ora} showed that if one considers realistic values of the $\beta$ the deviations due to scalarization are smaller than the uncertainties in microphysics and therefore there is no degeneracy between the two effects. Oscillations of rapidly rotating neutron stars were examined in \cite{Yazadjiev2017}. The results show that the deviations from pure Einstein's theory can be significant especially in the case of nonzero scalar field mass.

Additional ``ingredients'' to the scalarized neutron stars were also explored, such as the presence of anisotropic pressure \cite{Silva:2014fca} both in the nonrotating and slowly rotating regimes using a quasi-local equation of state similar to the pure general relativistic case \cite{Horvat:2010xf,Doneva:2012rd}. Depending on the ``sign'' of the anisotropy, i.e. whether the tangential pressure is larger than the radial one or the other way around, the deviations from pure general relativity due to the presence of nontrivial scalar field are either suppressed or magnified. This gives us hope that the binary pulsar observations can set constraints on the degree of anisotropy. 

An interesting extension of the above work is to consider not only one scalar field coupled to the metric, but multiple scalar fields instead. This problems is of course much more involved and the first steps in this direction were undertaken in \cite{Horbatsch:2015bua} (see also \cite{Damour1992}). The simplest case is to consider two real scalar fields instead of one, that after a complexification can be casted to the problem of adding one complex scalar field. The criterion for neutron star  scalarization  were examined in this theory and the 3+1 formulation of the field equations was derived.

So far all the presented results are for the case when the potential of the scalar field $V(\varphi)=0$. New and interesting effects arise, though, if we drop this assumption as commented in the previous subsection. For example the inclusion of scalar-field mass $m_\varphi$, via the specific form of the potential given by eq. \eqref{eq:Vphi}, leads to the fact that the scalar field is suppressed at length scales larger that its Compton wavelength $\lambda_\varphi$. Therefore, if properly chosen, the mass of the scalar field can help us reconcile the theory with the observations for a much larger range of parameters. One should note that the calculation of the neutron star models is numerically challenging because of the following reason. The presence of nonzero mass makes the field equation for the scalar field  \eqref{eq:FieldEqEF_phi} stiff because this equation admits both exponentially decreasing and exponentially growing solutions at infinity. Clearly, only the exponentially decreasing solutions is physically relevant, but it is very difficult in certain cases to converge numerically to it and special techniques should be used.

Let us consider the same class of scalar-tenor theories with coupling function $\alpha(\varphi)=\beta_\varphi \varphi$ and impose the constraint that the Comptop wavelength $\lambda_\varphi \ll 10^{10}{\rm m}$ (this is roughly the minimal observed orbital separation for close binary pulsars)  or equivalently $m_{\varphi} \gg 10^{-16} {\rm eV}$. Then the emission of scalar gravitational waves will be suppressed which means that practically no constraints can be imposed on the parameter $\beta$. Thus, the deviations from pure general relativity can be very large. Neutron star models in such class of scalar-tensor theories were considered for the first time in \cite{Popchev2015,Ramazanoglu:2016kul} for the static case and in \cite{Yazadjiev:2016pcb} for the slowly rotating case. Rapidly rotating models in massive scalar-tensor theories were examined in \cite{Doneva:2016xmf}.  Further studies on the astrophysical implications of neutron stars in massive scalar-tensor theories are needed because too large, negative $\beta$ would clearly lead to dramatic changes in the neutron star structure that can be tested with the present astrophysical observations. Considering other forms of the scalar field potential would be interesting as well. The gravitational radiation of compact binaries in another class of massive scalar-tensor theories, the massive Brans-Dicke theory, was considered in \cite{Berti2012,Alsing2012}.

\vspace{0.3cm}

\textbf{\emph{Neutron stars in $f(R)$ theories of gravity}} Neutron star models in $f(R)$ theories of gravity attracted considerable interest recently as a natural attempt to study the astrophysical applications of a class of alternative theories that gives promising results on cosmological scales, such as an alternative explanation of the accelerated expansion of the universe. Here we will not give a thorough review on the subject but instead focus mainly on the realistic astrophysically relevant models which were also used to construct the universal relations discussed in this chapter. We will concentrate on the particular case of $R^2$ gravity that is supposed to give the leading corrections for the neutron star structure. Moreover, the $f(R)$ theories of gravity are mathematically equivalent to a specific class of scalar-tensor theories with nonzero scalar field potential, as discussed in the previous subsection, which can simplify their treatment.       

The initial work on the subject was concentrated mainly on discussing the existence of solutions and building such solutions (see for example  \cite{Kobayashi:2008tq,Upadhye:2009kt,Babichev:2009td,Jaime:2010kn,Babichev:2009fi}). In the beginning there was some controversy on the question of whether compact star solutions exist in $f(R)$ theory but the overall studies showed that such stars can indeed be constructed. A drawback, though, is that solving the reduces field equations suffers from severe numerical instabilities. This can be easily demonstrated using the scalar-tensor formulation of $f(R)$ theory. Let us consider for example the case of $R^2$ gravity, i.e. when $f(R)=R+aR^2$. In this case the resulting scalar field potential will lead to a nonzero mass of the scalar field. Thus, similar to the massive scalar-tensor theories, the field equation for the scalar field \eqref{eq:FieldEqEF_phi} becomes stiff and thus leads to severe computational difficulties in certain cases. That is why, as a simplification, realistic neutron stars in $R^2$ gravity were first studied perturbatively \cite{Cooney:2009rr,Arapoglu:2010rz} (see also \cite{Alavirad:2013paa} for the case of logarithmic $f(R)$ theory), i.e. instead of solving the full field equations a perturbative expansion in the parameter $a$ was made. The studies in \cite{Yazadjiev:2014cza} went beyond the perturbative approach calculating for the first time sequences of realistic neutron star models in $R^2$ gravity and comparing them with the observations. The calculations were performed using the mathematical equivalence to a particular class of scalar-tensor theories. The results showed that the nonperturbative results are not only quantitatively but also qualitatively very different from the results in the perturbative approach. Thus the perturbative approach is not applicable for $f(R)$ theories and in order to obtain correct results one has to calculate the full field equations. Later the full unperturbed field equations derived and solved directly in $f(R)$ theories without going through scalar-tensor theories or different frames \cite{Yazadjiev:2015xsj,Astashenok2017}. The results showed that the two approaches for obtaining solutions in $f(R)$ gravity are equivalent.

\begin{figure}
 	\includegraphics[width=0.45\textwidth]{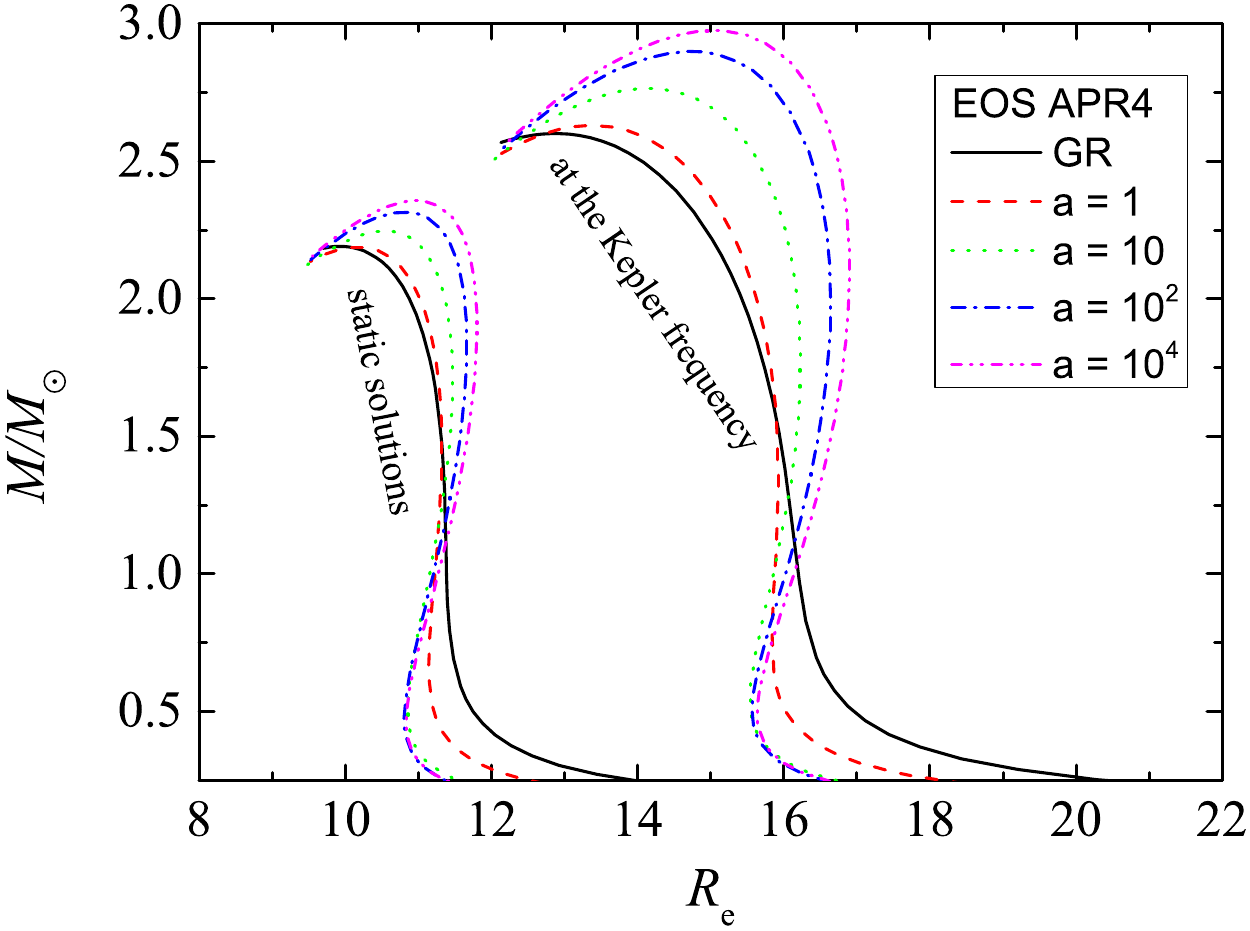}
 	\includegraphics[width=0.44\textwidth]{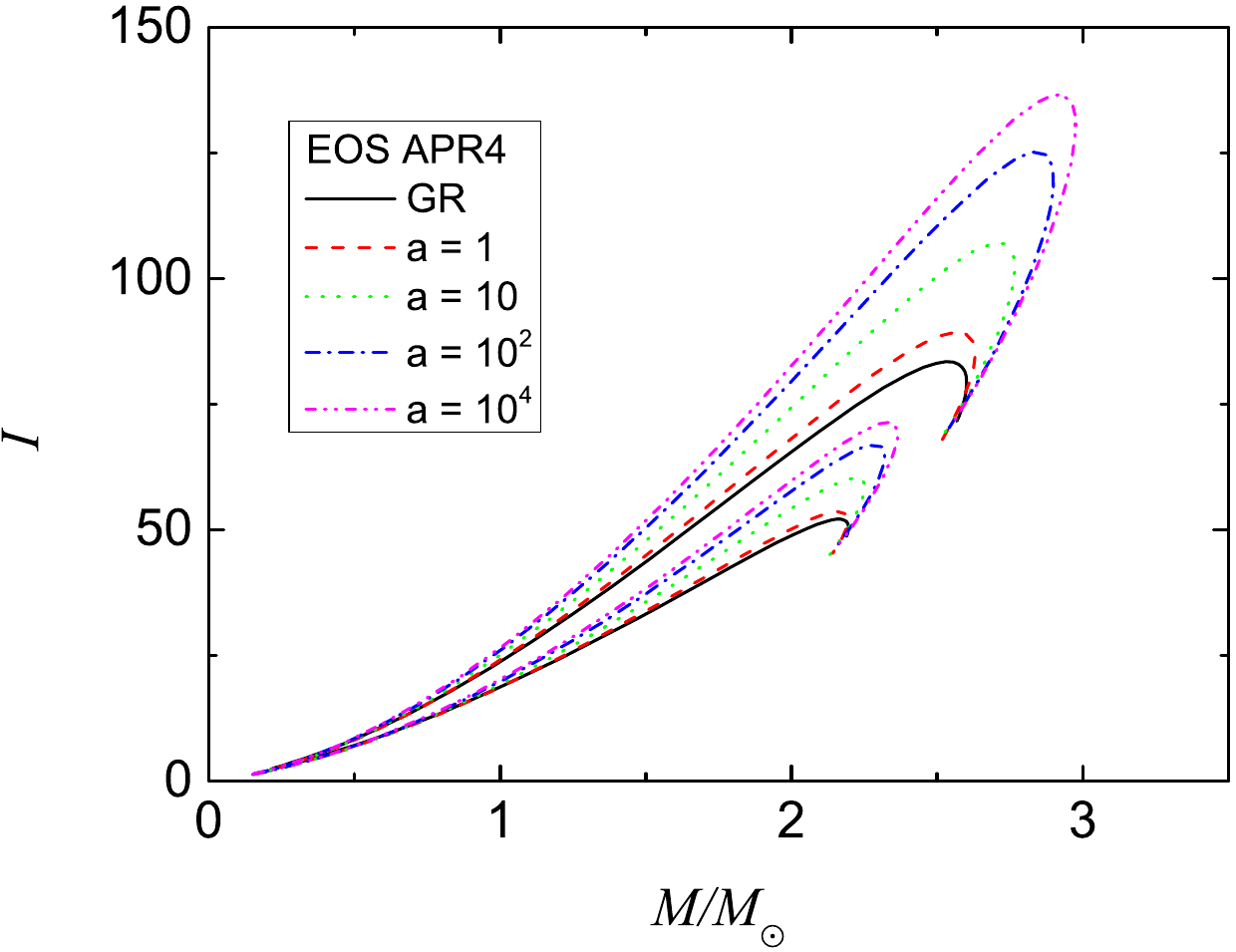}
 	\caption{ The neutron star mass as a function of the radius (left panel) and the moment of inertia as a function of the mass (right panel) for  static and rotating at the Kepler limit sequences of neutron stars in $R^2$ gravity. The results for several values of the parameter $a$ are plotted for the APR4 EOS. [Credit Ref.  \cite{Yazadjiev:2015zia}]}	\label{fig:NS_FR}
 \end{figure}

The results in   \cite{Yazadjiev:2014cza}  confirmed what was expected from the theory -- in the limit when $a \rightarrow 0$ the solutions converge to the general relativistic ones and in the limit $a \rightarrow \infty$ the solutions tend to the case of massless Brans-Dicke theory with coupling function $\alpha(\varphi) = -1/\sqrt{3}$. Therefore, the neutron star solutions in $R^2$ gravity are bounded between two limiting cases which  also puts an upper limit on the deviations from general relativity. This can be seen on Fig. \ref{fig:NS_FR} where different colors correspond to different dimensionless values of $a$ and the case of $a=10^{4}$ corresponds to nearly the maximum possible deviation from pure general relativity (we use the dimensionless parameter $a\rightarrow a/R_0^2$, where $R_0$ is one half of the solar gravitational radius $R_0 = 1.47664 {\rm km}$, i.e. the solar mass in geometrical units). The Gravity Probe B experiment imposes the following constraints on the values of  $a$, namely $a \lesssim 2.3 \times 10^5$ in the same dimension units (or $a \lesssim  5 \times 10^{11}{\rm m}^2$ in physical units) which means that all cases plotted on Fig. \ref{fig:NS_FR} fall into the allowed range of values of $a$. The studies for several equations of state in \cite{Yazadjiev:2014cza} showed, though, that the deviations due to the change of the parameters $a$ are of the same order of magnitude as the equation of state uncertainties which of course poses significant obstacle for testing the $f(R)$ theories. This is not the case, though, with the moment of inertia calculated in \cite{Staykov:2014mwa,Yazadjiev:2015zia} where the differences with pure general relativity are better pronounced. Thus, the future observations of the neutron star moment of inertia can be used to discriminate between different gravitational theories in the strong field regime. Rapidly rotating models in $R^2$ gravity were calculated in \cite{Yazadjiev:2015zia} and similar to the case of scalar-tensor theories, the rotation magnifies the deviations from pure general relativity considerably in comparison to the static case. The gravitational collapse in $f(R)$ theories was studied in \cite{Borisov:2011fu,Cembranos:2012fd} and the merger of neutron stars was examined in \cite{Sagunski2017}. The astrophysical implication of the constructed neutron star solutions, such as the orbital and epicyclic frequencies, were considered in \cite{Staykov:2015kwa} and the oscillations of neutron stars, including different asteroseismology relations, where examined in \cite{Staykov:2015cfa}. These studies showed that the deviations from  Einstein's gravity are non-negligible, even though in some cases they are within the equation of state uncertainty. This gives us hope that the $f(R)$ theories would be better constraint in the future when the astrophysical observations limit further the allowed range of  equations of state. 

 \vspace{0.3cm}
 
 \textbf{\emph{Neutron stars in EdGB gravity}}
 Neutron stars in the EdGB theories of gravity were constructed for the first time in  \cite{Pani:2011xm} both in the static and the slowly rotating cases. The coupling function between the scalar field and the Gauss-Bonnet scalar in eq. \eqref{eq:EdGB} is chosen to be
 \begin{equation} \label{eq:BDCouplingFunction}
 f_1(\varphi) = \alpha_{GB} e^{\beta\varphi},
 \end{equation}
 where $\alpha_{GB}$ and $\beta$ are constants. 
 
 An interesting fact is that in the decoupling limit the monopole scalar charge, i.e. the coefficient in front of the $1/r$ term in the asymptotic  of the scalar field at infinity, is zero unlike for example the massless scalar-tensor theories \footnote{The neutron stars, though, can still have nonzero higher  scalar multipoles and thus nontrivial scalar field}. Thus there would be no scalar dipole radiation and it is not possible to impose constraints on the theory from the binary pulsar observations \cite{Yagi2016}.

\begin{figure}
	\includegraphics[width=0.36\textwidth]{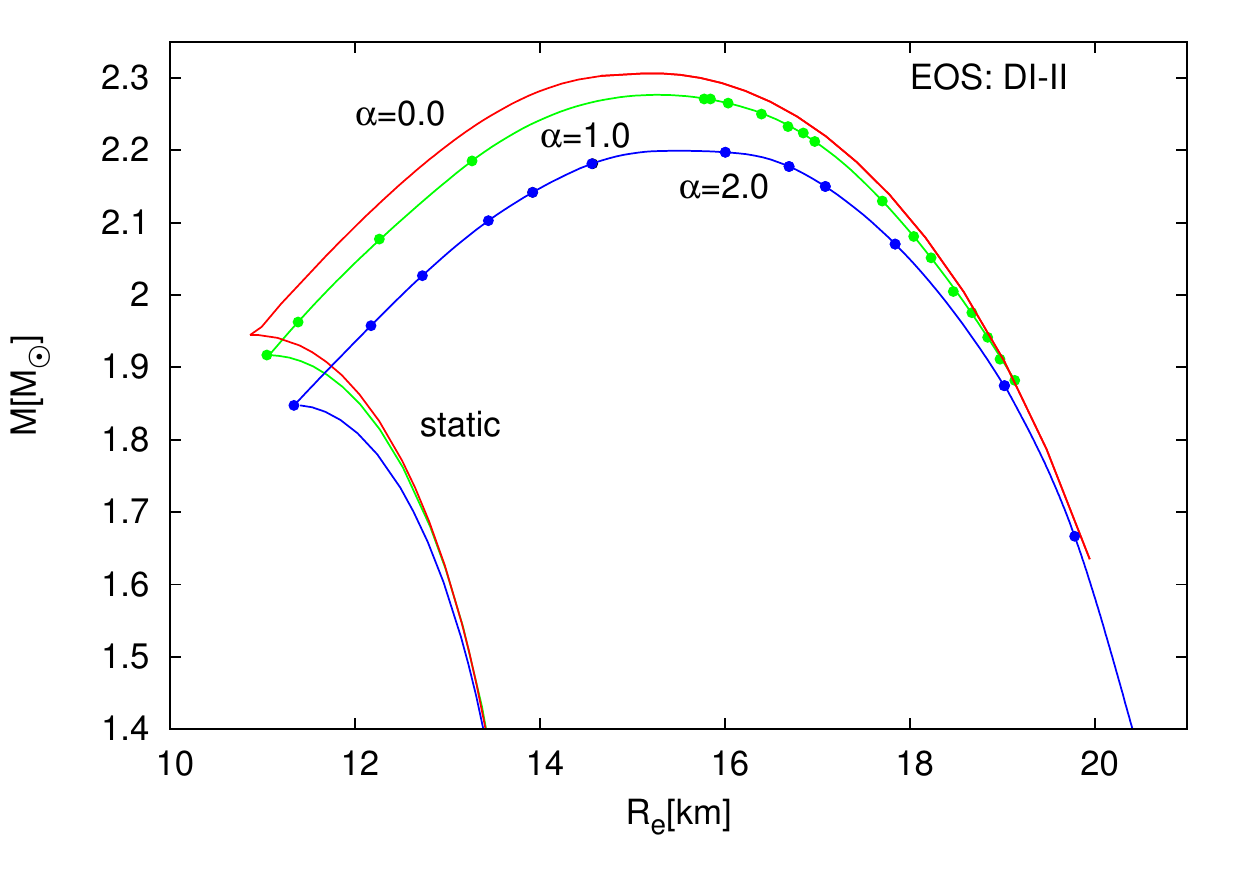}
	\caption{The mass-radius relation for different values of the Gauss-Bonnet coupling constant $\alpha=0,\; 1$ and $2$  ($\alpha\equiv \alpha_{GB}/M_\odot^2$)  for the EOS DI-II (the parameter $\beta$ in eq. \eqref{eq:BDCouplingFunction} is fixed to $\beta=-1$).
		The chosen values of $\alpha$ are below the  the upper bound obtained from low mass	x-ray binaries \cite{Yagi:2012gp} which would correspond to $\alpha = 12$ when converted to the dimensionless $\alpha$ used in the graph. For a given $\alpha$ the left boundary curve represents the sequence of static solutions,
		while the right boundary curve represents the sequence of neutron stars rotating at the
		Kepler limit. Both are connected by the secular instability line. [Credit Ref. \cite{Kleihaus:2016dui}]}	\label{fig:NS_EdGB2}
\end{figure}

In Figs. \ref{fig:NS_EdGB2}  the mass as a function of the central energy density and the moment of inertia as a function of the mass for neutron stars in EdGB gravity are plotted. As one can see, contrary to scalar-tensor and $f(R)$ theories, the maximum neutron star mass decreases with the increase of $\alpha_{GB}$ for fixed $\beta$. Even more, it was proven in \cite{Pani:2011xm} that for fixed $\alpha_{GB}\beta$ no neutron star solutions exist above some critical maximum central energy density (the exact limit is a quite lengthy expression and can be found in \cite{Pani:2011xm}). This can serve as a way to impose constraint on the theory if the nuclear matter equation of state is known with a good accuracy -- one should simply require that  $\alpha_{GB} \beta$ is below some critical values chosen in such a way that the maximum mass is above the two solar mass barrier \cite{Demorest:2010bx,Antoniadis:2013pzd}.
 
As we mentioned above, one of the hopes to test the deviations from pure general relativity is via the future observations of the neutron star moment of inertia. Unfortunately, the studies showed that for EdGB gravity this would not be possible since the moment of inertia deviate from the one in pure Einstein's theory by a few percents at most, that is inside the expected observational error. 

The axial quasinormal modes of neutron stars in EdGB were examined in \cite{Blazquez-Salcedo:2015ets}. The frequencies of the fundamental spacetime modes increase compare to pure general relativity. An interesting observation is that the universal (equation of state independent) relations for the oscillation modes that are available in Einstein's theory still hold for the  EdGB gravity that can be used to put constraints on the parameters of the theory. 
 
Rapidly rotating neutron star models in EdGB gravity were calculated in \cite{Kleihaus:2014lba,Kleihaus:2016dui}. The properties of the rotating compact stars are examined there in detail and the quadrupole moment is calculated as well. The mass-radius relation for rapidly rotating neutron stars is shown in Fig. \ref{fig:NS_EdGB2}.

 \begin{figure}
 	\includegraphics[width=0.30\textwidth]{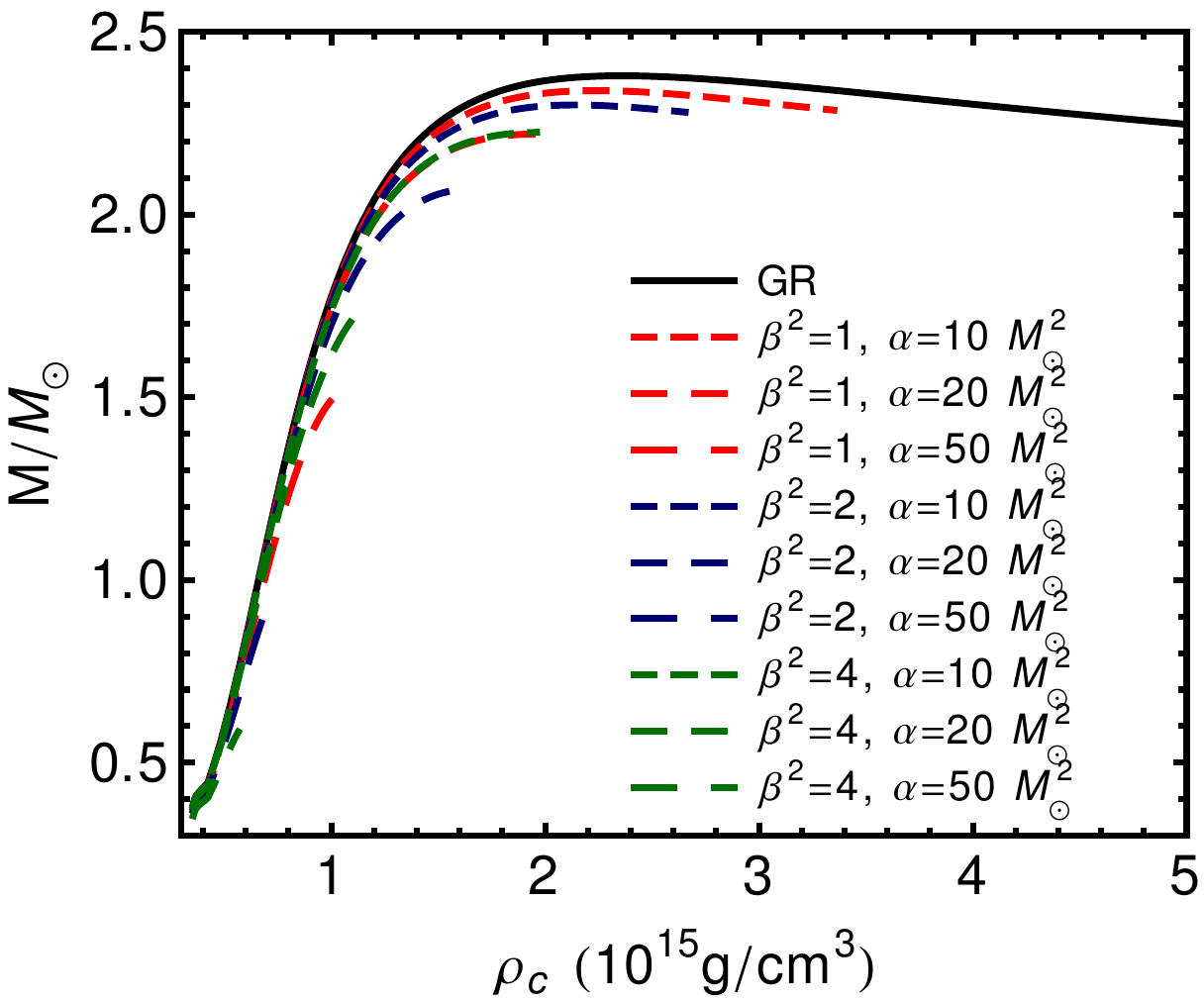}
 	\includegraphics[width=0.30\textwidth]{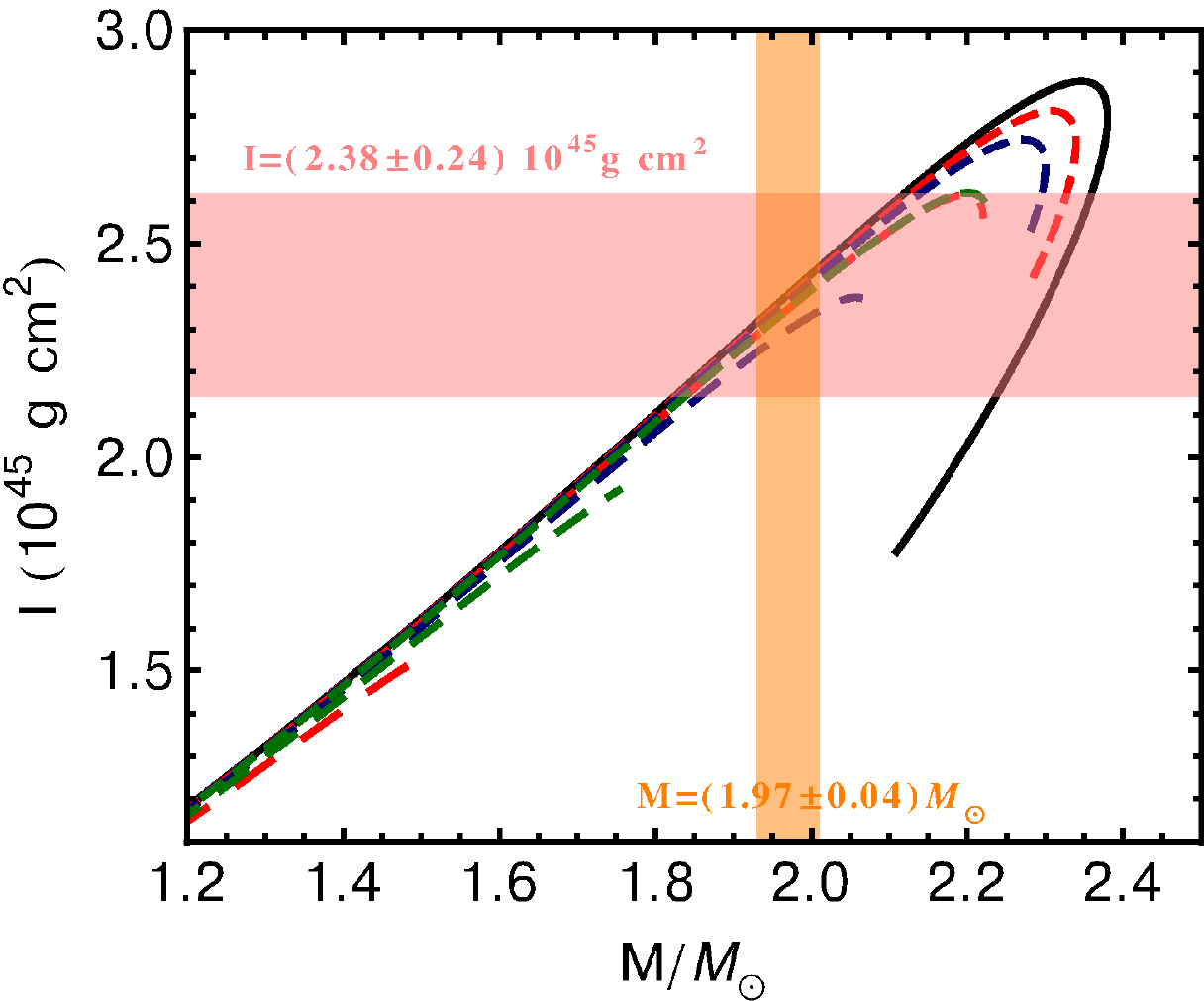}
 	\caption{The mass as a function of the central energy density (left panel) and the moment of inertia as a function of the mass (right panel) for neutron stars in EdGB gravity using the APR EOS. Models for different values of
 		the  Gauss-Bonnet coupling constants $\alpha$ (denoted by $\alpha_{GB}$ in the text) and $\beta$ are shown. In the
 		right panel the recent observation of a neutron star
 		with $M\approx 2M_\odot$ and a possible future observation of the
 		moment of inertia confirming general relativity within
 		$10\%$~\cite{Lattimer2005ApJ} are shown. [Credit Ref.  \cite{Pani:2011xm}]}	\label{fig:NS_EdGB1}
 \end{figure}

\vspace{0.3cm}
 
\textbf{\emph{Neutron stars in dCS gravity}}
As we commented above,  static neutron stars in dCS gravity are identical with the pure general relativistic ones and differences are present only in the rotating case. 
Up to now only models in the slow rotation approximation were calculated at first order in the rotation \cite{Yunes:2009ch,AliHaimoud:2011fw} and later at second order \cite{Yagi:2013mbt}. Up to leading order in rotation only the moment of inertia is affected by the dCS gravity, while the mass-radius relation remains the same as in Einstein's theory. 

\begin{figure}
	\includegraphics[width=0.45\textwidth]{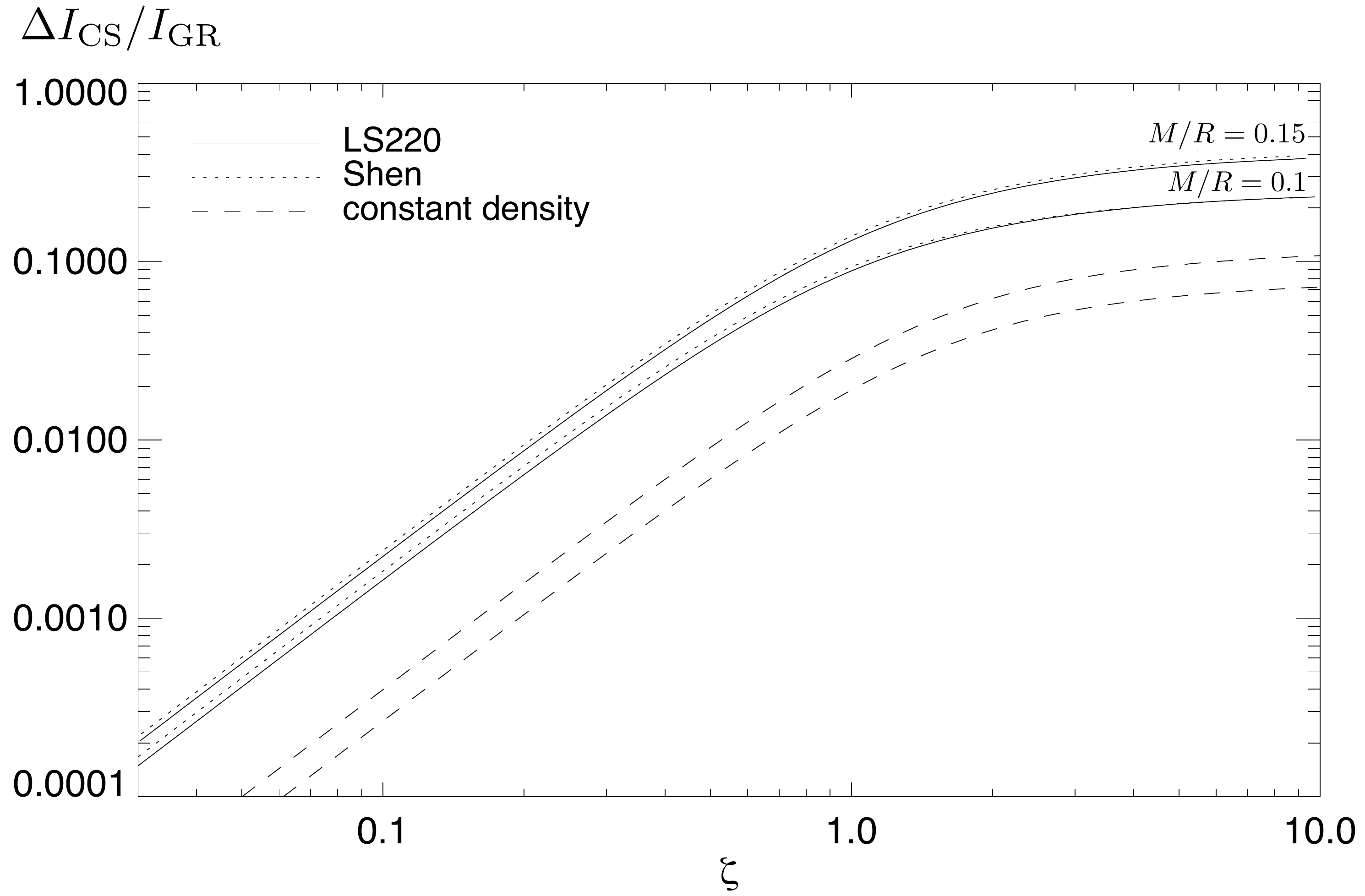}
	\caption{ Change in the moment of inertia $\Delta I_{\rm CS}/I_{\rm GR}$ induced by the CS modification as a function of the CS coupling strength $\zeta$($\zeta\equiv 4\sqrt{2}\alpha_{\rm CS} M/R^3$) , calculated using first order corrections in the rotational frequency for the two EOSs. The quantity $\Delta I_{\rm CS}/I_{\rm GR}$ is plotted for the values $M/R = 0.1$ and 0.15. The dashed lines show the analytic result  for a constant density nonrelativistic star. [Credit Ref. \cite{AliHaimoud:2011fw}]}	\label{fig:NS_dCS1}
\end{figure}

The change in the moment of inertia $\Delta I_{\rm CS}/I_{\rm GR}$ induced by the CS gravity as a function of the CS coupling strength is shown in Fig. \ref{fig:NS_dCS1}. As one can see, depending on the Chern-Simons coupling constant, the moment of inertia can deviate significantly from  general relativity and if we assume that the accuracy of the future observations of the neutron star moment of inertia are of the order of 10\%, the Chern-Simons coupling constant can be constrained several orders of magnitude better than the current estimates.
 
\begin{figure}
 	\includegraphics[width=0.45\textwidth]{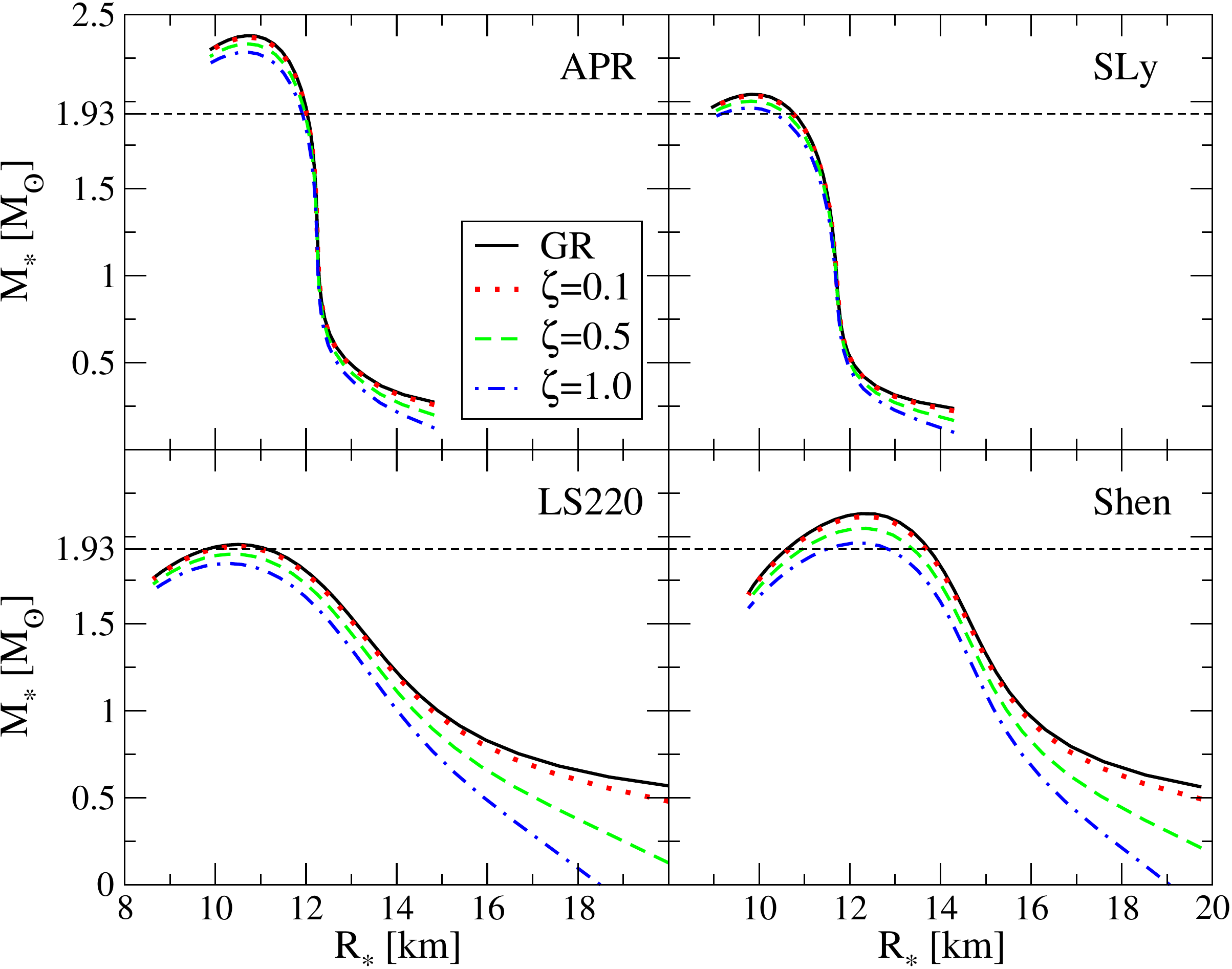}
 	\caption{ The mass-radius relations for neutron stars in pure general relativity (solid curves) and dCS with different
 		coupling strengths $\zeta$ ($\zeta\equiv 4\alpha_{\rm CS}^2 M^2/R^6$) calculated using second order corrections in the rotational frequency. Results for several equations of state are shown. The
 		mass-radius relations depend on the spin of the neutron star, which is set
 		to match that for PSR J1614-2230. [Credit Ref. \cite{Yagi:2013mbt}]}	\label{fig:NS_dCS2}
 \end{figure}

The calculation of neutron stars in  dCS gravity up to second order in rotation allowed to calculate the mass quadrupole moment and the rotational corrections to the mass and radius that is plotted in Fig. \ref{fig:NS_dCS2}. Unfortunately, it turns out that these corrections are inside the EOS uncertainty and they can not be used to test the  dCS gravity with the present observations. The corrections to the post-Keplerian parameters are also too small to be observable nowadays via double binary pulsars.
 
\subsection{Universal relations in alternative theories}
\label{sec:past3}
Neutron star universal relations offer a very important tool for testing alternative theories of gravity, because the equation of state uncertainty, that causes a lot of problems as we have already mentioned, is taken out of the picture. There are three classes of universal relation in generalised theories of gravity that have been discussed in the literature so far. These are the gravitational waves asteroseismology relations, the I-Love-Q relations, and relations between the moment of inertia and the compactness of neutron stars. In general relativity we also discussed the 3-hair relations for the multipole moments but this topic has not been tackled yet in alternative theories of gravity. Below we will comment on the three former classes in further detail.

The oscillations of neutron stars are directly related to the emitted gravitational wave signal. That is why relations that connect the neutron star parameters to the oscillations frequencies and damping times, the so-called gravitational wave astereseismology relations, were extensively studies in pure Einstein's theory. They can be used in practice when gravitational waves from oscillating neutron stars are observed in the future (see discussion in subsection \ref{sec:seismology}). The extension to alternative theories of gravity was done in \cite{Sotani:2004rq,Sotani:2005qx} for the case of massless scalar-tensor theories, in \cite{Staykov:2015cfa} for $f(R)$ theories in the static case. The results there show, that if we restrict ourselves to values of the parameters that are in agreement with the current observational constraints, the relations are as equation of state independent as in pure general relativity but the deviations from pure Einstein's theory are very small in most of the cases. For example in $R^2$ gravity one can use an expression like the one in eq.~\eqref{eq:fmode} with a small variation to the coefficients to fit the $f$-mode frequency in terms of $\eta$. Extensions to other alternative theories of gravity would be interesting, especially the massive scalar-tensor theory case that can produce very large differences. The first studies of oscillations modes of rapidly rotating neutron stars in alternative theories of gravity were performed very recently \cite{Yazadjiev2017} in the case of (massive) scalar-tensor theories and the results show that the deviations from pure general relativity can be large. Asteroseismology relations in the rapidly rotating case are still lacking.

Another set of relations that attracted a lot of attention recently are the I-Love-Q relations, that are discussed in a lot of detail in the rest of the chapter for the pure general relativistic case. A natural question is whether we can use them to constrain the strong field regime of gravity. If such universal relations hold in alternative theories as well and we are able to determine independently via observations two quantities from the I-Love-Q trio, then possible deviations from general relativity can be detected. With this motivation in mind, such relations were examined in  \cite{Yagi:2013bca,YagiYunes2013PRD} for the dCS gravity, in  \cite{Pani:2014jra,Doneva:2014faa} for massless and in \cite{Doneva:2016xmf} for massive scalar-tensor theories, in \cite{Kleihaus:2014lba} for the EdGB theory, in \cite{Sham:2013cya} for the Eddington-inspired Born-Infeld (EiBI) theory  and in \cite{Doneva:2015hsa} for $f(R)$ theories. From these studies only the papers \cite{Doneva:2014faa,Kleihaus:2014lba,Doneva:2015hsa,Doneva:2016xmf} include the rapidly rotating case. It turned out that for all these theories the equation of state independence is preserved up to a large extend. The deviations from pure general relativity, though, are almost negligible for a big portion of the cases if one considers the allowed range of parameters of the corresponding theory. These are the massless STT, the  EdGB and the EiBI theory. Larger differences on the other hand are observed for dCS gravity, massive STT and $f(R)$ theories. One should note that all these conclusions are true only for the normalized relations, if one considers the non-normalized relations then the deviations from pure general relativity can reach very large values. This means that the normalization that is used for the I-Love-Q relations is so good that not only the equation of state dependence is taken away, but in some cases also the dependence on the particular theory of gravity. Therefore, it is unclear so far up to what extend the I-Love-Q relations can be used to test the various alternative theories of gravity. 

We should note here that, as with I-Love-Q, 3-hair relations could also be used to test gravity modifications, but as we have mentioned earlier these relations have not been studied yet in alternative theories of gravity. It is possible that the 3-hair relations in alternative theories can be quite different from their general relativity counterparts. The caveat here is that it might not be possible to define multipole moments in all classes of alternative theories of gravity, the way that they are defined in general relativity. This has been done so far only in scalar-tensor theory for a massless scalar field \cite{Papas2015PhRvDmoments}, where possible astrophysical applications in terms of these moments have been explored \cite{PappasSotiriou2015MNRAS}. We should note here that the quadrupole moment has been calculated in some additional cases such as in the case of EdGB theory by Kleihaus et al. \cite{Kleihaus:2016dui,Kleihaus:2014lba} as well as in $R^2$ gravity by Doneva et al. \cite{Doneva:2015hsa}, where the I-Q relations in these theories were compared against the corresponding relations in general relativity. Another case where a general definition of multipole moments has been given, is the case of $f(R)$ theories by Suvorov~\&~Melatos \cite{Suvorov2016PhRvD}, although they investigate only Ricci flat cases where there are no extra degrees of freedom present and the corresponding multipole moments are equivalent to those of general relativity.

Finally, relations between different normalizations of the moment of inertia and the compactness of neutron stars, including also relations involving the maximum stellar mass, were examined for STT and $f(R)$ theories \cite{Staykov2016PhRvD} motivated by the work in pure general relativity \cite{Breu2016MNRAS,Lattimer2005ApJ}.\footnote{I-C relations have been also studied in Horndenski and beyond-Horndeski gravity \cite{2016PhRvD..93l4056M,2016CQGra..33w5014B,2017PhRvD..95f4013S}, as well as for theories with disformal couplings \cite{Minamitsuji2016}, which are not covered here.} The results showed that the equation of state universality is as good as in Einstein's theory of gravity but the deviations from general relativity can be large in certain cases,\footnote{For the I-C relations in EdGB, as we have noted earlier, this is not the case though. On the contrary we expect that the general relativistic relations to hold also in EdGB.} especially for the relations involving the maximum stellar mass, that can be used as a probe of the gravitational theories. 
Specifically, Staykov et al. \cite{Staykov2016PhRvD} studied neutron star models using the following theories: 1) a scalar-tensor theory (denoted as STT) with a massless scalar field ($V(\varphi)=0$) and a conformal factor, that relates the Einstein frame to the Jordan frame, of the form $A(\varphi)=e^{\beta \varphi^2/2}$, with a chosen value for $\beta = -4.5$, which is within the range allowed by observational constrains and produces spontaneous scalarisation, 2) an $f(R)$ theory with the specific choice for the Lagrangian, $f(R)=R+a R^2$, where the coupling parameter $a$ is in $km^2$ in geometric units and has the value $a=10^4 M_{\odot}^2$, where the mass of the Sun is $M_{\odot}=1.477km$, and finally 3) general relativity.     
For these theories and the various models that they constructed, they fitted the moment of inertia in terms of the compactness $\cC$ for two different choices of normalisation using two different polynomial forms following \cite{Lattimer2005ApJ} and \cite{Breu2016MNRAS}, i.e., 
\begin{equation} \label{eq:fit_tilde}
\tilde{I} = I/(MR^2) = \tilde{a}_0 + \tilde{a}_1 \cC + \tilde{a}_2 \cC^4, \quad \textrm{ and } \quad \bar{I} = I/M^3 = \bar{a}_1 \cC^{-1} + \bar{a}_2 \cC^{-2} + \bar{a}_3 \cC^{-3} + \bar{a}_4 \cC^{-4},
\end{equation}
respectively. The coefficients of the fits for the three theories and for three rotation rates are given in Table \ref{Tbl:IMRR}
%
\begin{table}[htb]
\begin{centering}
\begin{tabular}{ccccccccc}
\hline
\hline
\noalign{\smallskip}
\multicolumn{2}{c}{$\tilde{I} = I/(MR^2)$} & & & &  \multicolumn{2}{c}{$\bar{I} = I/M^3$} \\
\cline{1-2} \cline{6-7}
\quad & $\tilde{a}_0$ & $\tilde{a}_1$ & $\tilde{a}_2$ & \quad & $\bar{a}_1$ & $\bar{a}_2$ & $\bar{a}_3$ & $\bar{a}_4$ \\
\hline
\multicolumn{2}{c}{GR} \\
\cline{1-2}
slow. rot.     & 0.210    & 0.824     & 2.480    & &   1.165    & 0.0538    &   0.0259 &  $- 0.00144$   \\
$\chi$ = 0.2          & 0.211    & 0.788     & 3.135     &  & 1.077    & 0.100     &   0.0172 & $-9.830 \times 10^{-4} $  \\
$\chi$= 0.4          & 0.200   & 0.823    & 2.469    &  & 1.024    & 0.123      &    0.0129 & $-7.985 \times 10^{-4}$     \\
$\chi$ = 0.6          & 0.176    & 0.839     & 2.393     &  & 0.943    & 0.143      &   0.00714 & $-5.539 \times 10^{-4}$   \\
\multicolumn{2}{c}{STT} \\
\cline{1-2}
slow. rot.     & 0.201    & 0.897    &   0.603   & & 1.057    & 0.110    &  0.0173    & $-0.00103$   \\
$\chi$ = 0.2          & 0.196    & 0.909    &   0.224   & & 0.884   & 0.197    &   0.00270    & $-3.254 \times 10^{-4}$   \\
$\chi$ = 0.4          & 0.171   & 1.055    &  $-2.985 $   & & 0.664     & 0.313     &   $-0.0163$    & $ 5.647 \times 10^{-4}$  \\
$\chi$ = 0.6          & 0.127   & 1.256     &  $-7.190$   & & 0.257     & 0.513     &   $-0.0492$   & 0.00202   \\
\multicolumn{2}{c}{$f(\mathcal{R})$} \\
\cline{1-2}
slow. rot.     & 0.221   & 0.981     &  $ -0.755$    &  & 0.941    & 0.214      &   0.00521     & $-4.412 \times 10^{-4}$  \\
$\chi$ = 0.2          & 0.216   & 0.989     &   $-1.255$    &  & 0.853    & 0.243    &  0.00190    & $-3.841 \times 10^{-4}$  \\
$\chi$ = 0.4          & 0.208   & 1.011     &   $-1.748$    &   & 0.805   & 0.264      &  $-0.00201$    & $-1.978 \times 10^{-4}$  \\
$\chi$ = 0.6          & 0.201   & 0.998     &   $-2.143$    &  & 0.725    & 0.280      &  $-0.00407$     & $-1.706 \times 10^{-4}$  \\
\noalign{\smallskip}
\hline
\hline
\end{tabular}
\end{centering}
\caption{The fitting coefficients for the fit given by eqs. (\ref{eq:fit_tilde}). The results are given for the slowly rotating cases and rapidly rotating cases with $\chi = 0.2$, $\chi = 0.4$, $\chi = 0.6$, for the GR case, followed by the STT case and by the $f(R)$ case. The table is taken from \cite{Staykov2016PhRvD}.}
\label{Tbl:IMRR}
\end{table}

\begin{figure}[htb]
\includegraphics[width=7.1cm,clip=true]{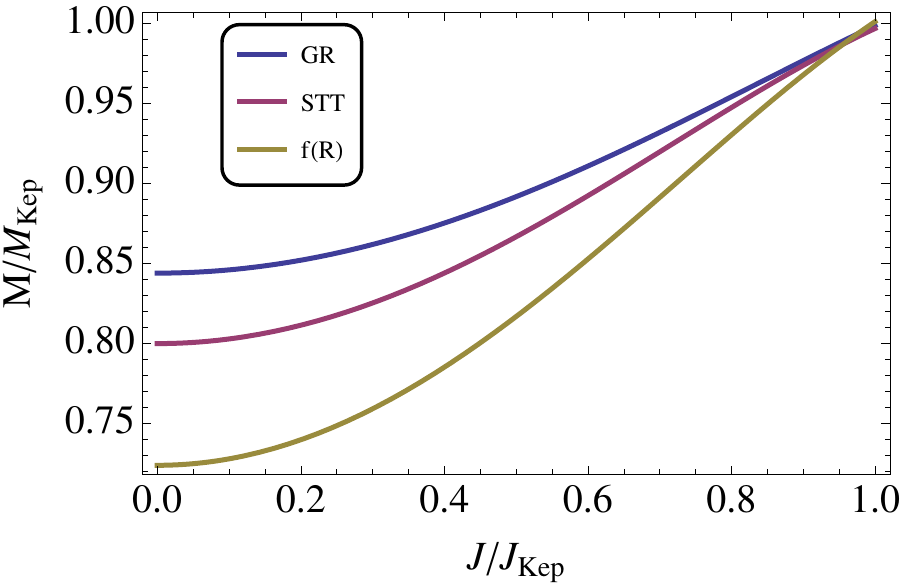}
\caption{\label{fig:max} The maximal mass normalised to the maximal Keplerian mass as a function of the angular momentum, normalised to the maximal Keplerian one for the three different theories. The curves are for the fit coefficients in Table \ref{Tbl:max}.
}
\end{figure}

For the relations that concern the maximum stellar mass in the various theories, Staykov et al. \cite{Staykov2016PhRvD} used the following two fitting functions, one where the mass is normalised with respect to the maximum non-rotating mass, $M_{\rm TOV}$, and another normalised with the maximum mass of the models rotating at the Kepler limit, $M_{\rm Kep}$, 
\begin{equation} \label{eq:Mmax_tov}
\frac{M}{M_{\rm TOV}} = 1 + a_1\left(\frac{J}{J_{\rm Kep}}\right)^2 + a_2\left(\frac{J}{J_{\rm Kep}}\right)^4, \quad \textrm{ and } \quad \frac{M}{M_{\rm Kep}} = a_0 + a_1\left(\frac{J}{J_{\rm Kep}}\right)^2 + a_2\left(\frac{J}{J_{\rm Kep}}\right)^4,
\end{equation}
where the normalisation of the rotation is done with the angular momentum at the Kepler limit $J_{\rm Kep}$. As one can see in Table \ref{Tbl:max}, where the various coefficients of the fits are given, there is no fit for the STT case in the $\frac{M}{M_{\rm TOV}}$ normalisation. This is because in this particular scalar-tensor theory the neutron star models get spontaneously scalarised, which means that the models of the theory are the same as the corresponding models in general relativity with a vanishing scalar field, until one gets to some critical value of central density and rotation and then the scalarised solutions appear. This causes a discontinuity in the behaviour of the $M/M_{\rm TOV}$ and the results are quite scattered, also because the point where spontaneous scalarisation starts depends on the equation of state. This is the reason why the second normalisation with respect to $M_{\rm Kep}$ was also considered in \cite{Staykov2016PhRvD}. This latter normalisation gives a nice behaviour without discontinuities and significantly less scattering in the STT case. The very interesting result is that the three different theories separate and are distinguishable as one can see in Figure \ref{fig:max}.

\begin{table}[htb]
\begin{centering}
\begin{tabular}{ccccccccc}
\hline
\hline
\noalign{\smallskip}
\quad & $a_0$ & $a_1$ & $a_2$ \\
\hline
\multicolumn{1}{c}{$M/M_{TOV}\left(J/J_{Kep}\right)$ } \\
\cline{1-2}
   GR & 1  & 0.251      &  $ -0.0626$      \\
$f(\mathcal{R})$     & 1  & 0.382     &  $ -0.127$     \\
STT   &  -- & -- & --\\

\multicolumn{1}{c}{$M/M_{Kep}\left(J/J_{Kep}\right)$} \\
\cline{1-2}
GR  & 0.844    & 0.203      &   $-0.0484$    \\
$f(\mathcal{R})$  & 0.800    & 0.290    &   $-0.0933$     \\
STT & 0.724  & 0.403  & $-0.126$  \\
\noalign{\smallskip}
\hline
\hline
\end{tabular}
\end{centering}
\caption{The fitting coefficients for the maximal mass for the fits (\ref{eq:Mmax_tov}). The table is taken from \cite{Staykov2016PhRvD}.}
\label{Tbl:max}
\end{table}
%
%

 
\section{Present challenges and future prospects}
\label{sec:future}


As we have seen, the topic of universal relations in general relativity is quite wide with a variety of relations having already been found. This does not mean that there are no more topics to be explored and progress to be made. One such case is the question of what happens in the presence of strong magnetic fields. We have seen already that there has been some work on I-Love-Q relations and magnetic fields \cite{Haskell2014MNRAS} but there has been no work on 3-hair relations and strong magnetic fields. In the I-Love-Q case we saw that strong magnetic fields and slow rotation can destroy universality, but it might be worth exploring the possibility that there is a more appropriate parameterisation that takes into account the magnetic field and in the end preserves universality. Universal relations that take into account magnetic fields could be useful in observationally constraining the strength of the magnetic field in neutron stars.   

There is also a lot of work to be done on the quasi-normal modes universal relations. So far the calculations have been done mainly using the Cowling approximation. It is very important therefore that the quasi-normal modes are calculated beyond that approximation and it is verified that the universal relations that have been found for the various modes and decay times still hold. This will enable the more accurate determination of the various neutron star parameters that can be extracted by the quasi-normal modes observations from gravitational waves.  Another question that could be of interest in the quasi-normal modes topic is to identify the theoretical/mathematical reason behind the existence of the universality in the modes. 

Further expanding on this, it would be interesting to know if the various universal relations can be traced to a common mechanism or not. 
We should note here that relations like the I-Love-Q, the 3-hair or the quasi-normal modes universal relations are relations between ``integral'' or average quantities of the entire star. The question then is, are there any other quantities that are local and satisfy universal relations or is it that such relations can exist only between particularly weighted average quantities?  
  
Finally, there is a lot of work to be done on the front of utilising universal relations for making measurements of astrophysical observables from neutron stars. Some work has already been done on using the I-Love-Q relations in order to break degeneracies in gravitational wave observations (see for example refs. \cite{Yagi:2013bca,YagiYunes2013PRD}) and there has been some preliminary work on using the 3-hair relations to model electromagnetic observables from X-ray binary systems in order to measure neutron star parameters \cite{Pappas2015MNRAS,Tsang2016ApJ,Pappas2017MNRAS}, but there is a lot more work to be done.  

Turning to neutron stars in alternative theories, even though the topic has been studied for several decades,   
there is still a lot to be done in the field. One of the main reasons for this is the complexity of the equations especially if one wants to consider not only static but also rapidly rotating neutron stars. Thus the studies have to be extended in two main directions -- constructing new neutron star models in alternative theories of gravity and studying their astrophysical implication in further detail. Below we will discuss both of them.

Constructing compact star models for a broader spectrum of alternative theories of gravity, such as more general or more realistic cases of tensor-multi scalar theories, Quadratic gravity, Horava gravity, Lorentz-violating theories, etc., is important since there are many extensions of Einstein's theory and the neutron stars offers a unique way of testing them. In addition,  for some alternative theories of gravity a perturbative approach was used for constructing the solutions. This is a drawback since some strong field effects can be omitted or even worse, the perturbative approach can be misleading in certain cases such as the $f(R)$ gravity where it was explicitly shown that the non-perturbative results are not only quantitatively but also qualitatively  different from the perturbative ones. As far as rotation is concerned, rapidly rotating neutron star solutions were constructed up to now only in three classes of alternative theories -- the scalar-tensor theories, $f(R)$ theories and Einstein-dilaton-Gauss-Bonnet theory. All these studies show that the rapid rotation offer new phenomenology that should be further explored in other generalized theories of gravity. In addition, the full stability of neutron stars in alternative theories of gravity has not been studied yet, not even in the case of scalar-tensor theories.

The astrophysical implications of neutron stars, such as stellar collapse, neutron star mergers, quasiperiodic oscillations, absorption lines, etc. are scarcely studied in most of the alternative theories of gravity, with the exception of scalar-tensor theories and some sectors for the quadratic gravity. One of the problems there is that the observations themselves either suffer from large uncertainties or even there are multiple astrophysical models explaining the same observations that makes testing the strong field regime of gravity extremely difficult. Nevertheless, neutron stars are ones of the very few astrophysical objects where the strong field regime of gravity can be explored and that is why further studies in this directions are needed. 

Last but not least, larger efforts should be put in the directions of breaking the degeneracy between the effects coming from the alternative theories of gravity and the uncertainties of the nuclear matter equation of state. Currently this degeneracy is plaguing a very large portion of the attempts to constrain the strong field regime of gravity. Some advance in this direction is actually expected to come in the near future since the astrophysical observations are narrowing the possible set of nuclear matter equations of state more and more. One thing we should always keep in mind, though, is that the interpretation of these observations is always done within general relativity, or even some type of Newtonian approximation. Therefore, an interesting study that deserved to be done is to explore whether the alternative theories of gravity could change the corresponding predictions.  

On this note, the study of universal relations in alternative theories of gravity is very important, because in principle such relations could be used to test deviations from general relativity while evading the equation of state degeneracies. So far work has been done on the I-Love-Q relations in alternative theories of gravity and in most of the cases the I-Love-Q relations have been found to be identical to those that hold in general relativity. Exception to this are the cases of dCS gravity, massive scalar-tensor theories and $f(R)$ theories. At a first glance therefore one would say that the applicability of I-Love-Q relations in testing theories alternative to general relativity is limited. Still, more work needs to be done before we can definitively decide on the extent of the usefulness of I-Love-Q relations as tests of our theories of gravity. 

Beyond the I-Love-Q relations, there are other relations, discussed in subsection \ref{sec:past3} that show more promise in distinguishing different theories. The exploration of these relations is only the beginning. There are many more relations discussed in section \ref{sec:past1} that should be also explored in other theories as well. Two important classes of relations that should be explored in alternative theories are the quasi-normal modes relations and the 3-hair relations. The quasi-normal modes relations are important to be studied in alternative theories especially in this moment in time since the opening of the gravitational waves observational window offers a unique opportunity to probe the structure of compact objects. So far, studies of gravitational waves emission in alternative theories are missing from the literature. These studies are very difficult but need to be made in order to better constrain possible deviations from general relativity.

3-hair relations in alternative theories will be easier to study and already some steps towards that direction are being made. For example multipole moments in some classes of alternative theories have been defined \cite{Papas2015PhRvDmoments,Suvorov2016PhRvD}, while there exist numerical codes that can calculate neutron star models in these theories \cite{Doneva:2013qva}. Again a possible issue might be to properly define the normalised quantities that will enter the various universal relations in the different theories. Also, it will be necessary to have a good correspondence between the quantities in different theories. At this point there is an issue that will have to be addressed in order to make progress. Multipole moments are defined in general relativity as asymptotic quantities (these are the Geroch-Hansen multipole moments) that characterise fields that exist on asymptotically flat spacetimes. This definition will not be always possible in all classes of alternative theories of gravity. Therefore one should be careful when talking about moments in other theories, making sure that the asymptotic moments are meaningful quantities. Furthermore, in the cases that asymptotic moments (a la Geroch-Hansen) cannot be defined, alternative quantities should be shot for. Possible alternatives could be source integrals like the ones given in the case of general relativity by G\"urlebeck \cite{Gurlebeck2014PhRvD} for example, or maybe other type of source integrals like those given by Hern\'andez-Pastora et al. \cite{HernandezPastora2016CQG}. In that case one would have to re-evaluate all the relevant quantities in general relativity as well and find the new relations that they will satisfy. In addition one will also have to find a way to relate these new quantities to astrophysical observables as well as some of the usual neutron star properties, such as the moment of inertia for example.\footnote{It is also worth noting that one could take a theory independent approach in studying neutron stars and universal relations in alternative theories of gravity, such as the post-TOV approach \cite{postTOV1,postTOV2}}. 

Finally there is an elephant in the room that we should address. As we place more constrains on the equation of state from astrophysics, the notion of universal relations will become meaningless. This is not expected to happen soon and probably there will be faster progress on universal relations than on the equation of state question, but eventually it will happen. Will universal relations become obsolete? The answer is, not necessarily. If the equation of state is determined, then it will be in all likelihood one that will preserve the universal relations. These relations can express quantities (neutron star parameters) in terms of other quantities and therefore will still be a tool for determining parameters that are difficult to measure directly. In addition, such relations can serve as consistency checks for our various astrophysical models that enter the ``measurement'' of different quantities from observations. Therefore they will continue to be a useful tool to do astrophysics. Furthermore, the class of these relations that can distinguish between different theories of gravity will be a particularly useful tool to do some fundamental physics by testing general relativity and even possibly selecting a likely alternative, or at least excluding unlikely ones.


\section{Conclusions}
\label{sec:conclusions}

This chapter has covered two quite wide topics, that of neutron star universal relations and that of neutron stars in alternative theories of gravity. We have tried to present the fullest possible spectrum of universal relations in general relativity, but inevitably some results were covered very briefly or not at all. Similarly, for neutron stars in alternative theories of gravity we have focused on the better studied classes of these theories, such as scalar-tensor gravity, $f(R)$, EdGB and CS. These are all theories for which neutron star universal relations have been considered. 
Our aim has been to present some of the basic ideas, giving some attention to the most important aspects of them, and then focus on the most important results. 

As it was discussed, the notion of neutron star universal relations is not new, but the field gained a lot of momentum the last few years and has opened the way to circumvent the uncertainties of the equation of state. With respect to isolated neutron stars, one could say that the main classes of universal relations are three. The first class are the I-Love-Q relations which relate the moment of inertia, the quadrupolar love number and the mass quadrupole of a neutron star. The second class is the 3-hair relations that relate the higher order multipole moments of the spacetime around neutron stars to the first three non-zero multipole moments, i.e., the mass, the angular momentum and the mass quadrupole. Finally the third class are the universal relations of the oscillation frequencies, which relate the real parts and the imaginary parts of neutron star QNMs to its other parameters such as the moment of inertia or the compactness or the average density. Relations like the I-Love-Q and the 3-hair relations, characterise both neutron and quark stars and can have many useful applications. They can be used to measure neutron star properties that are hard to measure directly or they can be used to break degeneracies between parameters that enter the description of observables. They could also be used to test our assumptions or our models of astrophysical mechanisms, if one could measure one member of the I-Love-Q trio in one way and infer another member from another observation and an assumed model for example. Furthermore, by measuring the neutron star parameters from different channels, with the help of the universal relations, we could use our knowledge of these parameters to solve the inverse problem and constrain initially and maybe finally identify the equation of state for matter at supra-nuclear densities. In addition, with the opening of the gravitational wave observational window, we expect to learn many things for the structure of neutron stars from the mergers of binary systems and the universal relations of the oscillation frequencies will play an important part since the fundamental frequencies of the remnants will be some of our primary observables. 
In addition, apart from using the various universal relations, one could also test for their validity. In this way one could test the assumptions that enter the various relations, such as the assumptions on the equation of state, i.e., whether our current models are correct or not. Even more, one could test one of the most fundamental assumptions of all, that of the theory of gravity that we are using to construct neutron stars.       

Modifications or extensions to general relativity are motivated for several reasons, such as cosmological and astrophysical considerations like the nature of dark matter and dark energy questions, as well as more theoretical considerations coming from attempts to create a theory of everything where additional degrees of freedom are introduced to the low energy description of such a theory.  The resulting modifications can have the form of additional scalar fields or modifications of the Einstein-Hilbert action with the introduction of higher powers or arbitrary functions of the Ricci scalar or various contractions of the Riemann tensor. Apart from the cosmological implications of such modifications, the implications for the structure of compact objects in general and neutron stars in particular are of extreme interest and this is why neutron stars have been extensively studied in such theories. Here we focused on the most widely studied classes of these theories, i.e., scalar-tensor gravity, $f(R)$, EdGB and dCS, that are also the ones for which there have been studies of various universal relations beyond general relativity. When studying neutron stars in these theories one needs to also take into account the constrains that exist on various scales: from laboratory  and solar system tests of gravity to constraints coming from astrophysical and cosmological observations, as well as viability restrictions to the theories. Although the aforementioned constrains can be very strong for some parameters of modified theories and the viability considerations limit the possible options, there is still the possibility to find theories that are viable and free of pathologies (like tachyonic instabilities and such), to which the various classes that we present here belong, and have neutron stars that are quite different from their general relativistic counterparts. One such example are neutron stars in massive scalar-tensor theory that exhibit spontaneous scalarisation and another example are neutron stars in $R^2$-gravity, a subclass of $f(R)$ theories. In both cases neutron stars can be quite different than their general relativistic counterparts, which makes them very interesting astrophysical targets. On the other hand, the deviations of neutron star mass and radius constructed in EdGB and dCS from their general relativistic counterparts are quite smaller making them more difficult to distinguish with astrophysical observations (we should note though that in dCS only slow rotation models have been studied so far). The main source of uncertainty again is the unknown equation of state and any changes in the structure and the parameters of neutron stars in modified theories needs to be measured against the range of these parameters that is possible within general relativity and comes from choosing different equations of state. 

This is where universal relations come into play. Comparing the relations in general relativity against their counterparts in modified theories of gravity has the potential to distinguish different theories even though the uncertainty due to the equation of state seems to give a mixed picture. Such success stories are the applications of the I-Love-Q relations in dCS and massive scalar-tensor or $f(R)$ theories, where the relevant relations have been found to differ from their general relativistic counterparts. In contrast, theories such as the massless scalar-tensor or EdGB seem to have the same relations as in general relativity, if we restrict ourselves to values of the parameters of the theory allowed by the observations. Apart from the I-Love-Q relations, some other classes of promising relations have emerged such as the relations that express the maximum neutron star mass in terms of the maximum non-rotating mass and the rotation rate, which seem to be more potent in distinguishing different theories of gravity. Finally, there is plenty of work to be done in other directions as well, such as the study of 3-hair relations in alternative theories, as well as extending work on the study of neutron star QNMs in alternative theories. 

In conclusion, with respect to universal relations, although their usefulness now that the equation of state is still uncertain is quite obvious, it is important to stress that these relations will still be useful even after the equation of state has been determined. This is because they will still express useful relations between different properties of neutron stars, that can be utilised in the analysis of observations and in measuring the full range of the relevant neutron star parameters.
%

\begin{acknowledgments}
We would like thank Kent Yagi for providing us data for some of the plots as well as useful comments on the manuscript. We would also like to thank for their many useful comments, Hector O. Silva, Kostas Glampedakis, Jutta Kunz and Stoytcho Yazadjiev.
 
G.P. is supported by FCT contract IF/00797/2014/CP1214/CT0012 under the IF2014 Programme.

D.D. would like to thank the European Social Fund, the ``Ministry of Science, Research and the Arts'' Baden-Wurttemberg, the Baden W\"urttemberg Foundation and the Bulgarian NSF Grant DFNI T02/6 for the support.

  \end{acknowledgments}

\appendix

\section{Additional material}

In this appendix we will complement the main text with some additional brief material on things that were mentioned in the main text but not elaborated on.

\subsection{Chandrasekhar-Friedman-Schutz (CFS) instability}
\label{sec:app:CFS}

Rotating stars are generally unstable to non-axisymmetric perturbations whenever there is a dissipation mechanism in action. For rotating neutron stars the emission of gravitational waves drives an instability, discovered by Chandrasekhar, Friedman, and Schutz \cite{Chandra1970PhRvL,Chandra1970ApJ,FriedmanSchutz1978ApJ...221,FriedmanSchutz1978ApJ...222,Friedman1978CMaPh}, which is active whenever a mode has a frequency that vanishes or is co-rotating in the frame of inertial observers while it is counter-rotating in the frame that rotates with the star.  

Assume a non-rotating star that has some non-axisymmetric modes. The eigenvalue problem admits solutions with frequencies $\sigma$ for both the ``forward'' and ``backward'' propagating modes. The forward modes have positive angular momentum and the backward modes have negative angular momentum. These modes of the star will emit gravitational waves with positive and negative angular momentum respectively, with respect to inertial observers at infinity. Therefore the modes are being damped by the emission.

When the star rotates with a rotation rate $\Omega$, the modes that are moving opposite to the sense of rotation of the star (and have frequency $\sigma_r$ in the rotating frame) will be dragged in the direction of rotation of the star and the frequency with respect to the inertial observers will be $\sigma_r-m\Omega$. If $\Omega$ or $m$ are large enough, then the mode will appear to be moving forward in the frame of inertial observers and therefore it will emit gravitational waves with positive angular momentum. But the mode has negative canonical angular momentum, i.e., with respect to the fluid the mode is moving backwards therefore the fluid in the presence of the mode has less angular momentum than what it would have without the mode. This means that the mode emits positive angular momentum making its angular momentum even more negative, therefore growing. The mode is unstable and is driven by the emission of gravitational radiation.  

\subsection{Cowling approximation}
\label{sec:app:Cowling}

When studying the non-radial oscillations of polytropic stars \cite{Cowling1941MNRAS},  T.~G.~Cowling introduced the approximation that the perturbations in the gravitational potential can be ignored in the calculation of the frequencies of the higher order fluid oscillations. The Newtonian approach by Cowling was later extended to relativistic stars by McDermott, van Horn, Scholl and Finn \cite{McDermottetal1983ApJ,Finn1988MNRAS}, where in this case the metric perturbations are ignored, since the metric is the analogue of the Newtonian gravitational potential. When the gravitational field is weak, the Cowling approximation gives accurate results for the fluid oscillations but for ``standard'' neutron stars the frequencies can be overestimated up to roughly 30\%. Nevertheless, it gives at least good qualitative results and it is often employed in more complicated cases such as rapid rotation \cite{Gaertig:2008uz,Gaertig:2010kc,Doneva:2013zqa} or alternative theories of gravity \cite{Sotani:2004rq,Staykov:2015cfa}. There is an associated approach to this which is called the ``inverse Cowling approximation'', where the fluid perturbations are assumed to be weakly coupled to the spacetime perturbations. In this approach the spacetime perturbations are studied independently and in this way one can derive the ``spacetime modes'' or ``$w$-modes'' \cite{Kokkotas1999LRR}.

\bibliography{biblio}

\end{document}